\documentclass[11pt]{article}
\usepackage[sc]{mathpazo} %Like Palatino with extensive math support
\usepackage{fullpage}
\usepackage[authoryear,sectionbib,sort]{natbib}
\usepackage[utf8]{inputenc}
\usepackage{lineno}
\usepackage{titlesec}
\titleformat{\section}[block]{\Large\bfseries\filcenter}{\thesection}{1em}{}
\titleformat{\subsection}[block]{\Large\itshape\filcenter}{\thesubsection}{1em}{}
\titleformat{\subsubsection}[block]{\large\itshape}{\thesubsubsection}{1em}{}
\titleformat{\paragraph}[runin]{\itshape}{\theparagraph}{1em}{}[. ]

\usepackage{amsmath}
\usepackage{amsthm, thmtools}
\usepackage{amssymb}
\usepackage{graphicx}   
\usepackage{enumitem}
\usepackage{listings}
\usepackage[hidelinks]{hyperref}
\usepackage{pgfplots}
\usepackage{tikz}
\usetikzlibrary{calc}
\usetikzlibrary{arrows.meta}
\usetikzlibrary{arrows}
\usepackage[makeroom,samesize]{cancel}
\usepackage{bm}

\usetikzlibrary{backgrounds}

\tikzset{
  invisible/.style={opacity=0},
  visible on/.style={alt=#1{}{invisible}},
  alt/.code args={<#1>#2#3}{%
    \alt<#1>{\pgfkeysalso{#2}}{\pgfkeysalso{#3}} % \pgfkeysalso doesn't change the path
  },
}

\tikzset{edge/.style = {->,> = latex'}}

\usepgfplotslibrary{dateplot}

\def\uplineage{
  \draw (1, 2.5) -- (1.5, 3);
  \draw (1.5,3) -- (1.6, 3.3);
  \draw (1.5,3) -- (1.8, 3.);

  \draw[dashed] (1.6, 3.3) -- (1.6, 3.6);
  \draw[dashed] (1.6, 3.3) -- (1.8, 3.4);
  
  \draw[dashed] (1.8, 3.) -- (1.95, 3.2);
  \draw[dashed] (1.8, 3.) -- (1.95, 2.9);
  \draw[rotate around= {-45:(1.55,3.05)}, dashed, purple] (1.55,3.05) ellipse (12pt and 22pt);

  \node[]  at (1.0, 3.3) {\tiny $\textcolor{purple}{L_m}$};

}

\def\lowlineage{
    
  \draw (1., 2.5) -- (1.2, 1.8);
  \draw (1.2, 1.8) -- (1., 1.6);
  \draw (1.2, 1.8) -- (1.3, 1.5);
  \draw (1.2, 1.8) -- (1.5, 1.7);

  \draw[dashed] (1., 1.6) -- (0.9, 1.4);
  \draw[dashed] (1., 1.6) -- (1.05, 1.4);
  \draw[dashed] (1.3, 1.5) -- (1.23, 1.31);
  \draw[dashed] (1.3, 1.5) -- (1.5, 1.3);
  \draw[dashed] (1.5, 1.7) -- (1.7, 1.7);
  \draw[dashed] (1.5, 1.7) -- (1.65, 1.5);

  \draw[rotate around= {15:(1.28,1.58)}, dashed, purple] (1.28,1.58) ellipse (12pt and 25pt);

  \node[] at (0.8, 1.1) {\tiny $\textcolor{purple}{L_1}$};

}

\newcommand{\R}{\mathbb{R}}

\newcommand{\CR}{\mathcal R}
\newcommand{\E}{\mathbb E}

\newcommand{\N}{\mathcal N}

\newcommand{\VV}{\mathcal V}
\newcommand{\W}{\mathcal W}

\usepackage{
nameref,
hyperref,
cleveref,
}

\declaretheorem[name = Proposition, refname = {proposition, propositions}, Refname = {Proposition, Propositions}]{prop}

\usepackage{setspace}

\title{Effects of phylogeny on coexistence in model communities}

\author{Carlos A. Serv\'an$^{1}$, Jos\'e A. Capit\'an$^{2\ast}$, Zachary R. Miller$^{1}$ \& Stefano Allesina$^{1,3}$}

\date{}

\begin{document}

\maketitle

\noindent{} 1. Department of Ecology \& Evolution, University of Chicago, Chicago, IL, USA;

\noindent{} 2. Complex Systems Group. Department of Applied Mathematics, Universidad Polit\'ecnica de Madrid, Av. Juan de Herrera, 6, 28040 Madrid, Spain;

\noindent{} 3. Northwestern Institute on Complex Systems, Evanston, IL, USA.

\noindent{} $\ast$ Corresponding author; e-mail: ja.capitan@upm.es.

\bigskip

\bigskip

\textit{Manuscript elements}: 7 figures in the main text, supplement provided in a separate document, separate repository with data and code.

\bigskip


\bigskip

\textit{Manuscript type}: Major Article. 

\bigskip

\noindent{\footnotesize Prepared using the suggested \LaTeX{} template for \textit{Am.\ Nat.}}

\newpage{}

\section*{Abstract}

Species' interactions are shaped by their traits. Thus, we expect traits -- in particular, trait (dis)similarity -- to play a central role in determining whether a particular set of species coexists. Traits are, in turn, the outcome of an eco-evolutionary process summarized by a phylogenetic tree. Therefore, the phylogenetic tree associated with a set of species should carry information about the dynamics and assembly properties of the community. Many studies have highlighted the potentially complex ways in which this phylogenetic information is translated into species' ecological properties. However, much less emphasis has been placed on developing clear, quantitative expectations for community properties under a particular hypothesis.
  
To address this gap, we couple a simple model of trait evolution on a phylogenetic tree with Lotka-Volterra community dynamics. This allows us to derive properties of a community of coexisting species as a function of the number of traits, tree topology and the size of the species pool. Our analysis highlights how phylogenies, through traits, affect the coexistence of a set of species.
  
Together, these results provide much-needed baseline expectations for the ways in which evolutionary history, summarized by phylogeny, is reflected in the size and structure of ecological communities.

\newpage{}

\section*{Introduction}

Understanding the connections between species' traits, interactions, and evolutionary histories has been an important, but elusive, goal for ecologists. Classic empirical and theoretical results~\citep{gause1932experimental, macarthur1964competition, macarthur1967limiting} engendered the principle of limiting similarity, which holds that the intensities of species interactions are controlled by trait, or niche, similarity, and that coexistence of competing species requires sufficient dissimilarity. It has also long been noted that species' traits, and consequently niches, are strongly influenced by phylogenetic history~\citep{webb2002phylogenies, wiens2010niche}. Together, these two ideas motivate the hypothesis that closely related species should share similar niches, and compete more strongly as a result~\citep{webb2000exploring, webb2002phylogenies}. Under this hypothesis, evolutionary history should predict the strength of species' interactions, and ultimately the likelihood of their coexistence.  While these ideas have found mixed support~\citep{cadotte2017phylogenies}, they serve as cornerstones of the young field of community phylogenetics~\citep{webb2002phylogenies, violle2011phylogenetic}. 

Guided by this logic, many studies have sought to link the phylogenetic structure of communities with empirical patterns of species abundance and coexistence. These efforts rely on a variety of tools developed to test whether a given mechanism of community assembly (e.g., competitive exclusion or environmental filtering) has acted in a community, by analyzing the signal it is expected to leave in the community's phylogenetic structure~\citep{silvertown2001phylogeny, freilich2015phylogenetic}. However, several influential critiques have been leveled at this kind of inference, on the grounds that phylogenetic relatedness might not translate into community patterns in straightforward ways~\citep{mayfield2010opposing, cadotte2017phylogenies}. For example, closely related species might share traits that lead to stronger competition---decreasing the chance they coexist---but also traits that increase their overall competitive ability relative to the broader community, which could simultaneously increase their probability of coexistence. In complex communities, it is also unclear how a large number of pairwise measures of evolutionary relatedness should be related to properties of the whole community, such as total biomass or coexistence.

It is now widely recognized that these and other issues complicate the project of relating evolutionary history to community assembly and co-occurrence, although the extent to which they limit inference has been hotly debated~\citep{mayfield2010opposing, mouquet2012ecophylogenetics, gerhold2015phylogenetic}. However, even leaving aside the potential effects of phylogeny on species' fitness differences, we lack a rigorous understanding of how shared evolutionary history should map into patterns of coexistence and abundance through niche overlap. A central prediction in community phylogenetics is that competitive exclusion should ``prune'' closely related species, producing a pattern of phylogenetic overdispersion~\citep{webb2000exploring, webb2002phylogenies}. But it is rarely clear how strongly or on which phylogenetic scales this pattern should manifest~\citep{swenson2006problem}. Additionally, phylogenetic structure might affect overall community richness (number of coexisting species), as well as patterns of biomass and abundance~\citep{Kraft07, cadotte2010phylogenetic}. While these patterns have been extensively scrutinized in empirical systems, we have surprisingly little quantitative theory to understand and predict them.

Here we take a step back to develop and analyze a quantitative model that helps clarify these relationships. Given a phylogeny summarizing the evolutionary history of a community, we aim to develop predictions for key ecological community properties. The link between phylogeny and community properties is mediated by ecologically-relevant traits, which we treat as the outcome of a stochastic evolutionary process. Thus, individual trait values are random variables in our framework, but systematic relationships between phylogeny and community properties emerge due to phylogenetic correlations between species' traits. 

More specifically, we use the framework of the well-known Lokta-Volterra model to construct an explicit link between phylogenetic relatedness and ecological interactions. We first connect phylogeny to species' traits, and then connect similarity in traits to the strength of interaction between any two species~\citep{bastolla2005biodiversity,maynard2018network}. Given a phylogenetic tree representing the evolutionary history of a regional pool of $n$ species, we assume that species interactions are determined by a set of $\ell \geq n$ traits, which have evolved independently on the tree via Gaussian processes such as Brownian motion~\citep{Kraft07, harmon2018phylogenetic} or Ornstein-Uhlenbeck processes~\citep{hansen1996translating}. The covariance between these traits controls the strength of the competitive effect between any two species. In this way, species that are more closely related tend to interact, on average, more strongly with each other than with distantly-related species. As we will show, the variance of the distribution of interaction strengths is controlled by the number of relevant traits $\ell$.

As noted above, species' overall competitive abilities, captured in our framework by intrinsic growth rates and self-regulation, could also reflect their evolutionary history, and exert effects on community patterns that are distinct from the effects of niche differences. While both kinds of phylogenetic effects (on competitive ability and niche differences) are likely present in real communities, we restrict our focus to niche differences, with the aim of developing clear, quantitative expectations for communities shaped primarily by limiting similarity. To separate the effect of phylogeny on interspecific interactions from its effect on overall competitive ability, we therefore assume that all species have identical intrinsic growth rates, and the same self-regulation (carrying capacity) in expectation~\citep{belyea1999assembly} (although we later relax this assumption by considering the effect of varying intrinsic growth rates, see Supplement, Section S7).

Having established a probabilistic model for trait evolution and a link between trait values and species interactions, we study a scenario where all species from the (regional) pool are present at arbitrary initial conditions in a local community, and dynamics follow the (generalized) Lotka-Volterra model. Unlike previous simulation-based studies~\citep{Kraft07,freilich2015phylogenetic}, we develop an analytical framework to characterize the resulting community of coexisting species, as a function of both the number of traits, $\ell$, and the tree structure. We focus on three biologically relevant quantities: community diversity, community biomass, and the abundance distribution. Having clear predictions for how these quantities depend on phylogeny is a key prerequisite to properly testing for phylogenetic structure in empirical communities. Our results also provide a potential way to infer important parameters, such as the number of traits $\ell$ that are relevant for species interactions in a natural community, as well as the phylogenetic tree structure most compatible with ecological interactions. Testing whether a community phylogeny inferred in this manner is concordant with molecular phylogeny, for example, could illuminate the evolutionary determinants of ecological interactions. Overall, our model clarifies how phylogenetic relatedness, modulated by the number of traits that
control species interactions, should be expected to shape multiple aspects of community assembly and structure.  

\section*{Model}

We consider a regional pool $\CR = \{s_i\}$ of $n$ species indexed by $1\leq i \leq n$, with a phylogeny $T_{\CR}$ describing the evolutionary history of the pool. Each species in the pool is defined by a set of fixed traits that have evolved over time. We focus on the diversity and ecological structure of a local community formed from the regional pool. Therefore, we assume that evolutionary processes, which have taken place in the pool, are separate from population dynamics in the local community, which occur on shorter time scales.

Each species is characterized by $\ell$ trait values, with $\ell \geq n$. For a given trait $k$, with $1 \leq k \leq \ell$, the values of $k$ for all species in the pool ($1 \leq i \leq n$) are collected in the trait vector $\bm{\tau}^k=(\tau_i^k)$. We assume each trait vector $\bm{\tau}^k$ is sampled independently from a multivariate normal distribution $\N(\bm{\mu}^k,\Sigma)$, with mean vector $\bm{\mu}^k$ and correlation matrix $\Sigma$. These assumptions imply that: (a) the values of distinct traits of a given species are independent, with no trade-offs or correlations between traits; and (b) the evolutionary processes for distinct traits are statistically equivalent. 

Because many functional traits of organisms are correlated with one another, traits in our model should be viewed as idealized trait values (uncorrelated at the species level), and $\ell$ as the effective number of independent traits relevant for interactions~\citep{laughlin2014intrinsic,mouillot2021dimensionality}. For instance, while we might measure concrete physical traits such as specific leaf area or leaf nitrogen content in plants, relationships between these traits make it more appropriate to consider abstract trait combinations, such as leaf investment (e.g. position along a leaf economics spectrum)~\citep{wright2004worldwide} as the relevant traits in our model. As another example, we discuss below (and illustrate in the Supplement, Section S1) how our model can be derived from underlying consumer-resource dynamics, where the $\ell$ traits are consumers' preferences or consumption rates for $\ell$ substitutable resources. Again, these rates may be high-level traits, representing a suite of physical adaptions to using different types of resources.

Each sampling of trait vectors defines a particular regional pool realization, all of them preserving the correlation structure $\Sigma$ among species. This correlation structure reflects the evolutionary history of the community, as we will describe below. The trait vectors in turn determine interspecific interactions in the local community. 

Drawing trait vectors from a multivariate normal distribution is equivalent to modeling stochastic evolution (with or without selection) of each trait on the phylogenetic tree $T_{\CR}$~\citep{hansen1996translating}. In our model, the tree structure is treated as a parameter, and we model the distribution of trait evolutionary trajectories compatible with the tree. Of course, the processes of speciation (generating the phylogeny) and trait evolution (generating the trait vectors) happen in concert, but in practice we often have access to a community phylogeny while lacking detailed knowledge of functional traits that control interactions. Thus, we aim to connect phylogeny to ecological properties by considering an ensemble of possible trait realizations compatible with a known phylogeny.

\begin{figure}[!t]
\centering
\includegraphics[width = \textwidth]{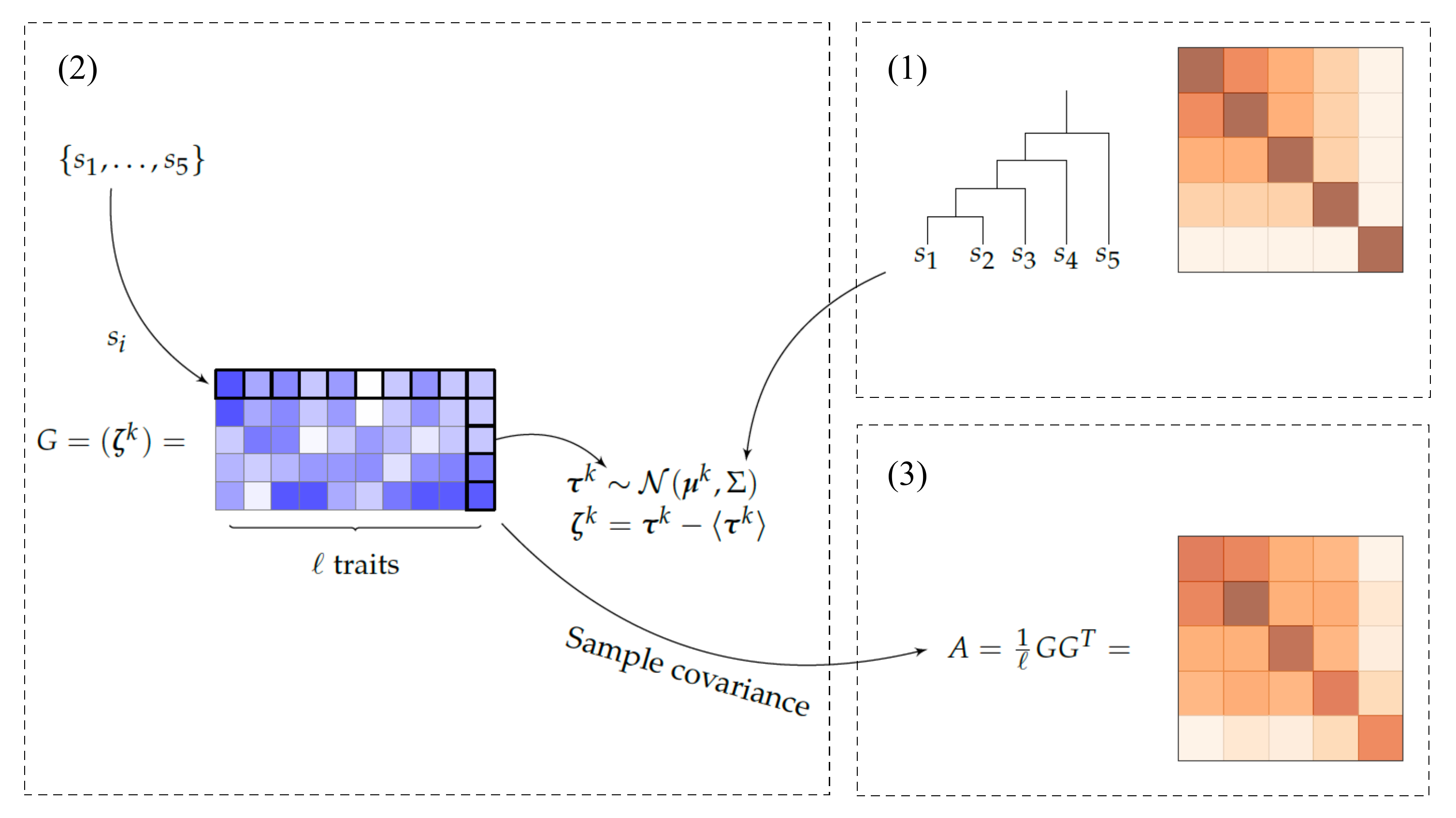}
\caption[Construction of the regional pool $\CR$ and interaction matrix $A$.]{\textbf{Construction of the regional pool $\CR$ and interaction matrix $A$}. (1) Matrix $\Sigma$ measures the shared evolutionary history between any two species $s_i$ and $s_j$ on the ultrametric tree $T_\CR$. (2) Each species in the pool $\CR$ is assigned $\ell$ trait values. The vector recording each species' value for trait $k$, $\bm{\tau}^k$, is sampled from $\N(\bm{\mu}^k,\Sigma)$, independently of all other traits. (3) The model then relates these trait values, contained in the matrix $G$, to the interactions between the species in the pool. Ultimately, the matrix of interactions $A$ is a noisy reflection of $\Sigma$, with the amount of noise determined by the number of traits, $\ell$.}
\label{fig:1}
\end{figure}

The phylogenetic tree defines a variance-covariance matrix $\Sigma$~\citep{harmon2018phylogenetic}, where each element $\Sigma_{ij}$ measures the shared evolutionary history (branch length) between species $s_i$ and $s_j$ (see~\Cref{fig:1,fig:2})~\citep{bravo2009estimating}. Assuming the tree $T_\CR$ is ultrametric, we can take $\Sigma_{ii} = 1$ for all $i$. Off-diagonal elements of the covariance matrix are computed as follows: For any $s_i$ and $s_j$, consider the paths ``backwards'' in time from each of these species to the ancestral species at the root of the tree; the time $t_{ij}$ at which these paths merge is the coalescence time between $s_i$ and $s_j$~\citep{wakely2016coalescent}. Then, $\Sigma_{ij} = 1 - t_{ij}$. In other words, $\Sigma_{ij}$ is the total time over which trait evolution for $s_i$ and $s_j$ was shared (see~\Cref{fig:2}).

%We call the times between branching events, $t_i - t_{i-1}$, inter-branching times. 

The simplest example of an evolutionary process consistent with these assumptions is one where each trait $k$ has an ancestral mean value of $0$ and evolves independently on the tree via Brownian motion. Then the value of trait vector $\bm{\tau}^k$ at the $n$ tips follows a multivariate normal distribution $\N(\bm{0},\Sigma)$ with $\Sigma$ generated by the tree. More generally, however, \cite{hansen1996translating} showed that any linear diffusion evolutionary process leads to a multivariate normal distribution for trait vectors at the tips of the tree. These so-called Gaussian processes include, in addition to Brownian motion, the well-known Ornstein-Uhlembeck (OU) process widely used as a model in evolutionary phylogenetics. Unlike Brownian motion, the OU process leads to non-zero expected values for traits, $\bm{\mu}^k\ne \bm{0}$, which can be interpreted as a selective force that pushes trait evolution toward an optimal value. Independent OU processes for each trait are also consistent with assumptions (a) and (b) above. In general, any linear diffusion process that evolves traits over the phylogenetic tree is compatible with our approach.

Next, we consider how a local community is formed from the regional pool. We model a scenario where all species from the pool enter some local habitat at the same time and at arbitrary initial densities~\citep{servan2018coexistence}. Population dynamics, as determined by the species' interactions and growth rates, will lead the community to an asymptotic state in which some of the species are locally excluded, while others coexist. To describe these local dynamics, we employ the Generalized Lotka-Volterra (GLV) model:

\vspace*{-2mm}
\begin{equation}\label{eq:covariance_model}
  \frac{d x_i}{dt} = x_i\bigg[r - \sum_{j=1}^n (\mu + A_{ij}) x_j\bigg].
\end{equation}
Here, $x_i$ is the density of species $i$ and $r$ is the intrinsic growth rate, assumed to be equal for all species. This assumption reflects our focus on species niche differences mediated by phylogenetic relationships~\citep{mayfield2010opposing}. However, in the Supplement (Section S7) we also consider variability in growth rates. 

\begin{figure}[!t]
\input{trees_picture}
\caption[Examples of ultrametric rooted phylogenies and induced correlation matrices]{\textbf{Examples of ultrametric rooted phylogenies and induced correlation matrices}. The perfectly unbalanced tree (left) has $n-1$ branching times $0 < t_1 < \ldots < t_{n-1}$ for a pool of $n$ species, where each new branching occurs in the ``leftmost'' lineage. The star tree (middle) exhibits a unique branching event which generates all $n$ species. For the perfectly balanced tree (right) we have $k = \log_2(n)$ branching times; at each branching time, all the tips present up to that point generate two new species (in this case $n = 2^k$ for some positive integer $k$). Then the correlation matrix $\Sigma$ associated with each tree is constructed (darker colors indicate larger values of $\Sigma_{ij}$).}
% For the simplest case of the star tree, this procedure yields $\Sigma_{ij} = \rho$ for any $i\not=j$ and $\Sigma_{ii} = 1$.}
\label{fig:2}
\end{figure}

Interaction coefficients are modeled as deviations $A_{ij}$ from a mean value $\mu > 0$. These deviations are controlled by trait similarity between species. In particular, we assume $A_{ij}$ is proportional to the sample covariance between the traits of $s_i$ and $s_j$. We define $\bm{\zeta}^k=\bm{\tau}^k-\langle \bm{\tau}^k \rangle = \bm{\tau}^k - \bm{\mu}^k$ to be the vector of centered trait values for the $k$-th trait, where the mean trait value across all species is removed. Then the matrix $A$ is given by

\vspace*{-2mm}
\begin{equation}\label{eq:A}
A_{ij} = \frac{1}{\ell} \sum_{k=1}^{\ell} \zeta_i^k \zeta_j^k,  \qquad
A = \frac{1}{\ell} GG^T.
\end{equation}
Here $G = (\zeta_i^k)$ is simply the species-by-trait matrix of centered trait values. Under this definition, the deviations $A_{ij}$ can be understood as the overlap between species trait deviations, calculated as the dot product between species' trait deviation vectors $\bm{\zeta}_i=(\zeta_i^k)$ and $\bm{\zeta}_j=(\zeta_j^k)$ (see~\Cref{fig:1}). Conceptually, this model dictates that pairwise species interactions are strengthened ($\vert A_{ij}\vert$ larger) if their trait deviation vectors are nearly parallel, and weakened as pairs of trait deviation vectors become more perpendicular. 

In the Supplement (Section S1), we show how the model defined by Eqs.~\eqref{eq:covariance_model}--\eqref{eq:A} can arise, for example, by assuming a separation of time scales for a consumer-resource model in which consumers share the same attack and death rates, but differ in their preferences for resources, which reflect evolutionary history. In this case, the number of traits $\ell$ takes the concrete meaning of the number of resources utilized by the $n$ consumers. In addition, the rows of the matrix $G$ correspond to resource preferences for each consumer -- these are the $\ell$ traits. The trait evolution process implies that more closely related consumers typically share more similar resource preferences and compete more strongly as a result.

Our model assumptions imply that $A$ is a symmetric and stable matrix belonging to the Wishart ensemble~\citep{wishart1928generalised,muirhead2009aspects}:

\vspace*{-2mm}
\begin{equation}\label{eq:Wishart}
A\sim \W_n(\ell^{-1}\Sigma, \ell).
\end{equation}
The Wishart distribution describes the probability of observing a given sample covariance matrix when sampling vectors from a zero-mean multivariate normal distribution. As described above, the trait vectors $\bm{\tau}^k$ are multivariate normal samples -- and thus $\bm{\zeta}^k$ are zero-mean normal samples -- for any Gaussian evolutionary process, such as Brownian motion or OU processes. As a sample from the Wishart ensemble, $A$ may contain both positive and negative elements. Because interactions are symmetric, the model accounts for a mixture of competitive interactions, if $A_{ij} = A_{ji}$ are positive (note the minus sign in the GLV dynamics), and facilitation, if $A_{ij}=A_{ji}$ are sufficiently negative. Given its many applications in statistics and other fields, the Wishart distribution has been studied extensively, allowing us to draw upon a large body of results to characterize the ecological dynamics in our model ~\citep{muirhead2009aspects,bodnar2008properties,bodnar2011product,kotsiuba2016asymptotic}.

The stability of $A$ has an important consequences for community assembly. As the GLV dynamics unfold, the community reaches a unique, globally-stable equilibrium, where some species go extinct, and the sub-community of coexisting species is characterized by feasibility and non-invasibility conditions~\citep{hofbauer1998evolutionary} (see also Section S3). The GLV dynamics lead to a unique final community where all surviving species have positive abundances (feasibility) and all of the excluded species have negative invasion growth rates (non-invasibility). Furthermore, the final community composition reached in our scenario where all species in $\CR$ are introduced simultaneously is the same as would be reached under sequential, one-at-a-time species invasions~\citep{servan2021tractable}. Thus, although we study simultaneous species invasions for simplicity, our results apply directly to the potentially more realistic process of bottom-up community assembly. 

In this setting, one can also prove that the effect of the mean interaction strength $\mu$ on the resulting community is very straightforward: $\mu$ does not affect the identity of the coexisting species, and simply rescales their densities by a constant (see Section S6 for details). Similarly, any choice of $r > 0$ only rescales equilibrium densities. Thus, without loss of generality, we can assume $\mu = 0$ and $r = 1$ so that the regional pool is completely characterized by the sample covariance matrix $A$.

Having established a simple model linking phylogeny to trait covariances to interaction strengths, our goal is to characterize the statistical properties of the equilibrium local community. To derive the distributions of richness, biomass, and relative abundances in this final community, as a function of the regional pool phylogeny, $T_\CR$, we study equilibrium solutions of Eqs.~\eqref{eq:covariance_model}--\eqref{eq:A}, imposing the feasibility and non-invasibility conditions.  

Below, we focus on results related to either arbitrary or idealized tree structures. To illustrate how these theoretical results apply to empirical tree structures, we also parameterize our model with phylogenies from an experimental grassland community used in the Biodiversity II experiment at Cedar Creek \citep{tilman2001diversity}, and for 94 species in the clade \emph{Senna} (Fabales; \citealt{Weber16442}) (see Supplement, Section S8). 

\section*{Results}

Within this framework, a particular phylogenetic tree defines an ensemble of regional pools with population covariance matrix $\Sigma$, from which we imagine sampling different pool realizations, each with distinct trait values leading to a sample covariance matrix $A$, according to Eq.~\eqref{eq:A}. For each pool, we obtain a stable local community according to the assembly procedure described above. Our basic aim is to answer the question: In local communities, once ecological dynamics have reached a steady state, what values would one expect, averaging over pool realizations, for fundamental ecological quantities? In particular, we focus on the number of species that coexist in the local community, the total community biomass, and the relative abundance distribution. We would like to understand how these properties depend on the tree (encoded in $\Sigma$), the number of traits, and the size of the pool. By deriving analytical predictions for how community size and structure depend on phylogeny, we provide a clear set of expectations for how shared evolutionary history shapes ecological dynamics, and lay a firm theoretical foundation for empirical tests of phylogenetic effects in ecological communities.

We consider three scenarios of increasing complexity. First, we consider the limit in which the number of traits, $\ell$, is very large relative to the size of the pool, $n$. Let $\gamma = \ell/n$ be their ratio. We call this situation the ``deterministic limit'', because in the limit $\gamma \to \infty$ we find that the sample covariance matrix $A$ converges to the population covariance matrix $\Sigma$, which is fixed. Thus, the properties of the community are determined solely by $\Sigma$, and there is no randomness.

Second, we let $\ell$ be finite and examine how varying $\ell$ and shared evolutionary history interact to shape community properties. In this case, $A$ is a random matrix, requiring a more complex analysis. Thus, to make the problem tractable, we consider the simplest non-trivial phylogeny: the ``star'' tree, where all species split from the ancestor at a single branching point. In this case, $\Sigma$ has a correspondingly simple structure, with $\Sigma_{ij}=\rho$ for $i\ne j$ (see \Cref{fig:2}). 

Finally, we consider more general phylogenetic structure with a finite number of traits. Here, we present results for small pools and an example empirical phylogeny, and in the Supplement we show how to calculate community properties for arbitrary phylogenetic trees.

\subsection*{Deterministic Limit} As the number of traits, $\ell$, becomes large relative to the size of the regional pool, $n$, the variance in interaction strengths decreases. Intuitively, as trait overlap depends on more and more traits, each evolving independently, the relationship between trait overlap and shared evolutionary history becomes more consistent, and less dependent on the stochastic trajectory of any single trait. In the limit $\gamma \to \infty$, the variance in niche overlap (and consequently interaction strengths) drops to zero, and each realization of the matrix $A$ becomes identical to $\Sigma$.

\noindent\textbf{Species coexistence.} Remarkably, in this limit we show that all members of the pool coexist in the local community, regardless of the tree topology or the size of the pool. This surprising behavior can be proved inductively. First, consider a very simple evolutionary scenario where all $n$ species diverge at time zero. In this special case, there is no shared evolutionary history, so the matrix $\Sigma$ is the identity matrix $I$. Coexistence of all $n$ species in the pool follows trivially, since $A_{ij} = 0$ for all $i \neq j$ and species do not interact with one another. Next, we recall that adding a constant value to the interaction matrix does not change the set of coexisting species (their densities are simply re-scaled). This corresponds to an evolutionary scenario where all species split at some time $t$, rather than time zero -- producing a ``star tree'' phylogeny (see \Cref{fig:2}). In this scenario, too, all species will coexist. Finally, we take the induction step: In an arbitrary tree, if $t_1$ is the time of the first branching event, then ``cutting'' the tree at this branching point generates two (or more) non-interacting sub-trees. Under the induction hypothesis, each of these sub-trees corresponds to a coexisting subset of species. ``Pasting'' these sub-trees together at the their roots preserves coexistence, since the sub-trees are still non-interacting (i.e., the corresponding interaction matrix has zero values for any pair of species not in the same sub-tree). We recover the full tree by adding branch length $t_1$ to the root. In terms of the interaction matrix, this amounts to adding a constant $t_1$, which does not change the set of coexisting species. Because any tree can be sequentially decomposed in this manner into a collection of star trees, we find that all species coexist, regardless of the full tree topology (see Figure~S1 and Section S2 for more details).

\noindent\textbf{Total biomass and abundance distribution.} In contrast to coexistence, which is guaranteed in this limit for any phylogeny, phylogenetic structure strongly influences the biomass and relative abundance distribution of a community. As illustrative examples, we consider two extreme tree topologies given by the ``perfectly unbalanced'' tree and the ``perfectly balanced'' tree (\Cref{fig:2}). In a perfectly unbalanced (or ``pectinate'') tree, only one lineage continues to speciate after each branching event. In a perfectly balanced tree, every extant lineage splits simultaneously at each branching event. These two topologies bookend the space of possible tree shapes~\citep{kirkpatrick1993searching}. For these two idealized cases, we are able to derive simple expressions for the individual biomass $x_i$ of each species $s_i$, where the index corresponds to the position in the ordered tips of the tree. In the deterministic limit, the vector of equilibrium abundances $\bm{x}=(x_i)$ satisfies the linear system

\vspace*{-2mm}
\begin{equation}\label{eq:abun}
    \Sigma \bm{x}=\bm{1},
\end{equation}
for $\bm{1}$ a vector of ones, and where $\Sigma$ is the correlation matrix of the tree as defined above. We consider the total community biomass $W(n)=\sum_{i=1}^n x_i$, which depends on $n$ because the local community contains all members of the pool. The relative abundance distribution is then given by the vector $\bm{x}/W(n)$.

\begin{figure}[!t]
\input{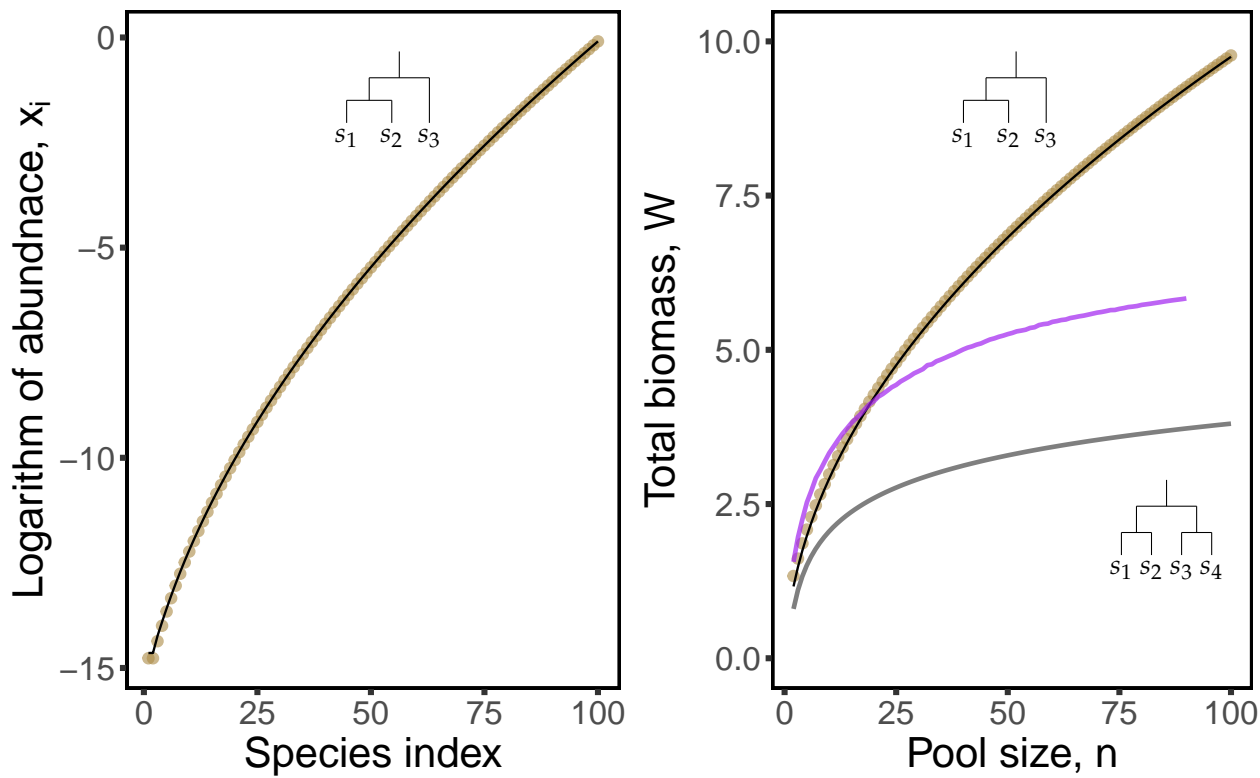}
\caption[Individual abundance and total biomass in the deterministic limit]{\textbf{Individual abundance and total biomass for the deterministic limit}. Log individual abundance (left) and total abundance (right) for communities with extreme tree topologies (depicted, for small pool sizes, next to each curve). Yellow dots mark the average values over simulations for a perfectly unbalanced tree with equal branch lengths. Black solid lines are the corresponding analytic predictions. In the right panel, the gray line represents the analytic formula for a perfectly balanced tree, $W(n) = (\log_2(n) + 1)/(2 - 1/n)$, which shows logarithmic growth. The purple line shows the scaling with size of total biomass averaged over $2000$ sub-communities sampled from the full \emph{Senna} phylogenetic tree (using equal inter-branching times for every sub-tree).}
\label{fig:3}
\end{figure}

Assuming equal time between each branching event (inter-branching times), the total biomass associated with a perfectly unbalanced tree is given by $W(n) \approx \sqrt{n} - 1/4$. In the perfectly balanced case, $W(n) = \frac{\log_2(n) + 1}{2 - 1/n}$ (see Section S2 for derivations). For the perfectly balanced case, each species necessarily has the same abundance, $x_i=W(n)/n$, by symmetry. On the other hand, the hierarchical nature of the perfectly unbalanced tree is reflected in the individual biomasses, with species that split from the rest early on having much higher abundances (\Cref{fig:3}). In Section S2, we show that these results are qualitatively unchanged if inter-branching times are exponentially or uniformly distributed, instead of constant. 

Interestingly, these results immediately indicate that asymmetric evolutionary histories promote higher community productivity: as a function of pool size $n$, total biomass in the perfectly unbalanced case (which grows as $\sqrt{n}$) is always greater than for the perfectly balanced case (which grows logarithmically). The uneven distribution of abundances for the perfectly unbalanced tree helps explain why total biomass is greater in this case: as $n$ grows, the early-diverging species interact less and less strongly with the rest of the community, so their abundance approaches carrying capacity (i.e., $x_i = 1$). In contrast, in the perfectly balanced case the abundance of all species is the same, equal to $W(n)/n \approx \log_2(n)/(2n)$, far less than $1$.

We also calculated the biomass scaling relationship and relative abundance distribution for an empirical tree topology in the deterministic limit by solving the linear system~\eqref{eq:abun} using the \emph{Senna} clade (and sub-clade) topology with equal inter-branching times. The \emph{Senna} phylogenetic topology falls between the two extremes of the perfectly unbalanced and balanced trees, but the same qualitative patterns emerge using this tree to parameterize the model (Figure 13). Averaging over the different topologies for sub-clades of a given size (always maintaining equal inter-branching times), we obtained an average relationship between biomass and pool size, $n$, which interpolates between the two extreme cases analyzed above for the perfectly balanced and unbalanced trees (\Cref{fig:3}, right panel).

\subsection*{Star phylogenies} 

As another informative case, we consider the simplest non-trivial phylogenetic structure -- the star tree -- and allow the ratio of traits to pool size, $\gamma = \ell/n$, to vary. Specifically, in this scenario, all $n$ species diverge at time $0 \leq \rho < 1$, so that $\Sigma_{ij} =\rho$ for all $i \neq j$. Thus, $\rho$ measures the amount of shared branch length in the phylogeny (see~\Cref{fig:2}). Now interaction strengths -- and thus all community properties -- become random variables, and we aim to characterize their distribution as a function of the amount of shared evolutionary history, $\rho$ (i.e., shared branch length), number of traits, $\ell$, and pool size, $n$.

\noindent\textbf{Species coexistence.} We have just seen that in the limit $\gamma \to \infty$, all $n$ species will coexist for any $\rho$. At the opposite limit, as $\gamma$ approaches 1, we know from classical ecological theory that at least $\ell = n$ traits must underpin interactions in order for $n$ species to coexist stably. This is the competitive exclusion principle~\citep{levin1970community,yodzisintroduction}, which states that there must be at least as many resources or regulating factors as species in any stably coexisting community (to make this parallel more concrete, recall the equivalence of our model framework with a standard consumer-resource model, Section S1). Now we ask: How does the fraction of coexisting species vary as $\ell$ ranges from $n$ to $\infty$? To answer this question, we exploit the fact that the interaction matrix $A$ is a sample covariance matrix following the Wishart distribution, so we can draw on tools developed in statistics and economics to explore how the limit of full coexistence is approached (see Section S3 for mathematical details).

For the star tree with $\rho > 0$, we find that when the number of traits $\ell$ is comparable to the number of species $n$ (i.e., $\gamma = \ell/n \approx 1$), full coexistence is almost never achieved for large enough communities (Figure~S3). Thus, while coexistence is guaranteed if interactions closely mirror phylogeny, so that $A = \Sigma$, when $A$ is a noisy sample from the Wishart ensemble with $\ell \approx n$, coexistence of the entire pool becomes highly unlikely.  

\begin{figure}[!t]
\includegraphics[width = \textwidth]{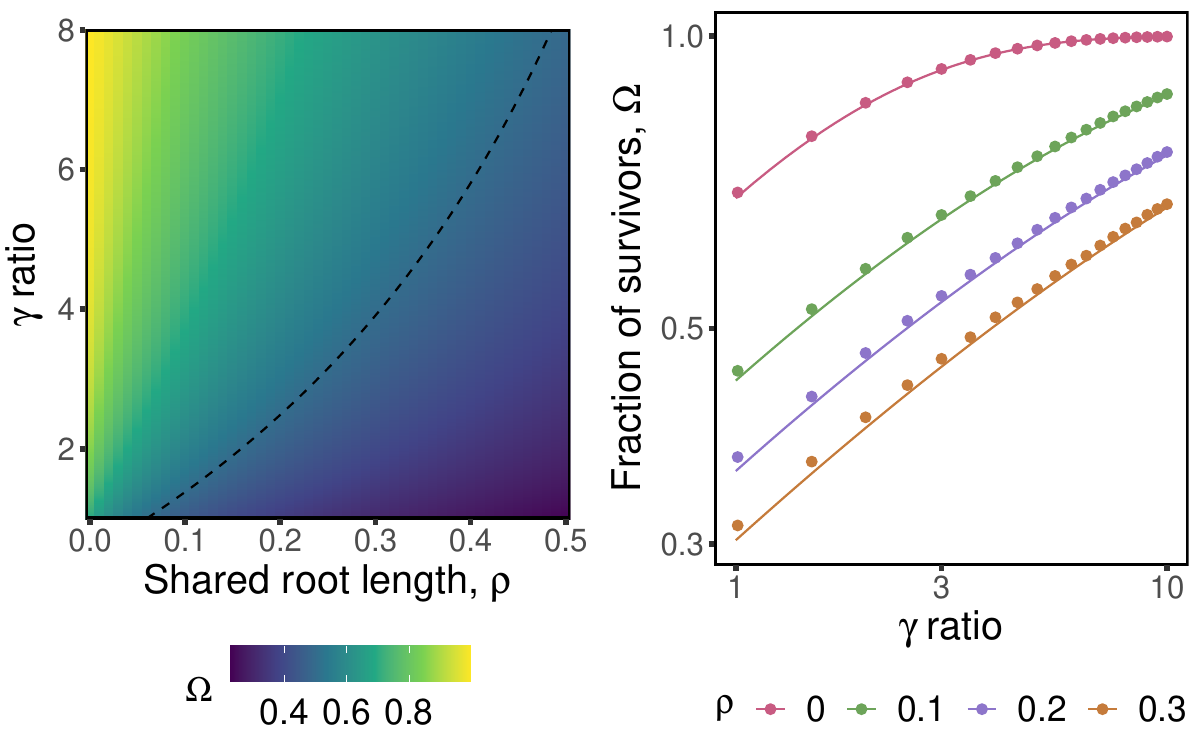}
\caption[Proportion of coexisting species $\Omega$ as a function of the shared branch length $\rho$ and the trait-species ratio $\gamma$ in star trees]{\textbf{Proportion of coexisting species $\Omega$ as a function of the shared branch length $\rho$ and the trait-species ratio $\gamma$ in star trees}. In the left panel, the dashed line marks parameters for which half of the species coexist in expectation. The ratio $\gamma=\ell/n$ needed to obtain a fixed $\Omega$ increases sharply with correlation $\rho$ (Eq.~\eqref{eq:half_species}). In the right panel, analytical approximations (solid lines, Eq.~(S108)) are compared with simulations (dots) for $n=50$ species (log-log scale). The classical competitive exclusion principle predicts that full coexistence is possible above $\gamma = \ell/n = 1$; we find that only a fraction of species survive, however, with fewer coexisting species as $\rho$ becomes larger.}
\label{fig:4}
\end{figure}

Nevertheless, the community does not collapse completely, and a non-vanishing fraction of species typically coexists in these cases~(\Cref{fig:4}). This fraction is greater than zero but less than one, demonstrating that $\ell$ traits are generally insufficient to support $\ell$ coexisting species, in contrast to a naive expectation based on the competitive exclusion principle (see also \citealp{cui2020effect}). The precise coexistence fraction, $\Omega$, depends on the proportion of shared evolutionary history, $\rho$. More shared history increases the correlation among species' traits, and therefore the strength of their interactions, reducing the fraction of species that are expected to coexist. In the Supplement, we derive a very good approximation (Eq.~(S108)) for $\Omega$ as a function of $\gamma$ and $\rho$. This relationship is illustrated in \Cref{fig:4}. Our theory shows, for example, that to observe at least half of the species coexisting (in expectation), these parameters must satisfy:

\vspace*{-2mm}
\begin{equation}\label{eq:half_species}
  2\gamma \geq 1 + \frac{n \rho}{\pi (1 - \rho)}.
\end{equation}
The quantity $\xi = \frac{\rho}{1-\rho}$ is the ratio of shared to unshared phylogenetic history (branch lengths) for any two species. It is a key quantity, in the sense that any two distinct pools $\CR$ and $\CR'$, of sizes $n$ and $n'$ will yield the same mean fraction of coexisting species for a given $\gamma=\ell/n$ if $n \xi = n'\xi'$.\bigskip

\noindent\textbf{Total biomass and abundance distribution.} As the ratio of traits to species, $\gamma$, and the trait correlation, $\rho$, vary, so does the distribution of total biomass $W$. Naively, one might expect that total biomass scales in a straightforward way with the number of coexisting species, following the relationships discussed above. However, the distribution of total biomass depends on $\gamma$ and $\rho$ in non-trivial ways even after conditioning on local community size. To explore these relationships, we derived an approximation for the mean of $W$, using the assumption that number of coexisting species is usually close to the mode (valid for large $n$). This approximation is given by Eq.~(S115) in the Supplement, and it closely matches results from simulations (see Section S4 and Figure S5 for exact results and the full distribution). 

\begin{figure}[!t]
\includegraphics[width = \textwidth]{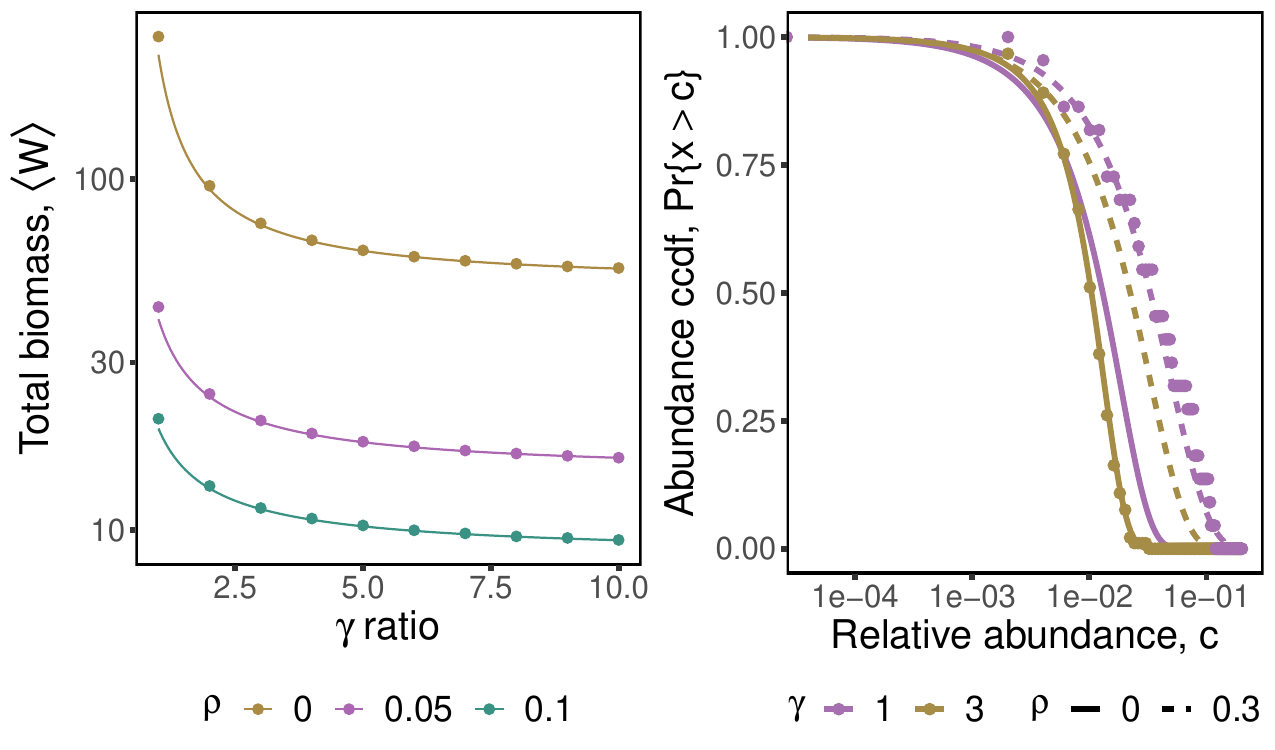}
\caption[Mean total biomass and relative abundance distribution for star phylogenies]{\textbf{Mean total biomass and relative abundance distribution for star phylogenies}. Left panel shows the total biomass, averaged over interaction matrix realizations, for the community of coexisting species (note the log-scale for the y-axis). Points represent simulations, and solid lines the corresponding analytical approximations for a pool of $n=50$ species (see Section~S6 for the effect of changing $\mu$). The survival function (complement of the cumulative distribution function) for relative abundance is plotted on the right panel (note the log x-axis), where again points stand for simulations and lines for analytical predictions (both for $n=100$). For clarity, we only show simulations for the parameters $\rho \in \{0, 0.3\}$ and $\gamma \in \{1,3\}$.}
\label{fig:5}
\end{figure}

We can understand the effect of $\gamma$ and $\rho$ on total community biomass by considering how these parameters affect the distribution of interaction strengths. Increasing $\rho$ increases the mean inter-specific interaction strength, driving a decrease in both the fraction of species that survive and the average total biomass (\Cref{fig:5}). The effect of $\gamma$ is more subtle. As we have discussed, $\gamma$ effectively controls the variance of the distribution of interaction strengths. When $\gamma$ is large the variance is small, and all interactions are competitive. For sufficiently small $\gamma$, interaction strengths are more variable, allowing for positive interactions, which greatly enhance total biomass. This shift in the probability of positive interactions drives a decrease in $W$ with increasing $\gamma$, even as the fraction of surviving species grows (\Cref{fig:5}).

We can similarly derive approximations (Eq.~(S120) in the Supplement, Section S5) for the cumulative distribution function of relative abundance under distinct values of $\rho$ and $\gamma$. The complement of this function, giving the proportion of species with abundance greater than a given value, is shown in \Cref{fig:5}, compared with simulations. This distribution becomes very peaked as $\gamma$ increases, consistent with the convergence to the deterministic limit, where all species are identical~(\Cref{fig:5}). Increasing $\rho$, however, tends to make the distribution flatter, even while decreasing overall biomass (compare panels in \Cref{fig:5}). Thus, with more shared evolutionary history, $\rho$, species abundances become smaller but much more variable, as a consequence of stronger interactions. As $\rho$ decreases, on the other hand, species interact more weakly and all species approach the same abundance.

\subsection*{More general tree structures} 

Considering more general tree structures, equivalent to imposing a more general covariance structure $\Sigma$, is challenging from a mathematical standpoint, due to the breaking of the statistical equivalence among species -- species in distinct parts of the tree have now different statistical properties. It is no longer straightforward to derive simple relationships between community properties and summary statistics for the tree. However, for a given tree structure we can numerically evaluate the probability of observing any particular sub-community using formulae derived in the Supplement. In particular, the probability that a particular subset of species forms the equilibrium local community can be found as the product of the probability of feasibility of the sub-community, Eq.~(S33), and the probability of non-invasibility by species not in the sub-community, Eq.~(S63). Moreover, we can also numerically evaluate Eq.~(S113) to obtain the average biomass for each species, as a function of the covariance matrix $\Sigma$ and the number of traits $\ell$, in a coexisting sub-community. Similarly, Eqs.~(S116)--(S117) yield the relative abundance cumulative distribution function for each individual species in a specific sub-community. 

Evaluating these formulae amounts to computing multidimensional Gaussian integrals, which can be done efficiently~\citep{genz1992numerical}. Therefore, key quantitative features can be calculated numerically for arbitrary tree structures, beyond the constant correlation case. These formulae could be used to investigate a variety of questions about how evolutionary history translates into ecological structure, removing the need to numerically integrate the model dynamics, which is computationally prohibitive for large communities. These formulae could also underpin statistical inference: from abundance or diversity data one could use our results to infer an effective number of traits (relevant to species interactions) or even the structure of the pool phylogeny by fitting these parameters to data.

To illustrate these ideas, in the the Supplement, Section~S8, we present an exploratory analysis of empirical diversity data and tree structures using our theoretical framework. We used data from the Biodiversity II biodiversity-ecosystem function experiments~\citep{tilman2001diversity}, which measured realized diversity (species richness) across different realizations of a pool of plant species. Following~\cite{lemos2024phylogeny}, we considered a subset of plots with minimal contamination by species outside of the species pool. Furthermore, we considered only plots seeded with the full pool of species, to match our theoretical scenario. The \emph{V.PhyloMaker} package in R~\citep{jin2019v} was used to obtain a phylogenetic tree for the species pool (which varied slightly across different years of the experiment). By averaging realized diversity across plots (experimental realizations of the local community), we estimated the proportion of coexisting species $\Omega$ and used it to estimate the number of traits $\ell$ characterizing each community. We find that the number of inferred traits varies from year to year, but the trait-species ratio $\gamma$ is typically between 1 and 10.

We then used the phylogenetic trees for the Biodiversity II species pools to test how well an ``averaged'' tree structure such as the star tree, captures the behavior of a more complex empirical phylogeny. We find that the dependence of both total biomass and realized diversity on the number of traits is similar using either the star tree or the true tree structure to parameterize the model (Figures S8 and S10). Additionally, we tested how realistic tree structures affect the convergence of species abundances to the deterministic limit  (obtained, accoding to Eq.~\eqref{eq:abun}, as $\bm{x}=\Sigma^{-1}\bm{1}$) as the number of traits increases. In Figure S12, we show simulated GLV dynamics for different values of $\gamma$; these simulations illustrate that the full coexistence regime is approached for large $\gamma$, but the convergence to the deterministic-limit abundances is slow.

\begin{figure}[!t]
\input{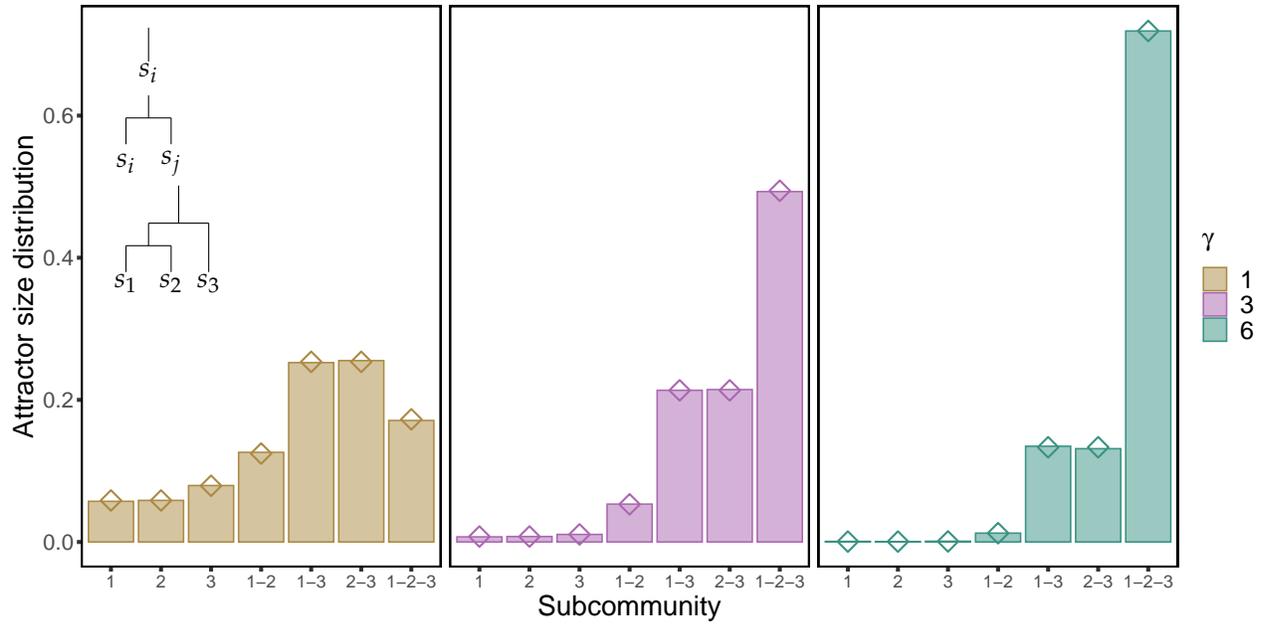}
\caption[Probability of observing sub-communities from a three-species pool]{\textbf{Probability of observing sub-communities from a three-species pool}. The pool phylogeny is the unique three-species (bifurcating) with equal branch lengths. The inset shows the tree sub-structures corresponding to different sub-communities. As the number of traits relative to the number of species ($\gamma$) increases, it becomes more likely to observe larger sub-communities, and for any fixed level of species richness the outgroup species $(s_3)$ is more likely than the other two to be present in the coexisting sub-community. Bars represent frequencies over 50000 simulations, and diamonds indicate the analytical predictions.}
\label{fig:6}
\end{figure}

Next, we explored how species' positions in the phylogeny shape their probability of survival. We first used Eqs.~(S33) and (S63) to compute the probability of observing each sub-community in a simple three-species community (\Cref{fig:6}). For $n = 3$, there is only one possible tree topology, and we consider the case where all branch lengths are equal. Mirroring our results for relative abundances in the deterministic limit, we find that sub-communities containing the outlier species, $s_3$, are always more likely to be observed than sub-communities of the same size in which $s_3$ is absent. This holds true for different values of $\gamma$, although the distribution of sub-communities shifts toward larger communities as $\gamma$ increases. These results provide a tractable example of phylogenetic overdispersion, and allow us to quantify the degree of overdispersion expected in a given scenario. Our formulae can be used to similarly compute an expected distribution of sub-communities for particular larger phylogenies of interest.

\begin{figure}[!t]
\includegraphics[width = \textwidth]{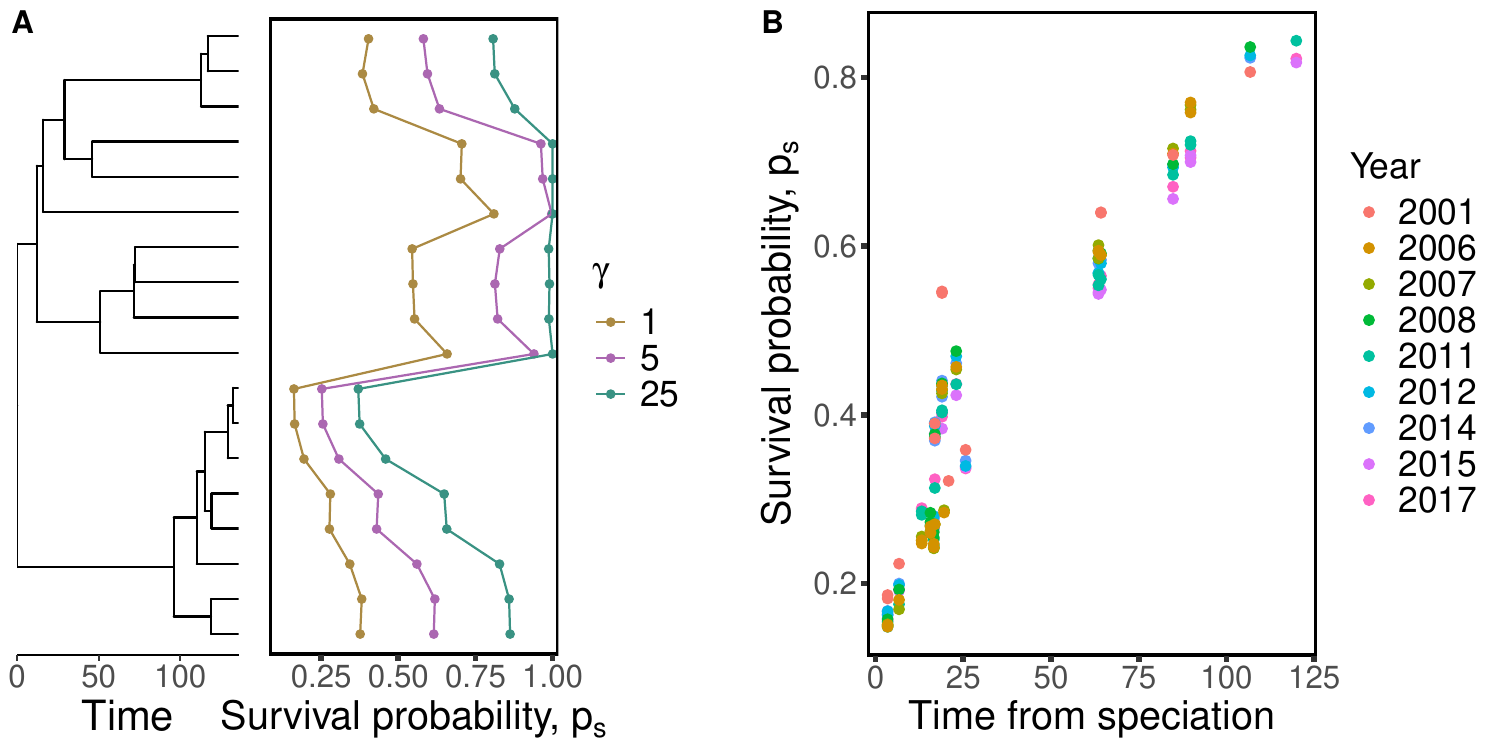}
\caption[Survival probability for Biodiversity II experiments]{\textbf{Survival probability for Biodiversity II experiments}. A. The probability of observing species in model realizations, for varying values of $\gamma$, reflect tree structure (2015 experiment). B. As panel A reveals, species that diverged early in the phylogeny (with largest time from speciation) are the ones more frequently represented in model communities. This pattern is consistent across experiments in different years. In this panel, survival probabilities are computed for $\gamma=1$.}
\label{fig:7}
\end{figure}

As an example, \Cref{fig:7} shows patterns of phylogenetic overdispersion predicted using the empirical tree for the 2015 Biodiversity II experiment species pool to parameterize our model (Section S8). We calculate the predicted probability of survival of each species in the tree by averaging over many realizations of the interaction matrix with different values of $\gamma$. In these simulations, we observe consistent over-dispersion across multiple cladistic scales: within each clade (defined by an internal node of the tree) the probability of survival is highest for the earliest diverging species. \Cref{fig:7}B shows that early diverging species are the most likely to be observed in the coexisting local community. Similarly, Figure S13 reproduces the same pattern for the much larger tree of the $\emph{Senna}$ clade. These results mirror our analytical findings for the perfectly unbalanced tree (in the determinisitic limit) and the three-species community, illustrating that our theoretical results for these simple cases carry over to more realistic tree structures.

\subsection*{Variability in growth rates} 

To test the robustness of our results, we relaxed the assumption of equal growth rates, assuming that $r_i$ are drawn from a normal distribution with a variance $\sigma^2$ (see Supplement, Section S7 for details). We considered two scenarios, both for a star tree topology characterized by an average correlation (equivalently, relatedness) $\rho$: (i) small variance and an arbitrary number of traits, and (ii) arbitrary variance in the deterministic limit. In the first case, our results remain valid by continuity with the $\sigma=0$ case. In the second case, we could apply previous results~\citep{servan2018coexistence} because the interaction matrix is deterministic in the limit $\gamma\to\infty$. Importantly, we find that full coexistence is no longer guaranteed if variability is large (and the distribution allows for negative values of $r_i$), unveiling a threshold in $\rho$ above which not all species coexist. Hoewver, if all the growth rates take positive values, then full coexistence is guaranteed (see Supplement, Section S6).

In addition, by allowing $\sigma$ to depend on $\rho$, we can also begin to explore fitness differences that are related to phylogeny in our framework. In the Supplement S7, we consider different scenarios that modulate how relatedness affects coexistence as a balance between species differences caused by variability in interactions (niche differences) or in growth rates (fitness differences). Figure S7 shows that if $\sigma$ increases with phylogenetic relatedness ($\rho$), average diversity decreases with $\rho$ and overdispersion patterns emerge. On the other hand, if $\sigma$ is a decreasing function of $\rho$, as is typically expected in nature~\citep{mayfield2010opposing}, our model can exhibit both species clustering patterns (species with similar growth rates are favored) when $\rho$ is close to one and overdispersion patterns for small $\rho$.

\section*{Discussion}

By considering community dynamics in a trait-based interaction model, we develop a theoretical link between the phylogeny of a regional species pool and key aspects of species coexistence in a local community. Importantly, in this framework the number of traits modulates how phylogenetic tree structure is reflected in community patterns. Although this quantity cannot be directly measured in most natural communities, it can be estimated indirectly~\citep{eklof2013dimensionality,laughlin2014intrinsic,mouillot2021dimensionality}, and in fact, our model framework provides another means to infer this quantity. Additionally, we establish a direct connection between our model and consumer-resource models (Section S1), showing how the number of traits $\ell$ can be interpreted as a number of resources or regulating factors, which can even be experimentally manipulated in some systems~\citep{van2019biodiversity}.

Our approach clarifies and quantifies long-standing expectations for how evolutionary history shapes community patterns \citep{silvertown2001phylogeny,webb2002phylogenies,violle2011phylogenetic,freilich2015phylogenetic}, and also makes new predictions about communities where interactions are structured by phylogeny. Fundamentally, our model provides a simple way to map phylogenies into community properties such as diversity, total community biomass, or species abundance distributions. It yields simple analytical predictions for these quantities in two cases -- for $\ell \gg n$ and for star phylogenies -- and provides insights and formulae that extend to more general scenarios. Most notably, we show that when the number of traits is large relative to the number of species, full coexistence of any pool of species is guaranteed by the tree-induced interaction structure. This result suggests that phylogeny organizes interactions in a way that promotes coexistence. However, when interactions reflect phylogeny imperfectly, because they depend on a finite number of stochastic traits, not all species coexist, and we quantify how the number of coexisting species depends on the traits-species ratio, $\gamma$, for the case of the star tree. 

At the other extreme, while $\ell \geq n$ is a well-known necessary condition for coexistence~\citep{yodzisintroduction,levin1970community}, we find that full coexistence is almost never achieved when the number of traits and species are equal (see also~\cite{Capitan2015} and~\cite{cui2020effect}). Yet, even when coexistence of all $n$ species is very unlikely, one typically observes coexisting communities of moderate size, as expected if interactions were purely random~\citep{servan2018coexistence,bunin2017ecological}. In this case, however, all species are not equally likely to persist in the final community, and we find that the probability a particular species remains extant is determined by its position in the phylogenetic tree, with species that diverged earlier in evolutionary time the most likely to persist in realized communities. Thus, our model provides an analytical framework for studying patterns of phylogenetic overdispersion in terms of both species' presence/absence and relative abundances. Beyond these two limiting cases, the general formulae derived in the Supplement pave the way to infer tree properties (encoded in matrix $\Sigma$), as well as the number of traits $\ell$ relevant to community assembly, from coexistence and abundance data. 

We model trait evolution with great generality by assuming trait values follow Gaussian drift processes on a phylogenetic tree, comprising both neutral evolution (Brownian motion) or selection scenarios (as in Ornstein-Uhlembeck processes). However, this approach is ultimately limited to trait distributions that are  multivariate normal. Additionally, our approach assumes an explicit separation between evolutionary processes at the regional level (which give rise to the phylogenetic structure) and ecological interactions (at the local level). By disallowing any feedback between species interactions and evolution, we remove the possibility for character displacement or other forms of coevolution. 

To remove this separation, future studies could leverage our theoretical foundation and model the tree generation process and ecological dynamics concurrently. For example, as in~\citet{maynard2018network}, one could ``run'' the dynamics after each speciation event, thereby pruning the community to obtain a new phylogeny for the next round of speciation and dynamics. In such a setting, similar to studies of community assembly~\citep{servan2021tractable} and the framework of adaptive dynamics~\citep{ad_coevo}, we would retain a separation of time-scales between the speciation events and the local community dynamics, but allow a feedback between the evolution of the tree structure and the ecological community. Our present results provide baseline expectations for this more complex evolutionary process: assuming that the number of traits is a constant $\ell$, in the early steps of the process the ratio of traits to species would be very high, and we expect that most speciation events occurring early on would not cause extinctions. In this case, the bulk of the phylogenetic structure would be built at the beginning of the process. It would be interesting to compare the structure of a tree evolved in this manner with the structure induced by dynamics in our model, starting with a large tree and letting species interactions prune the phylogeny all at once at the end of the branching process.

Our treatment of phylogenetically-structured interactions can also be viewed as an extension of recent results on large communities with random interactions~\citep{bunin2017ecological,servan2018coexistence,biroli2018marginally,barbier2018generic} to a case where interaction strengths reflect shared evolutionary history. Such models offer a way to relate community properties to ``summary statistics'' of species interactions, providing insights that are robust to the specific values of individual parameters. In this context, phylogeny is an informative summary of evolutionary history, capable of explaining aspects of community dynamics and structure that ultimately depend on the evolution of specific traits, modeled here as random processes. Thus, our analysis both leverages the power of random interaction models to link phylogeny and community properties, and advances the growing body of literature by incorporating an important type of biological structure.

Unlike many other models considered so far, tree-induced correlations provide a biologically-meaningful way to break statistical equivalence between species. Interestingly, despite the stronger correlation structure imposed in our model, our results closely resemble other random interaction models: full coexistence of large species pools is usually unlikely, but a moderate fraction of species coexist. We quantify both this probability of coexistence, as well as the mean number of coexisting species. Calculating distributions of community properties, such as richness or biomass, for arbitrary tree structures is possible by evaluating our integral formulas for all sub-sets of species, although this approach becomes burdensome for large species pools.

Owing to this random interaction perspective, our analytical predictions for diversity, biomass and relative abundance must be understood as averages over many realizations of the evolutionary processes assigning trait values to species. We treat these processes as random while fixing the correlation structure $\Sigma$ induced by a specific phylogenetic tree in order to ask how phylogenetic relationships are ``filtered'' through many possible trait realizations to impact dynamics. We expect, however, that averages across this ensemble will usually coincide with the evaluation of these quantities for a single, large realization of the species pool trait matrix. This equivalence, called the self-averaging property, is typical in random matrix theory~\citep{livan2018}.

Our approach could be extended in several additional ways. For example, our model only considers facilitative or competitive interactions. Using perturbation series expansions, it might be possible to incorporate trophic interactions~\citep{firkowski2022multi} or even higher-order interaction effects~\citep{letten2019mechanistic}, which may also be structured by phylogeny. Trophic interactions could be included by expressing the interaction matrix as a block-diagonal matrix, representing interactions within trophic levels, plus an off-diagonal block-structured matrix, standing for between-levels trophic interactions. The series expansion would be based on the interaction matrix between levels. Additionally, instead of assuming that the same tree structure controls the evolution of all $\ell$ traits, we could partition traits into $m$ classes and assume that the evolution of each class is determined by a distinct phylogenetic tree. These types of processes could arise when either admixture or incomplete lineage sorting lead to traits that cannot be explained by a single tree~\citep{nichols2001gene}. In such cases, $A$ would no longer follow the Wishart distribution but would rather be a sum of (possibly degenerate) Wishart matrices. 

Our framework assumes evolutionary processes that result in uncorrelated ``effective'' traits. Trait correlations could be incorporated by drawing trait vectors from a multivariate matrix normal distribution~\citep{sym15071445}. However, by introducing trait correlations, the sample covariance matrix $A$ will no longer be a realization of the Wishart ensemble, so further theoretical developments would be needed to rigorously include trait correlations. Lastly, our assumption of equal growth rates among species allowed us to examine how phylogenetic relatedness influences coexistence in a purely interaction-driven model. When variation in growth rates is included, we expect our results to hold for sufficiently small variance (Section S7). We also considered, in the deterministic limit, variability in growth rates under the influence of phylogeny (see also Section S7), finding that full species coexistence is no longer likely for large variability in growth rates. With further exploration of growth rate variation, potentially extending beyond the deterministic limit for interactions, our model offers a promising framework in which to investigate the the duality between ``competition'' and ``filtering'', related to the balance of niche and fitness differences, at the heart of community phylogenetics.

While there has been extensive discussion of the potential and possibly conflicting ways in which phylogeny could affect ecological differences, and thus interactions, among species~\citep{mayfield2010opposing, cadotte2017phylogenies}, much less has been said about the patterns one would observe under a particular hypothesis. In this work, we considered an idealized scenario where phylogenetic effects are realized exclusively through species niche differences, and where trait evolution is modeled by Gaussian processes. By linking phylogenies to this simple model of trait evolution and local community dynamics, we were able to fully characterize many global aspects of the community. We showed that the phylogenetic structure of the species pool and the number of traits determining competition affect these properties in concert. Our results provide a useful baseline prediction for the effect of phylogeny on community dynamics and coexistence.

\section*{Acknowledgments}
We thank P. Lemos-Costa and M.O. Carlson for comments on the manuscript. This work was supported by the National Science Foundation (DEB \#2022742 to SA) and by grant PRIORITY (PID2021-127202NB-C22) to JAC, funded by MCIN/AEI/10.13039/501100011033 and ``ERDF. A way of making Europe''. 

\section*{Statement of Authorship}

Conceptualization: CAS, JAC, ZRM, SA. Data collection: ZRM, SA. Data analysis and validation: CAS, JAC, ZRM, SA. Software development: CAS, JAC. Model analysis and simulation: CAS, JAC. Writing: CAS, JAC, ZRM, SA.

\section*{Data and Code Accessibility}

Data and code to reproduce results and figures can be found at the repository:\\
\href{https://doi.org/10.5061/dryad.cvdncjtck}{https://doi.org/10.5061/dryad.cvdncjtck}.

\newpage{}

\renewcommand{\thesection}{S\arabic{section}}
\renewcommand{\theequation}{S\arabic{equation}}
\renewcommand{\thefigure}{S\arabic{figure}}
\renewcommand{\thetable}{S\arabic{table}}
\setcounter{section}{0}
\setcounter{equation}{0}
\setcounter{figure}{0}
\section*{Supplementary Information}

\section{Motivation}
\subsection*{From consumer-resource dynamics to covariances}
We illustrate one particular setting where our model (Eq.~(1), main text) arises from neutral evolution of consumer preferences in consumer-resource model. Suppose we have a set of consumers, related by a particular evolutionary history, which differ only in the relative preference for each resource and assume that all resources have homogenous growth rates. Let $\bm{x} \in \R^n$, $\bm{y} \in \R^\ell$ be vectors denoting the density of consumers and resources. We model the dynamics as the MacArthur's consumer-resource model~\citep{macarthur1969species}:
\begin{equation}
  \begin{aligned}
    \frac{d\bm{x}}{dt} &= \bm{x} \circ (-d\bm{1}_n + \alpha \tilde{G} \bm{y}), \\
    \frac{d\bm{y}}{dt} &= \bm{y} \circ ( r\bm{1}_{\ell} - \bm{y} - \beta \tilde{G}^T \bm{x}),
  \end{aligned}
\end{equation}
where $\circ$ stands for the Hadamard (component-wise) matrix product, and $\bm{1}_k=(1,\dots,1)^T \in \R^k$ is a notation for a column vector whose entries are exactly $k$ ones.

By our assumptions, matrix $\tilde{G} \in \R^{n \times \ell}_+$ encodes the preference distribution (alternatively, the time allocation distribution) of the consumers over the resources, so that $\tilde{G}\bm{1}_{\ell} = \bm{1}_n$. Then by a separation of time scales, which implies that resource densities remain at equilibrium at all times, we can model the competition between the consumers as following competitive Lotka-Volterra dynamics~\citep{macarthur1969species}:
\begin{equation}\label{eq:LV-comp}
  \frac{d\bm{x}}{dt} = \bm{x} \circ( \alpha r \tilde{G}\bm{1}_{\ell} - d\bm{1}_n - \alpha \beta \tilde{G}\tilde{G}^T \bm{x}) = 
  \bm{x} \circ((\alpha r -d)\bm{1}_n - \alpha \beta \tilde{G}\tilde{G}^T\bm{x}).
\end{equation}

As long as $n\leq \ell$ (besides measure zero sets) we have that matrix $\tilde{A} := \tilde{G}\tilde{G}^T$ is positive definite. This property of $\tilde{A}
$ allows one to further transform the system \eqref{eq:LV-comp} without affecting the set of coexisting species. In particular we can perform the following operations (see section~\ref{section:lv_ops} for a more detailed discussion):
\begin{enumerate}
\item Rescale the growth rate, $\bm{v} = (\alpha r - d)\bm{1}_n$, by any positive constant.
\item Multiply $\tilde{A}$ by a positive, constant diagonal matrix.
\item Multiply both $\tilde{A}$ and $\bm{v}$ by a positive diagonal matrix.
\end{enumerate}

Using these operations we reduce the system to
\begin{equation}\label{eq:LV-almost}
  \frac{d\bm{x}}{dt} = \bm{x} \circ(\bm{1}_n - \tilde{G}\tilde{G}^T\bm{x}).
\end{equation}

To distinguish the effect of the mean of $\tilde{G}$, write $\tilde{G} = G + \frac{1}{n}\bm{1}_n\bm{1}_\ell^T$. Notice that this decomposition, together with the restriction $\tilde{G}\bm{1}_{\ell} = \bm{1}_n$, implies that $G\bm{1}_{\ell} = \bm{0}_n$, which means that the entries of $G$ have zero mean ---here $\bm{0}_k=(0,\dots,0)^T$ stands for a column vector formed by $k$ zeros. Then matrix $\tilde{A}$ can be decomposed as $\tilde{A} = GG^T + \bm{1}_n\bm{1}_n^T$. Because the system in \eqref{eq:LV-almost} has constant growth rates, one can show (section~\ref{section:lv_ops}) that, as long as $\ell > n$ (the strict inequality arising due to $G$ having rank $\ell-1$), the set of coexisting species for \eqref{eq:LV-almost} is invariant to the shift $\bm{1}_n\bm{1}_n^T$. Therefore the system reduces to:
\begin{equation}\label{eq:LV-final}
  \frac{d\bm{x}}{dt} = \bm{x} \circ(\bm{1}_n - GG^T\bm{x}) = \bm{x} \circ(\bm{1}_n - A \bm{x}),
\end{equation}
where we have defined $A:=GG^T$. This is the competitive, deterministic dynamics that we have assumed for consumers throughout this study. Observe that the set of coexisting species remains unchanged if we define interaction matrix $A=\frac{1}{\ell} GG^T$, as in the main text, because of the aforementioned invariant operations. Thus, the consumer-resource model implies a covariance matrix to represent inter-species interactions.

\subsection*{Modelling the covariance matrix}

From \eqref{eq:LV-final} we see that the interactions between species $A_{ij}$ are fully determined by the row vectors $\bm{G}_i$. Because each row $\bm{\tilde{G}}_i$ of matrix $\tilde{G}$ is a preference vector, then it lies on the standard $\ell-1$ dimensional simplex $\Delta^{\ell-1}=\{ \bm{\tilde{G}}_i\in \R^{\ell} \vert \sum_{j=1}^{\ell} \tilde{G}_{ij}=1, \text{ for }i=1,\dots, n\}$, which implies that $\bm{G}_i$ lies on a bounded subset of a linear subspace of $\R^{\ell}$ defined by the restrictions $\sum_{j=1}^{\ell} G_{ij}=0$ for $i=1,\dots, n$. By choosing a suitable (linear) coordinate system $\{\bm{w}_j\}_{j=1}^{\ell}$ we can express
\begin{equation}
  \begin{aligned}
    \bm{G}_i &= \sum_{j=1}^{\ell} c^j_i \bm{w}_j, \\
    A_{ij} &= \bm{G}_i\bm{G}_j^T = \sum_{k=1}^{\ell} c^k_i c^k_j.
  \end{aligned}
\end{equation}
Therefore, the entries of $A$ are fully determined by the coordinates of row vectors $\bm{G}_i$ on the basis $\{\bm{w}_j\}_{j=1}^{\ell}$.

To model coordinates $c^j_i$ we assume that each (rescaled) preference vector $\bm{G}_i$ is the result of a diffusion process starting at the origin of this space (this maps back to our $\bm{\tilde{G}}$ matrix as saying that every consumer has an \emph{homogeneous} preference for any resource). Assuming that each coordinate is independent and letting the diffusion time be small enough, then coefficients $c_i^j$ are normally distributed with zero mean, $c_{i}^j \sim \N(0, \sigma)$. The invariant properties of the model allow us to forget about the deviation $\sigma$ and simply model $c_{i}^j \sim \N(0, 1)$. This shows that $A$ satisfies the assumptions of model \eqref{eq:LV-final} explained in the main text for the Brownian motion case up to a change of number of traits from $\ell$ to $\ell-1$.

%%% Local Variables:
%%% mode: latex
%%% TeX-master: "sup_info"
%%% End:

%\section{Phylogenies and Covariance}
%
%\input{trees_picture}
%\textbf{Fig. S1: } Examples of ultrametric rooted phylogenies and its induced covariance matrices. The Perfectly
%hierarchical tree (left) has $n-1$ branching times $t_1 < \ldots < t_n$ for a pool of $n$ species, where each new branching happens to the
%``left'' and creates a new pair of species. The star tree (middle) posses a unique branching event which generates all
%the $n$ species. For the perfectly balanced tree (right) we have $n$ branching \emph{times} at each of which all the
%tips present up the that point generate two new species. Proceeding recursively, $n$ branching times generate $2^n$
%species and we have $n+1$ distinct branch lengths (i.e. the time between branching events). The covariance matrix associated with
%each tree is constructed as follows: For any $s_i$ take $\gamma_i$ to be the path ``backwards'' in time to the ancestral
%species at the root of the tree, then for any two $s_i, s_j$, let $t_{i,j}$ be the time at with $\gamma_i$ and $\gamma_j$
%merge, i.e. the time at which the most recent common ancestor of $s_i$ and $s_j$ splitted into distinct lineages. Then
%$\Sigma_{ij} = 1 - t_{i,j}$. In particular $\Sigma_{ij}$ is the total time for which the evolutionary process for $s_i$
%and $s_j$ is completely linked. For example, in the star tree $\Sigma_{ij} = \rho$ for any $i\not=j$ and $\Sigma_{ii} = 1$ given that
%each tree is ultrametric.

\section{Deterministic limit}
\subsection*{Full coexistence}
We provide more details for the proof that, in the deterministic limit, every subcommunity of the pool is feasible. Since every subcommunity has an interaction matrix induced by a tree, it is enough to show that feasibility is guaranteed whenever this is the case.

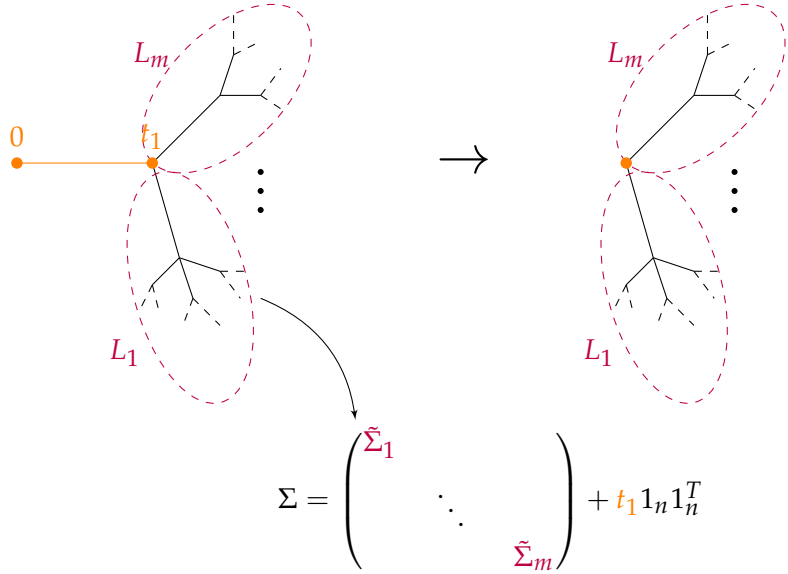
\begin{figure}[t!]
\begin{center}
\begin{tikzpicture}[scale=1.8, every node/.style={scale=1.8}]
  \draw[orange] (0,2.5) -- (1,2.5);

%  \draw[help lines] (0,0) grid (10,5);

  %% draw upper tree
  \uplineage;
  %% highlight lineage

  %% draw lower tree 
  \lowlineage
  %% highlight lineage
  \node[] at (1.8, 2.4) {$\vdots$};
  
  \fill[orange] (0,2.5) circle (1.2pt);
  \fill[orange] (1,2.5) circle (1.2pt);
  \node[] at (1., 2.7) {\tiny $\textcolor{orange}{t_1}$};
  \node[] at (0., 2.7) {\tiny $\textcolor{orange}{0}$};

  \node[] at (3.3, 2.5) {$\to$};
  
  %% collapsed tree
  \begin{scope}[shift = {(3.5,0)}]
    \uplineage[];
    \lowlineage[];
    \node[] at (1.8, 2.4) {$\vdots$};
  \end{scope}
  \fill[orange] (4.5,2.5) circle (1.2pt);

  \draw[edge,  every node/.style={sloped,anchor=north, auto=false}] (1.8, 1.5) to[bend left] (2.5,0.6);  
  \node[] at (3.5,0.0) {\tiny $\Sigma =
    \begin{pmatrix}
      \textcolor{purple}{\tilde{\Sigma}_1}\\ &\!\! \ddots\\ && \textcolor{purple}{\tilde{\Sigma}_m}
    \end{pmatrix} + \textcolor{orange}{t_1} 1_n1_n^T$};
\end{tikzpicture}

%%% Local Variables:
%%% mode: latex
%%% TeX-master: "paper"
%%% End:
\end{center}
\caption{\textbf{Schematic representation of the inductive step on the proof of full coexistence}. Starting with the tree $T$ (left), we shrink the ancestral branch up to the first splitting time $t_1$ to have a degenerate tree $\tilde{T}$ (on the right). $\tilde{T}$ splits at time $0$ into $m$ distinct subtrees induced by the lineages $L_i$ for $i = 1,\ldots, m$. The covariance matrix for $T$, $\Sigma$, is obtained from the covariance matrix $\tilde{\Sigma}$ of $\tilde{T}$ by ``adding back'' the ancestral branch. This
amounts to a constant rank-one update of $\tilde{\Sigma}$ which preserves feasibility.}\label{fig:S1}
\end{figure}

We proceed by induction on $n$, the number of species. For $n=1$ the claim holds trivially. Let $T$ be a phylogenetic tree (not necessarily ultrametric) for $n>1$ species, and $\Sigma$ its respective covariance matrix. Let $t_1$ be the time at which the first split happens, so that at $t_1$ the ancestral branch splits into $m \geq2$ lineages ($L_i$, with $i=1,\dots,m$) where each $L_i$ contains at most $n-1$ species. Lineages are defined by the condition that species $j,k \in L_i$ if and only if the shared branch length between both species $t_{j,k}$ satisfies $t_{j,k} > t_1$. That is, each lineage contains the subset of species whose shared evolutionary time is strictly greater than $t_1$. For each lineage $L_i$, take $T_i$ to be the subtree induced by $L_i$ up to this first branching point (see Figure~\ref{fig:S1}). To apply the inductive step we must reduce to the case of trees with strictly smaller number of species. One way to achieve this is as follows: Recall that for star-trees we can ``forget'' about the shared history by shrinking the ancestral branch to $0$ length, in terms of the covariance matrix this transforms a constant covariance-matrix with non-zero offdiagonal to the identity matrix. Here we can carry over the same process: By shrinking the ancestral branch segment between the root and the first split, we transform $T \to \tilde{T}$ where $\tilde{T}$ is a \emph{degenerate} tree in the sense that it splits into non-interacting subtrees. What are these subtrees? well if a pair of species $(i,j)$ share a non-trivial evolutionary history over $\tilde{T}$ we must have that $t_{i,j} > t_1$, thus our subtrees are precisely given by each of the lineages $L_i$ described above, i.e. they are given by $T_i$.

As we have non-interacting lineages, the induced covariance matrix $\tilde{\Sigma}$ is block-diagonal, where the blocks are given by $\tilde{\Sigma_i}$. Each $\tilde{\Sigma}_i$ comes from the relationships encoded in the respective $T_i$. As each lineage contains at most $n-1$ species we can apply our induction step on each of them. To conclude that coexistence holds in our original community just observe the following: $T$ is obtained from $\tilde{T}$ by adding a root segment of length $t_1$ (go from left to right in Figure~\ref{fig:S1}). In particular this says that the shared evolutionary times of all species increases by $t_1$, i.e. $\Sigma = \tilde{\Sigma} + t_1 \bm{1}_n\bm{1}_n^T$, so that $\Sigma$ is a constant rank-one update of $\tilde{\Sigma}$. Then by section~\ref{section:lv_ops}, the equilibrium associated to $\Sigma$ is feasible.

\subsection*{Perfectly hierarchical trees}
Consider a perfectly hierarchical tree $T_n$ with $n$ tips and branching times $t_0 = 0 < t_1< \ldots < t_n < 1$ (see Figures~1 and~2 of the main text), and let $\Sigma_n$ be its covariance matrix. Then it follows trivially that
\begin{equation}
  \Sigma_n = \begin{pmatrix} \tilde{\Sigma}_{n-1} & \bm{0}_{n-1}  \\ \bm{0}_{n-1}^T & s_1 \end{pmatrix} + t_1 \bm{1}_n\bm{1}^T_n, 
\end{equation}
where $s_i := \sum^n_{j=i +1} \Delta t_j$, for $\Delta t_j = t_{j} - t_{j-1}$ the time between two branching events---
the \emph{inter-branching time}. In this subsections we find accurate bounds for the total biomass and analyze the expected abundance distribution.

Define the vector of abundances $\bm{x}_n=(x_n^i)$ for a hierarchical tree $T_n$ with $n$ tips. In the deterministic limit,  this vector satisfies the linear system
\begin{equation}\label{eq:deteq}
\Sigma_n \bm{x}_n = \bm{1}_n.
\end{equation}
As in the proof of feasibility, $\bm{x}_n$ is given recursively by the updated equilibrium abundances $\bm{\tilde{x}}_{n-1}$ and $s_1^{-1}$ of the non-interacting subtrees $\tilde{T}_{n-1}$ and the one formed by the first species, respectively. Indeed, if we look for solutions of the form $\bm{x}_n=\begin{pmatrix}a\bm{\tilde{x}}_{n-1}\\x_n^n\end{pmatrix}$, where the vector of abundances $\bm{\tilde{x}}_{n-1}$ satisfies $\tilde{\Sigma}_{n-1}\bm{\tilde{x}}_{n-1}=\bm{1}_{n-1}$, $\tilde{\Sigma}_{n-1}$ being the covariance matrix of the subtree $\tilde{T}_{n-1}$, the equilibrium condition~\eqref{eq:deteq} for $\bm{x}_n$ reduces to a linear system for $a$ and $x_n^n$:
\begin{equation}
\begin{cases}
a+at_1 \bm{1}_{n-1}^T\bm{\tilde{x}}_{n-1}+t_1 x_n^n=1,\\
at_1 \bm{1}_{n-1}^T\bm{\tilde{x}}_{n-1}+(s_1+t_1)x_n^n=1.
\end{cases}
\end{equation}
The solution is $a=s_1x_n^n$, with $x_n^n=(s_1+t_1+s_1t_1 \bm{1}_{n-1}^T\bm{\tilde{x}}_{n-1})^{-1}$. Let $\tilde{W}_{n-1} := \sum_{i=1}^{n-1} \tilde{x}^i_{n-1}=\bm{1}_{n-1}^T\bm{\tilde{x}}_{n-1}$. Then $\bm{x}_n$ can be written in terms of $\tilde{W}_{n-1}$, $\bm{\tilde{x}}_{n-1}$, $s_0=s_1+t_1$, and $s_1$ as
\begin{equation}\label{eq:rec}
  \begin{aligned}
    x_n^n &= \frac{1}{s_0 + t_1\tilde{W}_{n-1}s_1}, \\
    x_n^i &= \frac{s_1\tilde{x}_{n-1}^i}{s_0 +t_1 \tilde{W}_{n-1}s_1}, \ \ 1\le i < n.
  \end{aligned}
\end{equation}
In particular, this implies the following recurrence for the total biomass, $W_n$:
\begin{equation}\label{eq:bio}
  W_n = \frac{1 + \tilde{W}_{n-1} s_1}{s_0 + t_1\tilde{W}_{n-1}s_1}.
\end{equation}

In the case of equal inter-branching times, $\Delta t_i = \frac{1}{n}$ for all $i=1,2,\dots, n$, observe that $s_0=1$, $s_1 = \frac{n-1}{n}$ and $\tilde{\Sigma}_{n-1} = \frac{n-1}{n} \Sigma_{n-1}$. Hence $\bm{x}_{n-1}=s_1\bm{\tilde{x}}_{n-1}$ and $W_{n-1}=s_1\tilde{W}_{n-1}$, so Eqs.~\eqref{eq:rec} and~\eqref{eq:bio} above reduce to:
\begin{equation}\label{eq:recursions_PH}
  \begin{aligned}
    x^n_n &= \frac{n}{n + W_{n-1}}, \\
    x^i_n &= \frac{n x^i_{n-1}}{n + W_{n-1}}, \ \ \ 1 \le i < n,\\ 
    W_n &= \frac{n(1 + W_{n-1})}{n + W_{n-1}}.
  \end{aligned}
\end{equation}
The following proposition provides accurate upper and lower bounds for total biomass in the limit of large number of species.

\begin{prop} Let
\begin{equation}\label{eq:varphi}
\varphi(n) := \frac{4n -1 - \sqrt{16n^2 + 1 -8 n\sqrt{n-1}}}{4\sqrt{n-1}}.
\end{equation}
Then, for equal branching times, it holds that $\sqrt{n} - \varphi(n) > W_n > \sqrt{n} - 1/4$  for $n\geq2$ and $\varphi(n) \to 1/4$ in the limit $n\to\infty$.
\end{prop}
\begin{proof}
Direct computation shows that the inequality holds at $n=2$ so we proceed by induction on $n$. 

Consider first the lower bound. Suppose it holds at $n-1$, then:
\[ W_n = \frac{n (1 + W_{n-1})}{ n + W_{n-1}} = n \left( 1 - \frac{n-1}{n + W_{n-1}}\right) > \frac{n (\sqrt{n-1} +3/4)}{n +
    \sqrt{n-1} -1/4}.\]

If the claim were not satisfied at $n$ we would have
\[ \sqrt{n}-1/4 \ge \frac{n (\sqrt{n-1} +3/4)}{n + \sqrt{n-1} -1/4}. \]
Rearranging terms, this gives the following chain of equivalent inequalities:
\begin{equation}
  \begin{aligned}
    n\sqrt{n} + \sqrt{n-1}\sqrt{n} + \frac{1}{16} &\ge  n\sqrt{n-1} + n + \frac{1}{4}(\sqrt{n-1} + \sqrt{n}), \\
    n(\sqrt{n} - 1) + \sqrt{n-1}\sqrt{n}(1 - \sqrt{n}) + \frac{1}{16} &\ge  \frac{1}{4}(\sqrt{n-1} + \sqrt{n}),  \\
    \sqrt{n}(\sqrt{n}-1)(\sqrt{n} - \sqrt{n-1}) + \frac{1}{16} &\ge \frac{1}{4}(\sqrt{n-1} + \sqrt{n}).
  \end{aligned}
\end{equation}
Multiplying both sides by $\sqrt{n-1} + \sqrt{n}$ we get
\begin{equation}\label{eq:in1}
    \sqrt{n}(\sqrt{n} -1) + \frac{1}{16}(\sqrt{n-1} + \sqrt{n}) \ge \frac{1}{4}(\sqrt{n-1} + \sqrt{n})^2= \frac{1}{4}(2n -1 + 2 \sqrt{n-1}\sqrt{n}).
\end{equation}
The last inequality implies
\[ \frac{3}{4} \ge \frac{7}{8} \sqrt{n},\]
which says $n \le 1$. This is a contradiction and we are done.

We proceed in the similar way for the upper bound. By induction hypothesis at $n-1$ we have
\[ W_n < \frac{n (\sqrt{n-1} + 1 - \varphi(n))}{n +  \sqrt{n-1} -\varphi(n)}.\]
If the inequality is not satisfied at $n$ then, a similar chain of inequalities yields
\begin{equation}
n-\sqrt{n} + \varphi(n)^2(\sqrt{n} + \sqrt{n-1}) \le \varphi(n)(2n -1 + 2\sqrt{n-1}\sqrt{n}).
\end{equation}
Note that the above restriction is exactly the same as~\eqref{eq:in1} with the inequality reversed and changing $\varphi(n)$ instead of $1/4$. Using that $\sqrt{n} > \sqrt{n-1}$, the last inequality implies
\[ n- \sqrt{n} + 2\sqrt{n-1} \varphi(n)^2 - (4n-1)\varphi(n) \le 0. \]
In particular, this means that $\varphi(n) \le u$ for $u$ the smaller root of the above quadratic equation,
\begin{equation*}
u := \frac{4n -1 - \sqrt{16n^2-8n + 1 - 8n\sqrt{n-1} + 8\sqrt{n-1}\sqrt{n}}}{4\sqrt{n-1}},
\end{equation*}
but with this definition and~\eqref{eq:varphi} it is easy to see that
\begin{equation*}
u  >  \frac{4n -1 - \sqrt{16n^2+ 1 - 8n\sqrt{n-1} }}{4\sqrt{n-1}} = \varphi(n),
\end{equation*}
which is again a contradiction and this completes the proof for the upper bound.

We have just proved that $\sqrt{n} - \varphi(n) > W_n > \sqrt{n} - 1/4$. In particular, this implies that $\varphi(n) < 1/4$. Taking the limit in the numerator of expression~\eqref{eq:varphi} it is easy to see that the leading order is
\begin{equation*}
\lim_{n\to\infty} 4n -1 - \sqrt{16n^2 + 1 -8 n\sqrt{n-1}}= \lim_{n\to\infty} \frac{(4n-1)^2-(16n^2 + 1 -8 n\sqrt{n-1})}{4n -1 + \sqrt{16n^2 + 1 -8 n\sqrt{n-1}}}= \lim_{n\to\infty} \sqrt{n-1},
\end{equation*}
which shows that 
\begin{equation}
\lim_{n\to\infty} \varphi(n)=\frac{1}{4}
\end{equation}
and the proof is complete.
%The above shows that in particular we have  $1/4 > \varphi(n)$. To show the last part of the claim we just bound
%$\varphi(n)$ by:
%\[ \varphi(n) > b(n) = \sqrt{n-1} - \sqrt{\frac{n}{n-1}(n - \frac{\sqrt{n-1}}{2} + \frac{1}{16n})}\]
%$b(n) \to 1/4$  as can be seen by:
%\[ b(n) = \frac{1 + \frac{n}{2}(\sqrt{n-1} - 4) - \frac{1}{16 (n-1)}}{(n-1)(\sqrt{n-1} + \sqrt{\frac{n}{n-1}(n -
%      \frac{\sqrt{n-1}}{2} + \frac{1}{16n})})} \]
%and dividing numerator and denominator by the leading order $n\sqrt{n}$ gives us the desired limit.
\end{proof}

Note that, for large communities, a very good approximation for the total biomass in a perfectly hierarchical tree is given by the formula $W_n=\sqrt{n}-\frac{1}{4}$.

The recursions in~\eqref{eq:recursions_PH} for individual abundances can be easily solved in terms of total biomass $W_n$ as
\begin{equation}
x_n^i = \prod_{j=i}^n \frac{j}{j+W_{j-1}}.
\end{equation}
This formula gives the abundance of the $i$-th species (in increasing order of the tips) for $i\ge 2$ (observe that the first two species have
the same abundance). Alternatively,
\[ \log(x^i_n) = \sum_{j=i}^n \log\left(\frac{j}{j + W_{j-1}}\right) = -\sum_{j=i}^n \log\left(1 + \frac{W_{j-1}}{j}\right). \]
Approximating $W_{j-1}$ by its lower bound, $W_{j-1}\approx \sqrt{j-1}-1/4$, we find
\begin{equation}
\log(x_n^i) \approx -\sum_{j=i}^n \log\left(1 + \frac{\sqrt{j-1} - 1/4}{j}\right).
\end{equation}
Cutting the series for $\log(1+x)$ at second order and considering only the leading term, with respect to $j$ for the quadratic term, we get:
\begin{equation}
\log(x_n^k) \approx -\sum_{j=k}^n \frac{\sqrt{j-1}}{j} - \frac{1}{4j} -\frac{1}{2}\frac{j-1}{j^2} \approx -\sum_{j=k}^n \frac{1}{\sqrt{j}} - \frac{3}{4j}.
\end{equation}
By the Euler-Maclaurin formula we obtain:
\begin{equation}
\log(x_n^k) \approx 2 (\sqrt{n} - \sqrt{j-1}) + \frac{3}{4} (\log(n) - \log(j-1)).
\end{equation}
and we can further refine the first terms $x_n^k$ for $k$ small by replacing the actual value $W_j$.
  
\subsection*{Perfectly balanced tree}

The total biomass for perfectly balanced trees is easier to derive because the covariance matrix has constant row sums in that case. To show this statement, order tree splits by the time they happen ($t_1<\ldots< t_q$). At each time $t_i$, the number of lineages doubles, so we get a total of $n=2^q$ species. As species interact by their shared evolutionary time, in this case each species shares the time with $2^{q-k}$ other species. Now let  $s_k = \sum_{i=1}^k \Delta t_i$, $\Delta t_i$ being the inter-branching time ---compare the different notation for $s_k$ here and in the previous subsection. Summing over all possible split times we get the sum over any row of $A$ (observe that $A_{ii}=1$),
\begin{equation}
 r_q = \sum_{j=1}^n A_{ij} =  1 + \sum_{k = 1}^{q} 2^{q-k}s_k,
\end{equation}
which is independent of $i$. Because row sums are constant, the vector or equilibrium abundances can be written as $\bm{x}_n=x\bm{1}_n$, and substitution into $\Sigma_n\bm{x}_n=\bm{1}_n$ yields $r_q x=1$. Therefore, individual abundances at equilibrium are constant and given by $x=r_q^{-1}$. Consequently, the total biomass at equilibrium, $W_q$, is simply given by
\begin{equation}
  W_q = \frac{2^q}{1 + \sum_{k = 1}^{q} 2^{q-k}s_k}.
\end{equation}

\begin{figure}[t!]
\centering
\includegraphics[width = 0.9\textwidth]{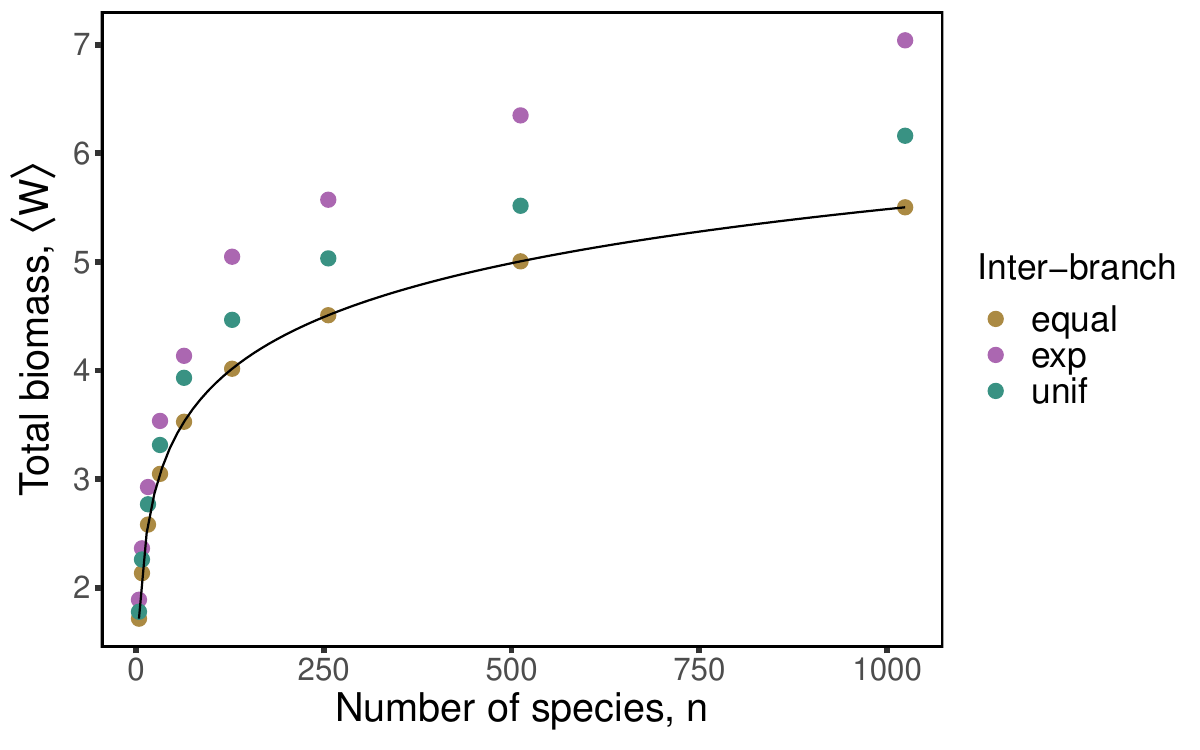}
\caption{\label{fig:S2}
\textbf{Total biomass for the perfectly balanced tree}. Dots mark the average values over simulations when sampling branch lengths from an exponential distribution with rate 1, a uniform $[0,1]$  distribution, and the case of equal branch lengths, for which the analytical prediction~\eqref{eq:biomass_PB} is shown with a solid line.}
\end{figure}

By our assumption of ultrametric trees, we have $s_k < 1$ (we need to add the tip lengths to sum up to one). In the particular case of equal inter-branching times, $\Delta t_i = \frac{1}{q+1}$, then $s_k = \frac{k}{q+1}$ and
\begin{equation}
  r_q = 1 + \frac{2^{q-1}}{q+1} \sum_{k =1}^q \frac{k}{2^{k-1}}.
\end{equation}
Observe that
\begin{equation}
  \sum_{k =1}^q \frac{k}{2^{k-1}}  = \frac{\partial}{\partial x}\left.\left(\frac{1 - x^{q+1}}{1-x}\right)\right|_{x = \frac{1}{2}} = 
  %\left.\frac{ n x^{n+1} - (n+1)x^n + 1}{(1-x)^2}\right|_{x = 1/2} =
  4\left(1 - \frac{1}{2^q}\left(q+1 - \frac{q}{2}\right)\right). 
\end{equation}
Thus,
\begin{equation}
  r_q = 1 + \frac{2^{q+1}-q - 2}{q+1} = \frac{2^{q+1} - 1}{q+1},
\end{equation}
and the total biomass reads
\begin{equation}
W_q = \frac{q+1}{2 - 2^{-q}}.
\end{equation}
Let $n=2^q$ be the number of species, then the number of tree splits is $q = \log_2(n)$. In terms of the number of species, the formula is given by
\begin{equation}\label{eq:biomass_PB}
W_n = \frac{\log_2(n) +1}{2 - 1/n},
\end{equation}
which grows logarithmically with $n$. Figure~\ref{fig:S2} compares the case of perfectly balanced trees for equal branching times with two cases, in which sampling times are drawn from exponential and uniform distributions.

%%% Local Variables:
%%% mode: latex
%%% TeX-master: "sup_info"
%%% End:

\section{Number of coexisting species}
\label{sec:Wish}

We have shown above that, in the $\ell \to \infty$ limit, full coexistence is guaranteed. To study species coexistence
for finite $\ell \ge n$ we use the fact that $A$ follows the Wishart distribution. As in~\cite{servan2018coexistence},
first we will compute the probability of the equilibrium point being feasible, i.e., where all species survive. Second,
since the attractor is unique (it is the only saturated equilibrium point that appears), we can calculate the
probability that the  equilibrium point cannot be invaded by the remaining species in the pool. Then we will show that the probability of feasibility and non-invasibility factors into the corresponding product, which yields the distribution of the number of species that coexist, as well as the expected number of species that survive.

Because matrix $A=GG^T$ is symmetric and positive definite, it is diagonally-stable~\citep{hofbauer:1998}, which implies that generalized Lotka-Volterra dynamics exhibits a single, globally stable fixed point~\citep{hofbauer:1998}, so there is a unique endpoint for the dynamics. Let us write the equilibrium abundances of the attractor, formed by $m$ survivors, as
\begin{equation}
\bm{x}_n=\begin{pmatrix} \bm{x}_m \\ \bm{0}_{n-m} \end{pmatrix},
\end{equation}
where, without loss of generality, we have located the survivors as the first $m$ species. Let $\{S\}_m$ denote the set of species that survive (i.e., the support of the endpoint). Therefore, the attractor can be fully characterized by two conditions~\citep{servan2018coexistence}:
\begin{itemize}
\item Define the vector $\bm{z}_n=\bm{1}_n-A\bm{x}_n=(x_n^i)$ with components $z_n^i$. Then it holds: first, $z_n^i = 0$ for all species $i\in \{S\}_m$, which simply states that equilibrium abundances of survivors satisfy the linear system $A_m\bm{x}_m=\bm{1}_m$, for $A_m$ the submatrix of $A$ restricted to the support $\{S\}_m$. Second, it also holds that $z_n^i < 0$ for all species $i\notin \{S\}_m$, i.e., the fixed point \emph{cannot be invaded} by the remaining species outside the endpoint. We have, therefore, a fixed point that cannot be invaded. 
\item The equilibrium point hast to be \emph{feasible}, i.e., $\bm{x}_m > \bm{0}_m$ ---here we use the notation that vectors $\bm{a}> \bm{b}$ if all inequalities are satisfied component-wise.
\end{itemize}
Since matrix $A$ belongs to the Wishart ensemble, these two conditions are to be understood in statistical terms. In the following subsections we are going to compute exact formulae for the probability that all the species in the pool form a \emph{feasible} attractor, and the probability that an endpoint formed by $m$ species remains \emph{non-invasible}. Using the properties of the Wishart ensemble~\citep{muirhead2009aspects}, we will calculate separately the probabilities of feasibility and non-invasibility, and with them we will obtain the distribution of the number of species that survive.

\subsection*{Probability of feasibility}
Let $n$ be the number of species in the pool, $\ell$ the number of traits, and define $\gamma :=\ell/n$ as the ratio between the number of traits and the size of the pool. An equilibrium point for the system such that all species coexist satisfies:
\begin{equation}
A \bm{x}_n  = \bm{1}_n, \text{ with }  x_n^i > 0 \text{ for all } i = 1,\ldots, n.
\end{equation}
The probability of feasibility is then the probability that $A^{-1}\bm{1}_n$ has all entries greater than 0. Observe that interaction matrix is defined as $A=\frac{1}{\ell}GG^T$ in the main text. Since rescaling by a positive constant in $A$ does not affect the condition for feasibility, we can forget about the rescaling by the number of traits $\ell$.

Let $A \sim \W_{n}(\Sigma,\ell)$ and $L_{n-1} = (I_{n-1}, \bm{0}_{n-1})$ be a rectangular $(n-1)\times n$ matrix with $0$ in its last column, $I_k$ being the $k\times k$ identity matrix. Then equation (2) of \cite{kotsiuba2016asymptotic} (similarly stated in the proof of Theorem~1 in \cite{bodnar2011product}) implies that
\begin{equation}\label{eq:t}
\widetilde{\bm{x}} := \frac{L_{n-1}A^{-1}\bm{1}_n}{\bm{1}_n^TA^{-1}\bm{1}_n} \sim t_{n-1}\left(\ell - n + 2, \frac{L_{n-1}\Sigma^{-1}\bm{1}_n}{\bm{1}_n^T \Sigma^{-1}\bm{1}_n}, \frac{ L_{n-1}R_{1}L^T_{n-1}}{(\ell - n +2) \bm{1}_n^T\Sigma^{-1}\bm{1}_n}\right),
\end{equation}
where $t_p(\nu,\bm{\mu},\Lambda)$ is a multivariate, $p$-dimensional $t$ distribution with $\nu$ degrees of freedom, localization vector $\bm{\mu}$ and dispersion matrix $\Lambda$~\citep{tong2012multivariate}. Matrix $R_1$ is given by
\begin{equation}
R_{1} = \Sigma^{-1} - \frac{\Sigma^{-1}\bm{1}_n \bm{1}_n^T\Sigma^{-1}}{\bm{1}_n^T\Sigma^{-1}\bm{1}_n}.
\end{equation}
Up to a normalization by a positive constant (which is precisely the total biomass, $\bm{1}_n^TA^{-1}\bm{1}_n$, given that $A$ is positive definite), vector $\widetilde{\bm{x}}=(\tilde{x}_i)$ precisely gives the abundances of the \emph{first} $n-1$ species. Moreover, the last (normalized) abundance is expressed as $1 - \bm{1}_{n-1}^T\widetilde{\bm{x}}$, so the probability of feasibility turns out to be
\begin{equation}
P_{\text{f}}(n) = \int d^{n-1}\widetilde{\bm{x}} f(\widetilde{\bm{x}}) \Theta(1-\bm{1}_{n-1}^T\widetilde{\bm{x}}) \prod_{i=1}^{n-1}\Theta(\tilde{x}_i),
\end{equation}
for $f(\widetilde{\bm{x}})$ the probability density function of the multivariate $t$ distribution defined in~\eqref{eq:t}.

Because a multivariate $t$ distribution is the ratio between a multivariate Gaussian and the square root of a chi-square distribution, it holds that if $\widetilde{\bm{x}}\sim t_{p}(\nu,\bm{\mu},\Lambda)$, then we have that $\widetilde{\bm{x}}=\bm{y}/\sqrt{u/\nu}+\bm{\mu}$, where $\bm{y}\sim \N(\bm{0},\Lambda)$ is a multivariate Gaussian and $u\sim \chi_{\nu}^2$, which is independent of $\bm{y}$. Therefore, conditioning on $u$, we find that $\bm{y}_u:=\widetilde{\bm{x}}\vert u \sim \N(\bm{\mu},\nu\Lambda/u)$ and we can transform the integral above to get
\begin{equation}\label{eq:pfeas}
P_{\text{f}}(n) = \int_{0}^{\infty}du\, g(\nu,u) \text{Pr}(\bm{y}_u > \bm{0}_{n-1}, \bm{1}_{n-1}^T\bm{y}_u < 1),
\end{equation}
where $u \sim \chi^2_{\nu}$, $g(\nu,u)$ is the corresponding pdf with $\nu=\ell - n + 2$, and the random variable $\bm{y}_u$ is distributed as a multivariate normal, 
\begin{equation}\label{eq:feasibility_pieces_general}
\bm{y}_u \sim \N\left(\frac{L_{n-1}\Sigma^{-1}\bm{1}_n}{\bm{1}_n^T\Sigma^{-1}\bm{1}_n},\frac{L_{n-1} R_1 L^T_{n-1}}{u \bm{1}_n^T \Sigma^{-1}\bm{1}_n}\right).
\end{equation}
In this way, all the dependence in the number of traits $\ell$ remains included in the chi-square
distribution. Eqs.~\eqref{eq:pfeas} and~\eqref{eq:feasibility_pieces_general} yield the probability of feasibility for
an arbitrary covariance matrix $\Sigma$. An explicit calculation of the probability of feasibility amounts to evaluating
the probability $ \text{Pr}(\bm{y}_u > \bm{0}_{n-1}, \bm{1}_{n-1}^T\bm{y}_u < 1)$. This can be done explicitly  for the
case of constant, non-negative correlation.

\subsubsection*{Constant, non-negative correlation}
\begin{figure}[t!]\label{fig:S3}
\centering
\includegraphics[width = 0.9\linewidth]{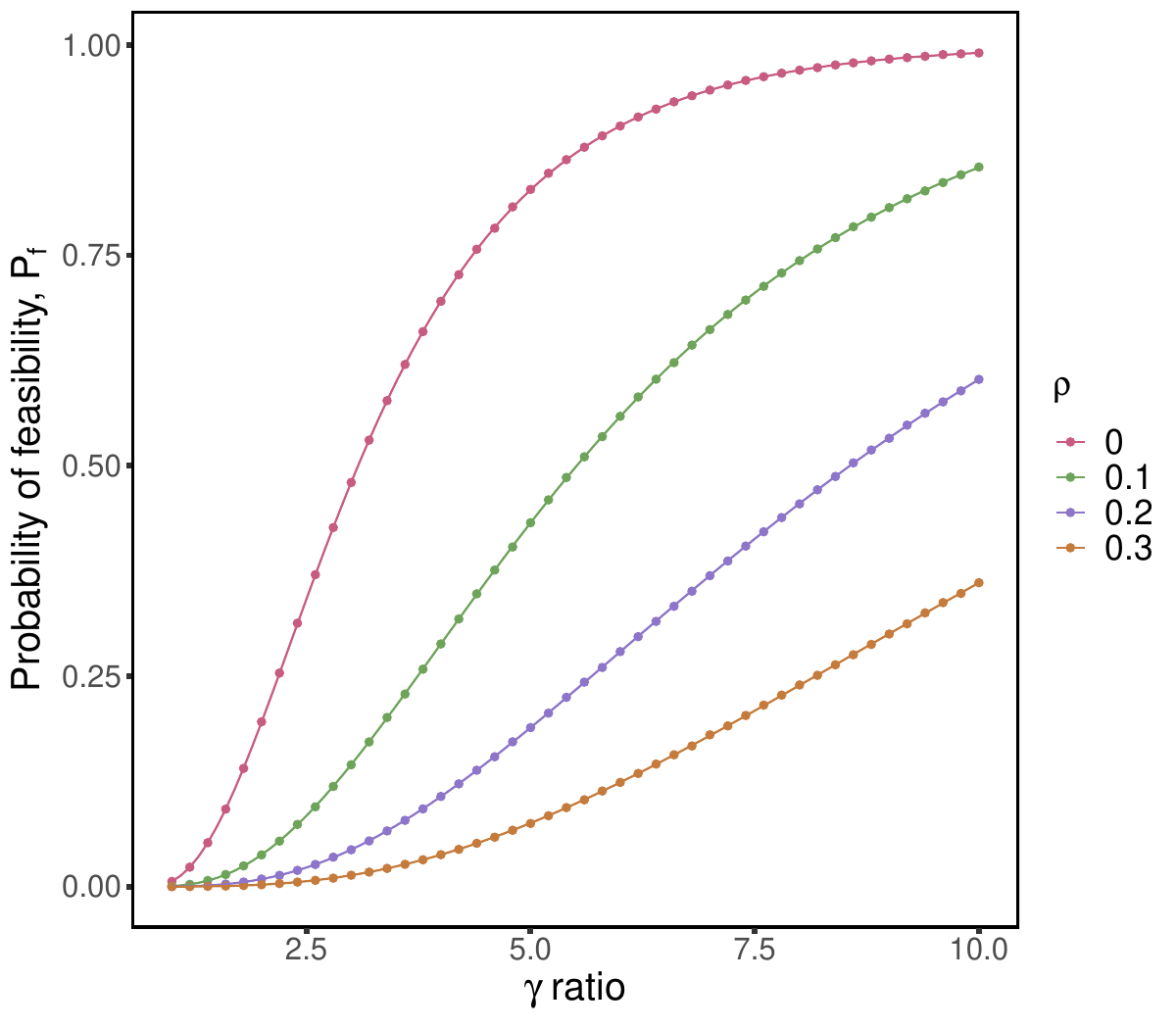}
\caption{\textbf{Probability of feasibility as a function of the ratio $\gamma$ of number of traits to number of species for different \emph{constant} correlation matrices.} The simulations were done with $n = 10$ species. Dots are simulations, solid lines are numerical evaluations of the exact formula \eqref{eq:feasibility_exact_constant_rho}. The larger the correlation, the slower curves approach to one in the deterministic limit $\gamma\to\infty$. }
% and the dotted is the normal approximation for $\rho = 0$.} 
\end{figure}

Consider the covariance matrix $\Sigma=(1-\rho)I_n+\rho\bm{1}_n\bm{1}_n^T$ with $\rho\ge 0$. Then~\eqref{eq:feasibility_pieces_general} simplifies to:
\begin{equation}\label{eq:yurho}
  y_u \sim \N\left(\frac{1}{n}\bm{1}_{n-1}, \frac{1 - \rho + n\rho}{u n (1 - \rho)}\left(I_{n-1} - \frac{1}{n}\bm{1}_{n-1}\bm{1}^T_{n-1}\right)\right).
\end{equation}
Let us define
\begin{equation}\label{eq:constants_feasibility}
    \alpha_u :=\frac{1 - \rho + n\rho}{u n (1 - \rho)} \text{ and }     \beta_u := \frac{\alpha_u}{n}.
\end{equation}
In this way, the covariance matrix $\Sigma_{u}$ in~\eqref{eq:yurho} can be expressed as $\Sigma_{u}=\alpha_u I_{n-1}-\beta_u \bm{1}_{n-1}\bm{1}_{n-1}^T$. $\Sigma_u$ has two eigenvalues, $\alpha_u$ and $\alpha_u + (n-1) \beta_u$. The first has multiplicity $n-1$, and the second $1$. Hence the determinant follows immediately,
\begin{equation}
  |\Sigma_u| = \alpha_u^{n-2}(\alpha_u - (n-1) \beta_u).
\end{equation}
The inverse can be easily calculated:
\begin{equation}
  \Sigma_u^{-1} = \frac{1}{\alpha_u}\left(I + \frac{\beta_u}{\alpha_u - (n-1)\beta_u}\bm{1}_{n-1}\bm{1}_{n-1}^T\right).
\end{equation}
Therefore we can write the pdf for the random variable $y_u$ as
\begin{equation}
  f_u(\bm{y}) = K e^{-\frac{1}{2}\left(\bm{y} - \frac{1}{n}\bm{1}_{n-1}\right)^T\Sigma_u^{-1}\left(\bm{y} - \frac{1}{n}\bm{1}_{n-1}\right)} = 
  K e^{-\frac{1}{2\alpha_u}\left(\left\|\bm{y}-\frac{1}{n}\bm{1}_{n-1}\right\|^2 + \frac{\beta_u}{\alpha_u - (n-1)\beta_u} (\bm{1}_{n-1}^T(\bm{y}-\frac{1}{n}\bm{1}_{n-1}))^2\right)}
\end{equation}
for $K = (2\pi)^{-(n-1)/2}|\Sigma_u|^{-1/2}$. First we have to compute the probability
\begin{equation}\label{eq:pu}
p(u):= \text{Pr}(\bm{y}_u > \bm{0}_{n-1}, \bm{1}_{n-1}^T\bm{y}_u < 1) = \int_{\R^{n-1}} d^{n-1}\bm{y}f_u(\bm{y})\Theta(1-\bm{1}^T_{n-1}\bm{y})\prod_{i=1}^{n-1}\Theta(y_i),
\end{equation}
with $\Theta(x)$ the Heaviside step function, defined as $\Theta(x)=1$ if $x\ge 0$ and $\Theta(x)=0$ if $x<0$. Thus after a change of variables $\bm{y}' = \bm{y} - \frac{1}{n}\bm{1}_{n-1}$, we have
\begin{equation}
p(u) = K \int_{\R^{n-1}}d^{n-1}\bm{y} e^{-\frac{1}{2\alpha_u}\left(\left\|\bm{y}\right\|^2 + (\bm{1}_{n-1}^T\bm{y})^2\right)} \Theta\left(\frac{1}{n}-\bm{1}_{n-1}^T\bm{y}\right)\prod_{i=1}^{n-1} \Theta\left(y_i + \frac{1}{n}\right),
\end{equation}
where we have omitted primes to ease notation and we have used~\eqref{eq:constants_feasibility} to see that
\begin{equation}
\frac{\beta_u}{\alpha_u - (n-1)\beta_u}=1.
\end{equation}
To simplify the term $(\bm{1}_{n-1}^T\bm{y})^2$ in the exponential, we introduce a Dirac's delta function,
\begin{equation}
p(u) = K \int_{\R^{n-1}}d^{n-1}\bm{y} \int_{\R}d\omega e^{-\frac{1}{2\alpha_u}\left(\left\|\bm{y}\right\|^2 + \omega^2\right)} \delta(\omega-\bm{1}_{n-1}^T\bm{y}) \Theta\left(\frac{1}{n}-\omega\right)\prod_{i=1}^{n-1} \Theta\left(y_i + \frac{1}{n}\right),
\end{equation}
and use its integral representation,
\begin{equation}
\delta(\omega-\bm{1}_{n-1}^T\bm{y})=\frac{1}{2\pi}\int_{\R}d\xi e^{-i\xi(\omega-\bm{1}_{n-1}^T\bm{y})}.
\end{equation}
This transformation, together with an interchange in the order of integration, yields the following expression for $p(u)$:
\begin{equation}
p(u) = \frac{K}{2\pi} \int_{\R}d\omega \int_{\R}d\xi \int_{\R^{n-1}}d^{n-1}\bm{y} e^{-\frac{1}{2\alpha_u}\left(\left\|\bm{y}\right\|^2 + \omega^2\right)+i(\bm{1}^T_{n-1}\bm{y}-\omega)\xi} \Theta\left(\frac{1}{n}-\omega\right)\prod_{i=1}^{n-1} \Theta\left(y_i + \frac{1}{n}\right).
\end{equation}
Apparently we are increasing the complexity of the integral, but rearranging terms we observe that
\begin{equation}
p(u) = \frac{K}{2\pi} \int_{\R}d\xi \int_{\R}d\omega e^{-\frac{\omega^2}{2\alpha_u} - i\omega\xi}\Theta\left(\frac{1}{n}-\omega\right)
\int_{\R^{n-1}}d^{n-1}\bm{y} e^{-\frac{\left\|\bm{y}\right\|^2}{2\alpha_u} +i \xi\bm{1}^T_{n-1}\bm{y}} \prod_{i=1}^{n-1}\Theta\left(y_i + \frac{1}{n}\right),
\end{equation}
and the integral over $\bm{y}$ factorizes,
\begin{equation}
p(u) = \frac{K}{2\pi} \int_{\R}d\xi \int_{-\infty}^{1/n}d\omega e^{-\frac{\omega^2}{2\alpha_u} - i\omega\xi}
\left(\int_{-1/n}^{\infty}dy e^{-\frac{y^2}{2\alpha_u} +i y \xi} \right)^{n-1}.
\end{equation}
Now, in the integral over $\omega$, change to the variable $\omega'=-\omega$ to get
\begin{equation}
p(u) = \frac{K}{2\pi} \int_{\R}d\xi \int^{\infty}_{-1/n}d\omega e^{-\frac{\omega^2}{2\alpha_u} + i\omega\xi}
\left(\int_{-1/n}^{\infty}dy e^{-\frac{y^2}{2\alpha_u} +i y \xi} \right)^{n-1} =  
\frac{K}{2\pi} \int_{\R}d\xi \left(\int_{-1/n}^{\infty}dy e^{-\frac{y^2}{2\alpha_u} +i y \xi} \right)^{n}.
\end{equation}
Let 
\begin{equation}
\Phi(x):=\frac{1}{2}\left(1+\text{erf}(x/\sqrt{2})\right)
\end{equation}
be the cdf of the standard Gaussian distribution, which can be extended to the complex plane. Then it holds that
\begin{equation}
\int_{-1/n}^{\infty}dy e^{-\frac{y^2}{2\alpha_u} +i y \xi} = \sqrt{2\pi\alpha_u}\,
e^{-\frac{\alpha_u\xi^2}{2}}\Phi\left(\frac{1/n+i\alpha_u\xi}{\sqrt{\alpha_u}}\right).
\end{equation}
Therefore, the sought probability can be written as
\begin{equation}
p(u) = \frac{K(2\pi\alpha_u)^{n/2}}{2\pi} \int_{\R}d\xi e^{-\frac{n\alpha_u\xi^2}{2}}\Phi\left(\frac{1/n+i\alpha_u\xi}{\sqrt{\alpha_u}}\right)^n.
\end{equation}
An alternative way to express the integral over $\xi$ it is to consider a path $\Gamma$ in the complex plane such that $\Gamma=\{z\in\mathbb{C} \vert z = x_0+i\xi\}$ and then reducing the result to the limit $x_0\to 0$, so that the integral over the imaginary axis is well defined. In practice, this amounts to change to the variable $\zeta=i\xi$. Consequently, an equivalent form of writing this equation is
\begin{equation}
p(u) = -i\sqrt{\frac{n\alpha_u}{2\pi}} \int_{\Gamma}d\zeta e^{\frac{n\alpha_u\zeta^2}{2}}\Phi\left(\frac{1/n+\alpha_u\zeta}{\sqrt{\alpha_u}}\right)^n,
\end{equation}
where we have used that $K=\sqrt{n}(2\pi\alpha_u)^{-(n-1)/2}$ in this case. Finally, according to~\eqref{eq:pfeas}, in the case of constant, positive correlation the probability of feasibility is given by a two dimensional integral,
\begin{equation}\label{eq:feasibility_exact_constant_rho}
P_{\text{f}}(n) = -i\sqrt{\frac{n}{2\pi}} \int_{0}^{\infty}du\, g(\nu,u) \sqrt{\alpha_u}
\int_{\Gamma}d\zeta e^{\frac{n\alpha_u\zeta^2}{2}}\Phi\left(\frac{1/n+\alpha_u\zeta}{\sqrt{\alpha_u}}\right)^n,
\end{equation}
where $g(\nu,u)$ is the pdf of the chi-square distribution with $\nu=\ell - n + 2$ degrees of freedom. Figure~\ref{fig:S3} compares this exact formula with numerical simulation for different values of the correlation.

%%% Local Variables:
%%% mode: latex
%%% TeX-master: "sup_info"
%%% End:

\subsection*{Probability of non-invasibility}
In this subsection we compute the probability that an attractor formed by $m \le n$ species cannot be invaded by the remaining $n-m$ species. Let $A \sim W_{n}(\Sigma, \ell)$. Observe that for invasibility the rescaling of interaction matrix as $A=\frac{1}{\ell}GG^T$ does not matter. Partition matrices $A$ and $\Sigma$ in four blocks as follows:
\begin{equation}
  A = \begin{pmatrix}
    A_{11} & A_{12} \\
    A_{21} & A_{22}
  \end{pmatrix}
  ,
  \ \ \ \Sigma = \begin{pmatrix}
    \Sigma_{11} & \Sigma_{12} \\
    \Sigma_{21} & \Sigma_{22}
  \end{pmatrix},
\end{equation}
where $\Sigma_{11}$ refers to the species that belong to the support $\{S\}_m$ of the attractor, $\Sigma_{22}$ is related to those species outside the attractor, and off-diagonal matrices are formed by the corresponding rows and columns in $\{S\}_m$ and $\{S\}_n\setminus\{S\}_m$, and \emph{vice versa}. The exact same notation applies to blocks in $A$.

Then by theorem 3.2.10 of~\cite{muirhead2009aspects} we have that
\begin{equation}
  A_{21} \vert A_{11} \sim \N(\Sigma_{21}\Sigma_{11}^{-1}A_{11}, \Sigma_{22.1} \otimes A_{11}),
\end{equation}
where $\Sigma_{22.1} = \Sigma_{22} - \Sigma_{21}\Sigma_{11}^{-1}\Sigma_{12}$ is the Schur complement of $\Sigma_{22}$, $\otimes$ is the tensor product of matrices, and the normal distribution appearing is meant to be understood as the distribution of the \emph{flatten} matrix $A_{21}$. By the properties of the normal distribution it follows that
\begin{equation}\label{eq:cond_wishart}
  \begin{aligned}
    A_{21}A_{11}^{-1}\vert A_{11} &\sim \N(\Sigma_{21}\Sigma_{11}^{-1}, \Sigma_{22.1} \otimes A^{-1}_{11}),\\
    A_{21}A_{11}^{-1}\bm{1}_m\vert A_{11} &\sim \N(\Sigma_{21}\Sigma_{11}^{-1}\bm{1}_m, \bm{1}_m^TA^{-1}_{11}\bm{1}_m\Sigma_{22.1}).
  \end{aligned}
\end{equation}
In order to get the last line, we first transpose the matrix, then notice that the $\bm{1}_m^T$ operator acts on the vector of elements of the matrix as $I_m \otimes \bm{1}^T$. Hence by the property $(A \otimes B)(C\otimes D) = AC\otimes BD$ of the tensor product the second statement above follows. 

As mentioned at the begining of  Sec.~\ref{sec:Wish}, the probability that the attractor cannot be invaded by any species in $\{S\}_n\setminus\{S\}_m$ coincides with the probability that $\bm{z}=\bm{1}_{n-m}-A_{21}A_{11}^{-1}\bm{1}_m<\bm{0}_{n-m}$. Define $W:=\bm{1}_m^TA_{11}^{-1}\bm{1}_m$ and $f_W(w)$ as the pdf of the random variable $W$, which is non-negative. Then
\begin{multline}
P_{\text{ni}}(m,n)  = \int_0^{\infty} dw f_W(w)\,\text{Pr}(\bm{z} < \bm{0} \vert W=w)  \\ = \int_0^{\infty} dw f_W(w) \int_{\VV^+_w} dA_{11} \text{Pr}(A_{11}\vert W=w) \text{Pr}(\bm{z} < \bm{0} \vert A_{11}, W=w),
\end{multline}
where $\VV^+$ is the set of positive definite symmetric matrices and $\VV^+_w$ the set conditional to $W=\bm{1}_m^TA_{11}^{-1}\bm{1}_m = w$. Using that $\bm{z}=\bm{1}_{n-m}-A_{21}A_{11}^{-1}\bm{1}_m$ and~\eqref{eq:cond_wishart}, the conditional variable $\bm{z}\vert A_{11},W=w$ is distributed as
\begin{equation}\label{eq:zcon}
\bm{z}\vert A_{11},W=w \sim \N\left(\bm{1}_{n-m}-\Sigma_{21}\Sigma_{11}^{-1}\bm{1}_m, w\Sigma_{22.1}\right),
\end{equation}
which does not depend explicitly on $A_{11}$. Neither does $\text{Pr}(\bm{z} < \bm{0} \vert A_{11}, W=w)$, so we can factor this probability out of the integration over $A_{11}$. In this way, we can write
\begin{equation}\label{eq:exact_inv}
P_{\text{ni}}(m,n) = \int_0^{\infty} dw f_W(w) Q^{-}_{n-m}\big(\bm{1}_{n-m}-\Sigma_{21}\Sigma_{11}^{-1}\bm{1}_m, w\Sigma_{22.1}\big),
\end{equation}
because $\int_{\VV^+_w} dA_{11} \text{Pr}(A_{11}\vert W=w)=1$. In~\eqref{eq:exact_inv} we have defined $Q^-_p$ as the probability that a multivariate Gaussian variable with the specified parameters is contained in the fully negative orthant,
\begin{equation}\label{eq:Qminus}
Q^{-}_{p}\big(\bm{\mu}, \Lambda\big):=(2\pi)^{-p/2}\vert\Lambda\vert^{-1/2}\int_{\R_-^n}d\bm{y}e^{-\frac{1}{2}(\bm{y}-\bm{\mu})^T\Lambda^{-1}(\bm{y}-\bm{\mu})}.
\end{equation}

Corollary 3.2.6 in~\cite{muirhead2009aspects} implies that $A_{11}\sim \W_m(\Sigma,\ell)$. Therefore, theorem 3.2.12 in the same reference holds, which ensures that
\begin{equation}
W^{-1}\bm{1}_m^T\Sigma_{11}^{-1}\bm{1}_m = \frac{\bm{1}_m^T\Sigma_{11}^{-1}\bm{1}_m}{\bm{1}_m^T A_{11}^{-1}\bm{1}_m} \sim \chi^2_{\ell-m+1}.
\end{equation}
This means that
\begin{equation}
g(\nu',w)=-w^{-2}\bm{1}_m^T\Sigma_{11}^{-1}\bm{1}_m f_W\big(w^{-1}\bm{1}_m^T\Sigma_{11}^{-1}\bm{1}_m\big),
\end{equation}
for $g(\nu',w)$ the pdf of a $\chi^2_{\nu'}$ distribution with $\nu'=\ell-m+1$ degrees of freedom. Now, making the change of variable $w'=w^{-1}\bm{1}_m^T\Sigma_{11}^{-1}\bm{1}_m$ in~\eqref{eq:exact_inv} we finally get
\begin{equation}\label{eq:exact_inv_gen}
P_{\text{ni}}(m,n) = \int_0^{\infty} dw g(\nu',w) Q^{-}_{n-m}\big(\bm{1}_{n-m}-\Sigma_{21}\Sigma_{11}^{-1}\bm{1}_m, w^{-1}\bm{1}_m^T\Sigma_{11}^{-1}\bm{1}_m\Sigma_{22.1}\big).
\end{equation}
%Therefore we have a density function for $w$:
%\begin{equation}
%  g(w) = h(\frac{w}{1^T\Sigma_{11}^{-1}1}) \frac{1}{1^T\Sigma_{11}^{-1}1}
%\end{equation}
%Where $h$ is the density function of an \emph{Inverse Chi square} with $k-m +1$ degrees of freedom.
%By a change of variables $w' = \frac{1^T\Sigma_{11}^{-1}1}{w}$, we have :
%\begin{equation}\label{eq:exact_inv_gen}
%  \begin{aligned}
%    P_{ni}(m) &= \int_{\R_+} dw \frac{1}{w^2}h(1/w) O^{-}(n-m, 1-\Sigma_{21}\Sigma_{11}^{-1}1, \frac{1^T\Sigma_{11}^{-1}1}{w} \Sigma_{22.1}) \\
%    &=\int_{\R_+} dw f(w) O^{-}(n-m, 1-\Sigma_{21}\Sigma_{11}^{-1}1, \frac{1^T\Sigma_{11}^{-1}1}{w} \Sigma_{22.1}) 
%  \end{aligned}
%\end{equation}
%For $f$ the density function of a \emph{Chi square} distribution with $k-m+1$ degrees of freedom\footnote{notice that this form can be derived directly if we condition on $\frac{1^T\Sigma_{11}^{-1}1}{W}$ instead of $W$.}

As for the case of feasibility,~\eqref{eq:exact_inv_gen} is an exact formula for the probability that an endpoint composed by $m$ species cannot be invaded by the remaining $n-m$ species. Similarly, the multidimensional integral associated to $Q^{-}_{n-m}$ can be reduced to a single integral in the case of constant, non-negative correlation, as we show in the following subsection. Thus, in that particular case, the probability of non-invasibility is expressed as a double integral. 

\subsubsection*{Constant, non-negative correlation}

In the case of constant, non-negative correlation, \eqref{eq:exact_inv_gen} simplifies to:
\begin{equation}\label{eq:invaux}
P_{\text{ni}}(m,n) = \int_{0}^{\infty} dw g(\nu',w)Q_{n-m}^{-}(\bm{\mu},\Sigma_w)
\end{equation}
with
\begin{equation}
\begin{aligned}
\bm{\mu}&=\frac{1-\rho}{1 -\rho + m\rho}\bm{1}_{n-m},\\
\Sigma_w&=\frac{m (1-\rho)}{w(1 -\rho + m\rho)} \left(I_{n-m} + \frac{\rho}{1 -\rho + m\rho} \bm{1}_{n-m}\bm{1}_{n-m}^T\right).
\end{aligned}
\end{equation}
Now focus on the probability $Q_{n-m}^{-}$. Making the substitution $\bm{y}'=k\bm{y}$ in~\eqref{eq:Qminus} it is easy to show that
\begin{equation}
Q_{p}^{-}(\bm{\mu},\Lambda)=Q_p^-(\bm{\mu}/k,\Lambda/k^2).
\end{equation}
Therefore, for $k=\frac{m (1-\rho)}{1 -\rho + m\rho}$ we recover Eq.~\eqref{eq:invaux} with $\bm{\mu}$ and $\Lambda$ given by
\begin{equation}\label{eq:muSigma}
\bm{\mu}=\frac{1}{m}\bm{1}_{n-m}, \ \ \ 
\Sigma_w=\frac{1 -\rho + m\rho}{m w (1-\rho)} \left(I_{n-m} + \frac{\rho}{1 -\rho + m\rho} \bm{1}_{n-m}\bm{1}_{n-m}^T\right).
\end{equation}
Now let us write $\Sigma_w:=\alpha_wI_{n-m}+\beta_w\bm{1}_{n-m}\bm{1}_{n-m}^T$, with $\alpha_w := \frac{1 -\rho + m\rho}{m w (1 - \rho)}$, $\beta_w := \frac{\rho \alpha_w}{1 -\rho + m\rho}$. As we did for the probability of feasibility, the probability $Q_{n-m}^{-}$ can be written as a one-dimensional integral. For that is crucial that, contrary to what happened in the case of feasibility, correlations given by $\Sigma_w$ are positive ---notice the plus sign in~\eqref{eq:muSigma}. This is due to the special structure of $\Sigma_w$, which implies that the correlation between any two distinct $y_i$, $y_j$ in~\eqref{eq:Qminus} is constant and given by $\lambda=\frac{\rho}{1+m\rho} \ge 0$. Hence, the following result of~\cite{tong2012multivariate} (section 8.2.5) applies:
\begin{prop}
Let $\bm{x}$ be distributed according to $\N(\bm{\mu},\Sigma)$ such that covariance matrix entries satisfy $\Sigma_{ii}=\sigma_i^2$ and $\Sigma_{ij}=\sigma_i\sigma_j\lambda$. Then, the joint probability that $\bm{x}\in C:=\{\bm{x}\in\R^n \vert b_i\le x_i \le a_i, i=1,\dots, n\}$, where $-\infty \le b_i < a_i \le \infty$ for i=1,\dots, n, is expressed as
\begin{equation}
{\normalfont \textrm{Pr}}(\bm{x}\in C) =\int_{-\infty}^{\infty} dy\phi(y)\prod_{i=1}^n
\left[\Phi\left(\frac{(a_i-\mu_i)/\sigma_i+\sqrt{\lambda}y}{\sqrt{1-\lambda}}\right)-
\Phi\left(\frac{(b_i-\mu_i)/\sigma_i+\sqrt{\lambda}y}{\sqrt{1-\lambda}}\right)\right]
\end{equation}
for $\phi(z)$ and $\Phi(z)$ the pdf and cdf, respectively, of a univariate standard normal distribution.
\end{prop}
In our particular case $\sigma_i^2=\frac{1+m\rho}{wm(1-\rho)}$, $\lambda=\frac{\rho}{1+m\rho}$, $b_i=-\infty$, $a_i=0$ and, according to~\eqref{eq:muSigma}, $\mu_i=\frac{1}{m}$ for $i=1,\dots,n-m$. Therefore, putting all the pieces together, we can write
\begin{equation}\label{eq:exact_inv_constant_rho_pre}
P_{\text{ni}}(m,n) = \int_0^{\infty} dw g(\nu',w) \int_{-\infty}^{\infty} dy \phi(y) \Phi\left(\frac{-1/m + y \sqrt{\beta_w}}{\sqrt{\alpha_w}}\right)^{n-m}.
\end{equation}
As for the probability of feasibility, in the case of constant, non-negative correlation we can reduce it to a two-dimensional integral.

Notice the resemblance between the expressions for feasibility and non-invasibility ---Eqs.~\eqref{eq:feasibility_exact_constant_rho} and~\eqref{eq:exact_inv_constant_rho_pre}. In the case of $\rho > 0$, by changing $y \to y' \frac{\alpha_w}{\sqrt{\beta_w}}$, we can make the resemblance stronger:
\begin{equation}\label{eq:exact_inv_constant_rho}
P_{\text{ni}}(m,n) = \sqrt{\frac{1 -\rho +m\rho}{2\pi\rho}} \int_0^{\infty} dw g(\nu',w) \sqrt{\alpha_w}  \int_{-\infty}^{\infty} dy 
e^{-\frac{(1 - \rho +m\rho)\alpha_w y^2}{2\rho}} \Phi\left(\frac{-1/m + y \alpha_w}{\sqrt{\alpha_w}}\right)^{n-m}.
\end{equation}
Observe that the number of degrees of freedom of the $\chi^2_{\nu'}$ distribution here is $\nu'=\ell-m+1$. Notice also that the change of variables leading to~\eqref{eq:exact_inv_constant_rho} does not apply for $\rho=0$. This case is trivial, however, and will not be discussed explicitly.
  
%%% Local Variables:
%%% mode: latex
%%% TeX-master: "sup_info"
%%% End:

\subsection*{Independence of feasibility and invasibility}
\label{sec:independence_feasibility_invasibility}
In this section we show that the joint probability of feasibility and non-invasibility factors into the product of the two probabilities calculated above. For that purpose, it suffices to show that
\begin{equation}\label{eq:indep}
\text{Pr}\big(\bm{z} < \bm{0}_{n-m} \vert A_{11}^{-1}\bm{1}_m > \bm{0}_m\big)  = \text{Pr}(\bm{z} < \bm{0}_{n-m} ).
\end{equation}
For that purpose we can calculate
\begin{multline}
\text{Pr}\big(\bm{z} < \bm{0}_{n-m} \vert A_{11}^{-1}\bm{1}_m > \bm{0}_m\big) = \int_0^{\infty} dw\, g_W(w)\, \text{Pr}\big(\bm{z} < \bm{0}_{n-m} \vert  A_{11}^{-1}\bm{1}_m > \bm{0}_m, W=w \big)\\
= \int_0^{\infty} dw\, g_W(w) \int_{\mathcal{G}^+_w} dA_{11} \text{Pr}\big(\bm{z} < \bm{0}_{n-m}\vert A_{11},W=w\big)
\text{Pr}\big(A_{11}\vert A_{11}^{-1}\bm{1}_m > \bm{0}_m, W=w \big),
\end{multline}
where $W=\bm{1}_m^TA_{11}^{-1}\bm{1}_m$ as for the calculation of $P_{\text{ni}}$, and $g_W$ is the pdf of the random variable $W \vert A_{11}^{-1}\bm{1}_m > \bm{0}_m$. In the second line we have introduced an integral over the set $\mathcal{G}^+_w$ of symmetric matrices and positive definite that verify the conditions $A_{11}^{-1}\bm{1}_m > \bm{0}_m$ and $W=\bm{1}_m^TA_{11}^{-1}\bm{1}_m=w$. As before, by~\eqref{eq:zcon} we can factor the probability $\text{Pr}\big(\bm{z} < \bm{0}_{n-m}\vert A_{11}, W=w\big)$ out, so we get
\begin{equation}
\text{Pr}\big(\bm{z} < \bm{0}_{n-m} \vert A_{11}^{-1}\bm{1}_m > \bm{0}_m\big) 
= \int_0^{\infty} dw g_W(w) Q^{-}_{n-m}\big(\bm{1}_{n-m}-\Sigma_{21}\Sigma_{11}^{-1}\bm{1}_m, w\Sigma_{22.1}\big),
\end{equation}
which coincides with~\eqref{eq:exact_inv_gen} except for the probability density $g_W$. In the last step we have used the normalization condition $\int_{\mathcal{G}^+_w} dA_{11} \text{Pr}(A_{11}\vert A_{11}^{-1}\bm{1}_m > \bm{0}_m, W=w)=1$.

Observe that the condition $A_{11}^{-1}\bm{1}_m > \bm{0}_m$ is equivalent to the conditions $\bm{1}^T_{m-1}\widetilde{\bm{x}}<1$ and $\widetilde{\bm{x}}>\bm{0}_{m-1}$, for $\widetilde{\bm{x}}$ the vector of the first $m-1$ relative abundances defined in~\eqref{eq:t}. Let $R:=\{\bm{v}\in\R^{m-1}\vert \bm{1}^T_{m-1}\bm{v}<1,\bm{v}>\bm{0}_{m-1}\}$ the set of vectors satisfying the two last conditions. Then it is easy to see that
\begin{multline}
g_W(w) = \frac{d}{dw}\text{Pr}\big(W < w \vert A_{11}^{-1}\bm{1}_m > \bm{0}_m\big) \\
= \frac{d}{dw}\text{Pr}\big(W < w \vert \widetilde{\bm{x}} \in R\big) = \frac{d}{dw} \text{Pr}(W<z) = f_W(w).
\end{multline}
The last equality in the chain above follows because $W$ and $\widetilde{\bm{x}}$ are independent random variables ---see the proof of theorem~1 in~\cite{bodnar2011product}. 

This shows that the probability of observing and endpoint with $m$ survivors can be factored as the probability of feasibility~\eqref{eq:pfeas} times the probability~\eqref{eq:exact_inv_gen} that the attractor cannot be invaded by the remaining $n-m$ species in the pool.

\subsection*{Distribution of the number of coexisting species}

Due to the independence shown in the previous section, the probability that the system settles in a subset $\{S\}_m\subset\{1,\ldots,n\}$ formed by $m$ species is simply
\begin{equation}\label{eq:exact_coexistence}
\text{Pr}(\{S\}_m|{n, \ell, \Sigma}) = \binom{n}{m} P_{\text{a}}(m,n) = \binom{n}{m} P_{\text{f}}(m)P_{\text{ni}}(m,n),
\end{equation}
because all subsets with cardinality $m$ are statistically equivalent.

Assuming constant and non-negative correlation, in Figure~S5 we compare numerical integration of Eqs.~\eqref{eq:feasibility_exact_constant_rho} and~\eqref{eq:exact_inv_constant_rho_pre} appearing in \eqref{eq:exact_coexistence} with simulations.

\begin{figure}[t!]
\includegraphics[width = 0.99\linewidth]{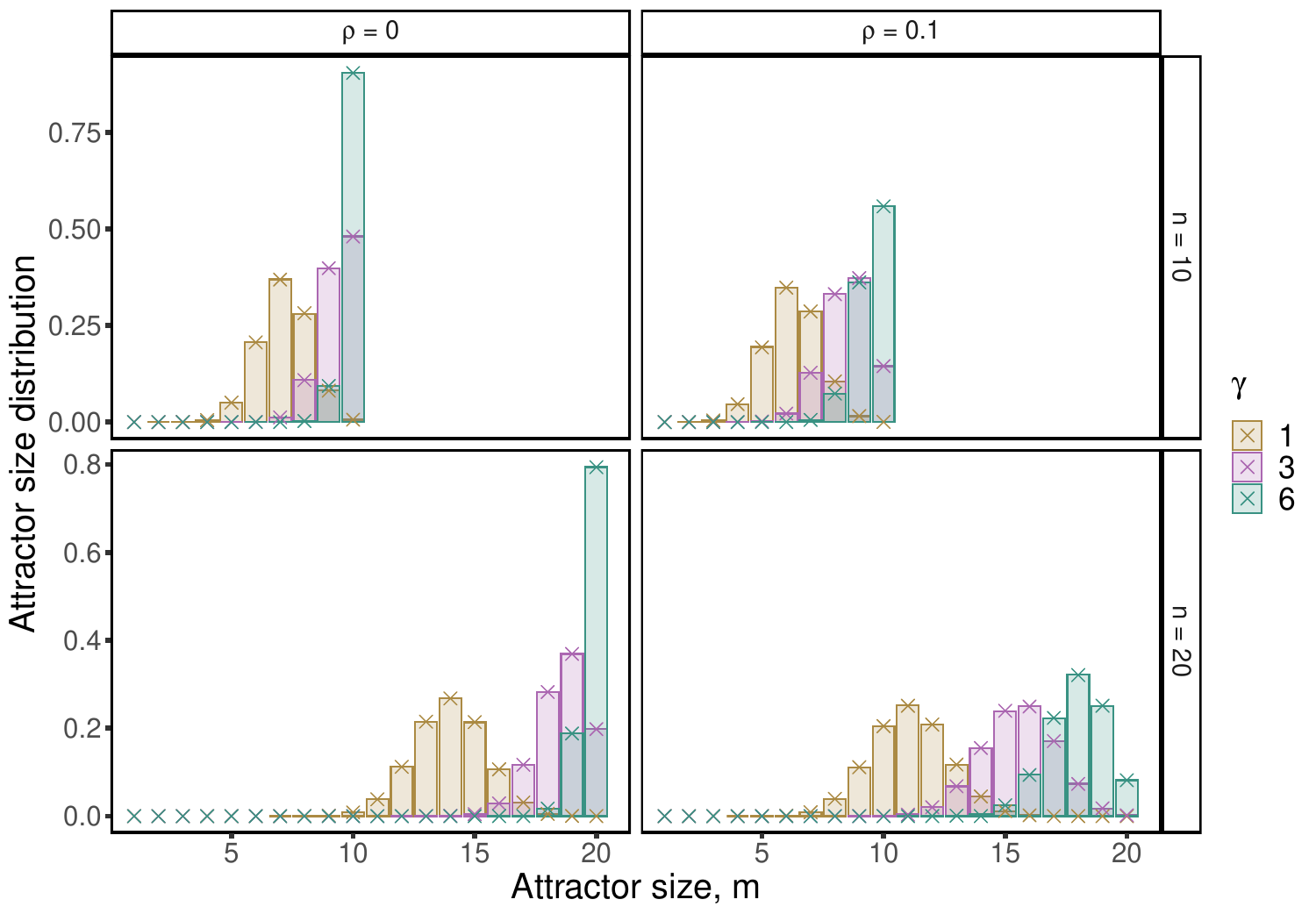}
\caption{\textbf{Distribution of the set of coexisting species as a function of the ratio $\gamma$ of number of traits to number of species for different \emph{constant} correlation matrices.}
The simulations were done with $n = 10$ and $20$ species. Bar are simulations, crosses are numerical evaluations of formula~\eqref{eq:exact_coexistence}.}
\end{figure}

\subsection*{Average number of species}
In this section we will focus on the case of constant correlation. Our aim is to approximate the integrals for feasibility and invasibility in the large number of species limit by a saddle point technique. With these approximations, we provide an analytical way to compute the probability of coexistence $\text{Pr}(\{S\}_m|{n, \ell, \rho})$ ---cf. Eq.~\eqref{eq:exact_coexistence}--- as well as an approximation for the average fraction of species
\begin{equation}\label{eq:average}
\Omega(n,\ell,\rho):=\frac{1}{n}\sum_{m=0}^n \binom{n}{m} mP_{\text{a}}(m,n).
\end{equation}

We distinguish the cases $\rho > 0$ and $\rho = 0$ for invasibility. For $\rho > 0$ we use expression \eqref{eq:exact_inv_constant_rho}. Let us define $q:=m/n$ as the fraction of survivors, and recall that $\ell = n\gamma$. Also let
\begin{equation}
\lambda_q := m w \alpha_w = 1 + \frac{m\rho}{1 - \rho} = 1 + \frac{nq\rho}{1 - \rho} .
\end{equation}
In terms of $\lambda_q$, the probability of non-invasibility reads
\begin{equation}
P_{\text{\text{ni}}}(m,n) =  \frac{\lambda_q}{\sqrt{2 \pi (\lambda_q - 1)}} \int_0^{\infty} dw g(\nu,w) w^{-1/2}
\int_{-\infty}^{\infty} dy e^{-\frac{y^2 \lambda_q^2}{2 w(\lambda_q - 1)}} 
\Phi\left(-\sqrt{\frac{w}{m \lambda_q}} + y \sqrt{\frac{\lambda_q}{mw}}\right)^{n - m}.
\end{equation}
Now we make a change of variables,
\begin{equation}\label{eq:inv_rescaling}
  \begin{aligned}
    w' &= \sqrt{\frac{w}{m}},\\
    \frac{y'}{w'} &= \frac{y}{\sqrt{w m}}.
  \end{aligned}
\end{equation}
Then the integral becomes
\begin{equation}\label{eq:invaux-1}
P_{\text{ni}}(m,n) = \frac{2 \lambda_q}{\sqrt{2 \pi (\lambda_q - 1)}} \int_0^{\infty} dw m^{3/2} g(\nu',mw^2) \int_{-\infty}^{\infty} dy
e^{-\frac{my^2 \lambda_q^2}{2 w^2(\lambda_q - 1)}} \Phi\biggl(-\frac{w}{\sqrt{\lambda_q}} + \frac{y}{w} \sqrt{\lambda_q}\biggr)^{n - m}.
\end{equation}

Recall that the probability density function $g(\nu',x)$, for $\nu'=\ell-m+1$, is:
\begin{equation}
g(\nu,x) = \frac{x^{(\ell - m -1)/2}e^{-x/2}}{2^{(\ell-m+1)/2}\Gamma((\ell-m+1)/2)}
\end{equation}
Hence the integral~\eqref{eq:invaux-1} is
\begin{equation}\label{eq:inv_saddle}
\begin{aligned}
P_{\text{ni}}(m,n) =& \frac{\lambda_qm}{\sqrt{\pi(\lambda_q - 1)}} \frac{(m/2)^{(\ell-m)/2}}{\Gamma((\ell -m +1)/2)} 
\int_0^{\infty} dw w^{\ell-m-1} e^{-mw^2/2} \\
\times & \int_{-\infty}^{\infty} dy e^{-\frac{my^2 \lambda_q^2}{2 w^2(\lambda_q - 1)}} 
\Phi\biggl(-\frac{w}{\sqrt{\lambda_q}} + \frac{y}{w} \sqrt{\lambda_q}\biggr)^{n - m} \\
=& \frac{\lambda_qm}{\sqrt{\pi(\lambda_q - 1)}} \frac{(m/2)^{(\ell-m)/2}}{\Gamma((\ell -m +1)/2)} \int_0^{\infty} dw w^{-1}
\int_{-\infty}^{\infty} dy e^{n F_{\text{ni}}(w, y)},
\end{aligned}
\end{equation}
where the exponent $F_{\text{ni}}(w,y)$ has been defined as
\begin{equation}
F_{\text{ni}}(w,y) := (\gamma - q)\log (w) - \frac{q w^2}{2} - \frac{q y^2 \lambda_q^2}{2 w^2 (\lambda_q - 1)} + 
(1- q)\log\Phi\biggl(-\frac{w}{\sqrt{\lambda_q}} + \frac{y}{w} \sqrt{\lambda_q}\biggr).
\end{equation}

Now we evaluate the double integral in the limit $n\to\infty$ via a saddle-point technique. For that purpose, since the exponential becomes peaked around the maximum of the exponent, we calculate the equations to be satisfied by the critical point. Taking derivatives of the exponent we get
\begin{equation}
\begin{aligned}
\frac{\partial F_{\text{ni}}}{\partial y} &=  -\frac{q y \lambda_q^2}{w^2 (\lambda_q - 1)} +  \frac{(1-q)\sqrt{\lambda_q}}{w}
\frac{\phi\Bigl(-\frac{w}{\sqrt{\lambda_q}} + \frac{y}{w} \sqrt{\lambda_q}\Bigr)}
%\frac{e^{-\frac{1}{2}\big(\frac{w}{\sqrt{\lambda_q}} - \frac{y}{w} \sqrt{\lambda_q}\big)^2}}
{\Phi\Bigl(-\frac{w}{\sqrt{\lambda_q}} + \frac{y}{w} \sqrt{\lambda_q}\Bigr)}, \\
\frac{\partial F_{\text{ni}}}{\partial w} &= \frac{\gamma -q}{w} - qw + \frac{qy^2 \lambda_q^2}{w^3(\lambda_q -1)} - 
(1-q)\left(\frac{1}{\sqrt{\lambda_q}} +\frac{y \sqrt{\lambda_q}}{w^2}\right)
\frac{\phi\Bigl(-\frac{w}{\sqrt{\lambda_q}} + \frac{y}{w} \sqrt{\lambda_q}\Bigr)}
{\Phi\Bigl(-\frac{w}{\sqrt{\lambda_q}} + \frac{y}{w} \sqrt{\lambda_q}\Bigr)} .
\end{aligned}
\end{equation}
Therefore at a critical point $(w^{\star}, y^{\star})$ we have the following conditions:
\begin{equation}\label{eq:inv_crit_cond}
\begin{aligned}
-\frac{q y \lambda_q^{3/2}}{w (\lambda_q - 1)} + 
(1-q)\frac{\phi\Bigl(-\frac{w}{\sqrt{\lambda_q}} + \frac{y}{w} \sqrt{\lambda_q}\Bigr)}
{\Phi\Bigl(-\frac{w}{\sqrt{\lambda_q}} + \frac{y}{w} \sqrt{\lambda_q}\Bigr)} = 0, \\
\gamma -q - qw^2 - \frac{q y\lambda_q}{\lambda_q - 1} = 0.
\end{aligned}
\end{equation}

Similarly we can rewrite the integral for the probability that an endpoint formed by $m$ species is feasible, see Eq.~\eqref{eq:feasibility_exact_constant_rho}, as 
\begin{equation}\label{eq:feas_saddle_init}
P_{\text{f}}(m) = -i\sqrt{\frac{\lambda_q}{2 \pi}} \int_0^{\infty}du g(\nu,u) u^{-1/2} \int_\Gamma d\zeta e^{\frac{\lambda_q\zeta^2}{2 u}} 
\Phi\biggl(\sqrt{\frac{u}{m \lambda_q}} + \zeta \sqrt{\frac{\lambda_q}{m u}}\biggr)^{m},
\end{equation}
where now the number of degrees of freedom is $\nu=\ell-m+2$. 

Following essentially the same procedure as before, i.e. making a change of variables and replacing the density function for the $\chi^2_{\nu}$ distribution we get 
\begin{equation}\label{eq:feas_saddle}
P_{\text{f}}(m) = -im^{3/2}\sqrt{\frac{\lambda_q}{2 \pi}} \frac{(m/2)^{(\ell - m)/2}}{\Gamma((\ell - m)/2 + 1)} 
\int_{-\infty}^{\infty}du  \int_\Gamma d\zeta e^{n F_{\text{f}}(u,\zeta)},
\end{equation}
with the exponent
\begin{equation}
F_{\text{f}}(u, \zeta) := (\gamma- q)\log(u) - \frac{q u^2}{2} + \frac{q\lambda_q \zeta^2}{2u^2} + 
q\log\Phi\biggl(\frac{u}{\sqrt{\lambda_q}} + \frac{\zeta}{u} \sqrt{\lambda_q}\biggr).
\end{equation}
Similarly, the conditions satisfied by the critical point $(u^{\star}, \zeta^{\star})$ are
\begin{equation}\label{eq:feas_crit_cond}
  \begin{aligned}
    \frac{\zeta\sqrt{\lambda_q}}{u} + \frac{\phi\Bigl(\frac{u}{\sqrt{\lambda_q}} + \frac{\zeta}{u} \sqrt{\lambda_q}\Bigr)}
    {\Phi\Bigl(\frac{u}{\sqrt{\lambda_q}} + \frac{\zeta}{u} \sqrt{\lambda_q}\Bigr)} &= 0, \\
    \gamma - q - q u^2 - q \zeta &= 0.
  \end{aligned}
\end{equation}

Notice that the product of the densities of the $\chi^2$ distributions in each integral ---Eqs.~\eqref{eq:inv_saddle} and~\eqref{eq:feas_saddle}--- introduce an extra term which scales exponentially with $m=nq$, namely
\begin{equation}
  \frac{m^{\ell-m}}{2^{\ell-m}\Gamma((\ell -m)/2 +1)\Gamma((\ell-m)/2 + 1/2)} = \frac{m^{\ell-m}}{\Gamma(\ell -m +1)}.
\end{equation}
Using the Stirling's asymptotic form of the gamma function we get
\begin{equation}
\frac{m^{\ell-m}}{\Gamma(\ell -m +1)} \sim \frac{e^{n(\gamma -q) (1 + \log q - \log (\gamma - q))}}{\sqrt{2 \pi n(\gamma - q)}}.
\end{equation}
Let
\begin{equation}
F_{\text{e}}(q) := (\gamma -q) (1 + \log q - \log (\gamma - q))
\end{equation}
and
\begin{equation}
F_{\text{c}}(q) := -q\log q - (1-q) \log(1-q),
\end{equation}
$F_{\text{c}}(q)$ being the exponent appearing in Stirling's asymptotic formula for the binomial coefficient $\binom{n}{nq}$. Consequentely the probability that the system settles in an endpoint with $m = nq$ species is given, up to a normalization factor, by:
\begin{equation}\label{eq:coexistence_prob_asymptotic}
\text{Pr}(\{S\}_m|{n, \ell, \rho}) =  \binom{n}{m} P_{\text{a}}(m,n) \sim \exp\{n(F_{\text{f}}(u^{\star},\zeta^{\star},q) + F_{\text{ni}}(w^{\star},y^{\star},q) + F_{\text{e}}(q) + F_{\text{c}}(q))\}.
\end{equation}
Observe that critical point coordinates $u^{\star}$, $\zeta^{\star}$, $w^{\star}$ and $y^{\star}$ depend implicitly on $q$ through~\eqref{eq:inv_crit_cond} and~\eqref{eq:feas_crit_cond}. Observe that one can use the asymptotic expansion~\eqref{eq:coexistence_prob_asymptotic} to obtain numerically the distribution of the number of survivors, $\text{Pr}(\{S\}_m|{n, \ell, \rho})$, up to a normalization factor. The calculation amounts to solve numerically the non-linear systems~\eqref{eq:inv_crit_cond} and~\eqref{eq:feas_crit_cond}.

We are now ready to provide an analytical approximation for the mean fraction of survivors $\Omega$,
cf. Eq.~\eqref{eq:average}. In the limit of large pool size $n$, we can approximate the mean of the distribution
$\Pr(\{S\}_m|m, \ell, \rho)$ by its mode, which is easier to compute. In fact, to calculate the mode of the distribution $q$ in the large $n$ limit we need to find the $q^{\star}$ value that maximizes the exponent in~\eqref{eq:coexistence_prob_asymptotic}. Due to the critical point conditions for $(u^{\star}, \zeta^{\star})$ and $(w^{\star}, y^{\star}$), $q^{\star}$ satisfies
\begin{equation}
\frac{\partial F_{\text{f}}}{\partial q} + \frac{\partial F_{\text{ni}}}{\partial q} + \frac{\partial F_{\text{e}}}{\partial q} + \frac{\partial F_{\text{c}}}{\partial q} = 0.
\end{equation}
Evaluated at the critical points $(u^{\star}, \zeta^{\star})$ and $(w^{\star}, y^{\star}$), the derivatives read
\begin{equation}
  \begin{aligned}
    \frac{\partial F_{\text{ni}}}{\partial q} &= -\log(w) - \frac{w^2}{2} - \frac{y^2 \lambda_q}{2 w^2} + \frac{y}{2} 
    - \log\Phi\left(-\frac{w}{\sqrt{\lambda_q}} + \frac{y}{w} \sqrt{\lambda_q}\right), \\
    \frac{\partial F_{\text{f}}}{\partial q} &= -\log(u) -\frac{u^2}{2} + \lambda_q \frac{\zeta^2}{2 u^2} + \frac{\zeta(\lambda_q -1)}{2 \lambda_q} 
    + \log\Phi\left(\frac{u}{\sqrt{\lambda_q}} + \frac{\zeta}{u} \sqrt{\lambda_q}\right), \\
    \frac{\partial F_{\text{e}}}{\partial q} &= \log\left(\frac{\gamma - q}{q}\right)  + \frac{\gamma - q}{q}  = \log\left(\frac{\gamma - q}{q}\right) 
    + \frac{u^2}{2} + \frac{w^2}{2} + \frac{q \zeta}{2} + \frac{q y \lambda_q}{2 (\lambda_q-1)},\\
    \frac{\partial F_{\text{c}}}{\partial q} &= \log(1-q)-\log q.
  \end{aligned}
\end{equation}
Therefore the condition for $q^{\star}$ reduces to
\begin{equation}\label{eq:qaux}
  -\log\left(\frac{q w u}{\gamma - q}\right) +\frac{\lambda_q}{2} \left( \frac{\zeta^2}{u^2} - \frac{y^2}{w^2}\right) + 
  \frac{2\lambda_q - 1}{2}\left(\frac{y}{\lambda_q - 1} + \frac{\zeta}{\lambda_q}\right)  + 
  \log\frac{(1-q)\Phi\Bigl(\frac{u}{\sqrt{\lambda_q}}+\frac{\zeta}{u} \sqrt{\lambda_q}\Bigr)}
  {q\Phi\Bigl(-\frac{w}{\sqrt{\lambda_q}} + \frac{y}{w} \sqrt{\lambda_q}\Bigr)} = 0.
\end{equation}
A direct calculation shows that, at $wu = \frac{\gamma - q}{q}$, the terms up to the last logarithm vanish. We now show
that the last one can be written as $\big(wu - \frac{\gamma -q}{q}\big) h$ for some function $h$.

Indeed, using conditions~\eqref{eq:feas_crit_cond} and~\eqref{eq:inv_crit_cond} we have
\begin{equation}
\frac{(1-q)\phi(-w,-y,q)}{q\Phi(-w,-y,q)} - \frac{\phi(u,\zeta,q)}{\Phi(u,\zeta,q)} = \frac{(u+w)\sqrt{\lambda_q}}{u w}\left(\frac{\gamma - q}{q} - uw\right),
\end{equation}
where we have used the abbreviations $\Phi(u,\zeta,q):=\Phi\Bigl(\frac{u}{\sqrt{\lambda_q}}+\frac{\zeta}{u} \sqrt{\lambda_q}\Bigr)$ and $\phi(u,\zeta,q):=\phi\Bigl(\frac{u}{\sqrt{\lambda_q}}+\frac{\zeta}{u} \sqrt{\lambda_q}\Bigr)$ to simplify notation. Therefore,
\begin{equation}
\frac{(1-q)\Phi(u,\zeta,q)}{q\Phi(-w,-y,q)} = \frac{\phi(u,\zeta,q)}{\phi(-w,-y,q)}
+ \frac{(u+w)\Phi(u,\zeta,q)\sqrt{\lambda_q}}{u w\phi(-w,-y,q)}\left(\frac{\gamma - q}{q} - uw\right).
\end{equation}
Letting $\mu_q := (\gamma - q)/q$, it holds that
\begin{equation}
\frac{\phi(u,\zeta,q)}{\phi(-w,-y,q)} = e^{(\mu_q^2 - (uw)^2)((\lambda_q - 1)^2u^2-\lambda_q^2w^2)/(2\lambda_q u^2w^2)}.
\end{equation}
Now, due to the series representation of the exponential function we have
\begin{equation}
\frac{\phi(u,\zeta,q)}{\phi(-w,-y,q)}  = 1 + (\mu_q - uw) h(u,w),
\end{equation}
where
\begin{multline}
h(u,w) := \frac{q(u+w)\Phi(u,\zeta,q)\sqrt{\lambda_q}}{u w\phi(-w,-y,q)} \\
+ \sum_{j = 1}^{\infty} \frac{1}{j!}(\mu_q - uw)^{j-1}\left((\mu_q + uw)\frac{(\lambda_q - 1)^2u^2-\lambda_q^2w^2}{2\lambda_q u^2w^2}\right)^j.
\end{multline}
Thus, the claim follows by using the series expansion of $\log(1 +x)$. Therefore, all the terms in~\eqref{eq:qaux} vanish at $uw = \mu_q$.

We have just shown that the last logarithm in~\eqref{eq:qaux} is equal to zero. Consequently $q^{\star}$ satisfies
\begin{equation}
\frac{(1-q)\Phi\Bigl(\frac{u}{\sqrt{\lambda_q}}+\frac{\zeta}{u} \sqrt{\lambda_q}\Bigr)}{q\Phi\Bigl(-\frac{w}{\sqrt{\lambda_q}} + \frac{y}{w} \sqrt{\lambda_q}\Bigr)} = 1.
\end{equation}
At the point $uw = \mu_q$ we can write
\begin{equation}\label{eq:uwaux}
\frac{u}{\sqrt{\lambda_q}}+\frac{\zeta}{u} \sqrt{\lambda_q} = \frac{\lambda_q w - (\lambda_q - 1) u}{\sqrt{\lambda_q}} = \frac{w}{\sqrt{\lambda_q}} - \frac{y}{w}\sqrt{\lambda_q},
\end{equation}
which in turn implies that
\begin{equation}
\Phi\left(\frac{\lambda_q w - (\lambda_q - 1) u}{\sqrt{\lambda_q}}\right) = q^{\star}.
\end{equation}
Let $\hat{q} := \Phi^{-1}(q^{\star})=\sqrt{2}\text{erf}^{-1}(2q^{\star}-1)$, for $\text{erf}^{-1}$ the inverse error
function. Then it holds that $(\lambda_q w - (\lambda_q - 1) u)/\sqrt{\lambda_q}=\hat{q}$ and using
\cref{eq:feas_crit_cond} we can solve for $u^{\star}, w^{\star}$ in terms of $\hat{q}$, yielding
\begin{equation}\label{eq:sol_sp}
  \begin{aligned}
    u^{\star} &= \sqrt{\lambda_q}\left(\frac{\phi(\hat{q})}{q^{\star}} + \hat{q}\right), \\
    w^{\star} &= \frac{1}{\sqrt{\lambda_q}} \left((\lambda_q - 1)\frac{\phi(\hat{q})}{q^{\star}} +\lambda_q \hat{q}\right).
  \end{aligned}
\end{equation}
The final condition for $q^{\star}$ at the saddle point reduces to substitute the expressions above into the condition $uw = \mu_q$, which finally reads
\begin{equation}\label{eq:saddle_point_mode_cond}
\frac{\gamma}{q^{\star}} = 1+ \left(\frac{\phi(\Phi^{-1}(q^{\star}))}{q^{\star}} + \Phi^{-1}(q^{\star})\right) 
\left(\frac{\phi(\Phi^{-1}(q^{\star}))}{q^{\star}}(\lambda_{q^{\star}} - 1) +\Phi^{-1}(q^{\star})\lambda_{q^{\star}}\right).
\end{equation}

The case $\rho = 0$ for invasibility is similar, and simpler.

\subsubsection*{Level Curves}
Eq.~\eqref{eq:saddle_point_mode_cond} gives a very good approximation to the level curves on the $(\rho, \gamma)$ plane mapping to constant mean fraction of survivors $q=m/n$. This implicit condition can be rewritten equivalently as
\begin{equation}
\gamma = q + \Phi^{-1}(q)H(q) + \frac{n \rho H(q)^2}{1-\rho},
\end{equation}
where $H(q) := \phi(\Phi^{-1}(q)) + q \Phi^{-1}(q)$. This condition is compared with simulation results in Figure~4 of the main text (right panel).

%%% Local Variables:
%%% mode: latex
%%% TeX-master: "sup_info"
%%% End:

\section{Total biomass distribution at endpoints}
\begin{figure}[t!]
\includegraphics[width = 0.99\linewidth]{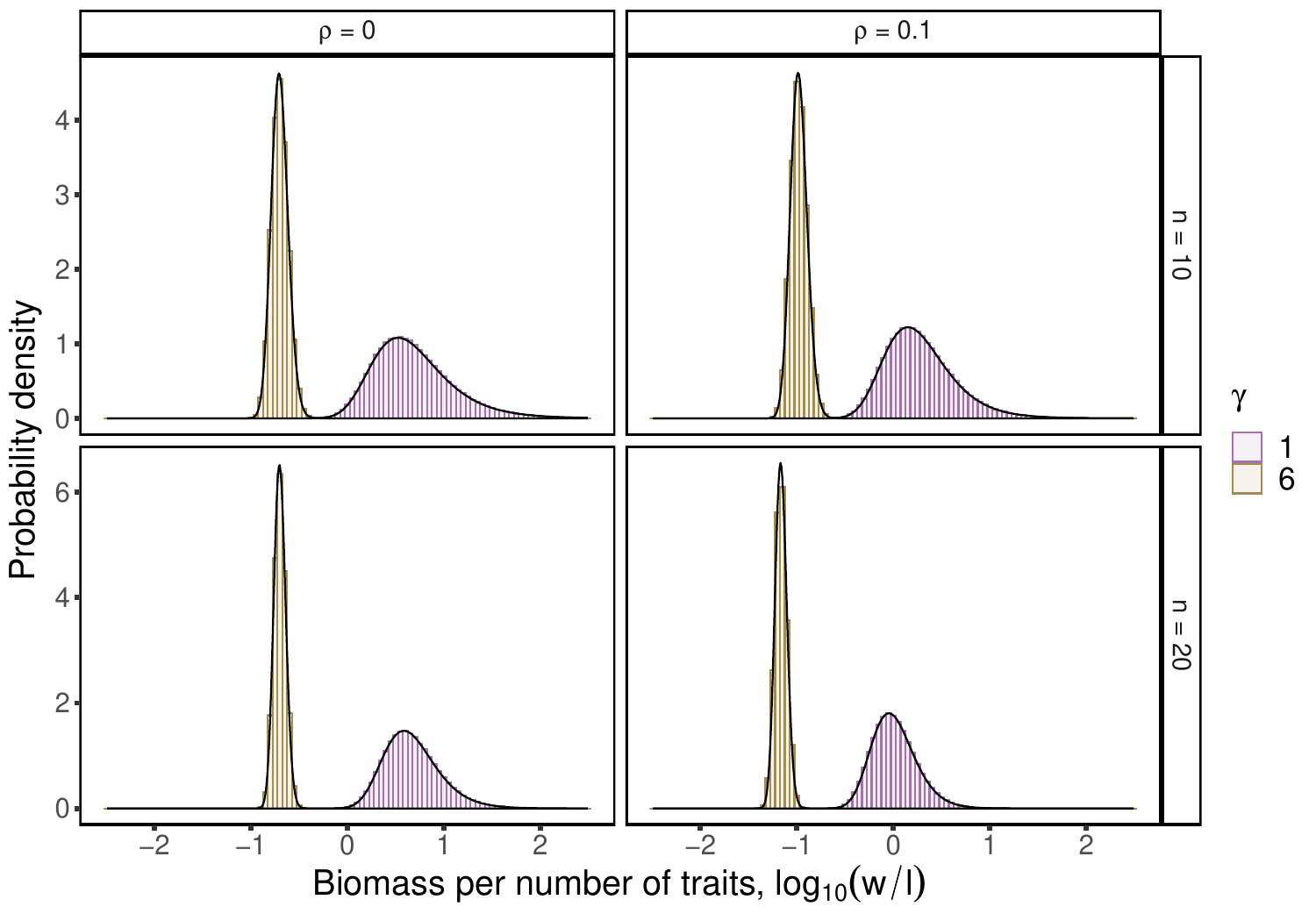}
\caption{\label{fig:biomass}
\textbf{Distribution of the total biomass $w$ of the survival community as a function of the ratio $\gamma$ of number of traits $k$ to number of species $n$ for different \emph{constant} correlation matrices.} The simulations were done with $n = 10, 20$ species. Histograms are simulations and black lines are the numerical integration of Eq.~\eqref{eq:biomass_density_attractor}.}
\end{figure}

The proof of independence of invasibility and feasibility (\cref{sec:independence_feasibility_invasibility}) also shows that, for any fixed size $m$ of a subset of species and total biomass $w$, we have that $\text{Pr}(\bm{z}_{n-m} < \bm{0}_{n-m} \vert \bm{x}_m > \bm{0}_m, W=w) = \text{Pr}(\bm{z}_{n-m} < \bm{0}_{n-m} \vert W = w)$. This remark, together with the independence of $W$ and $ \bm{x}_m > \bm{0}_m$ (feasibility), helps us derive the distribution of total biomass. To simplify notation we do not rescale the interaction matrix by $\ell$ (as shown in \cref{section:lv_ops} this would amount to a rescaling of total biomass $w \to \ell w$). The cdf for the random variable $W$ is precisely
\begin{equation}
 \text{Pr}(W < w) = \sum_{m=0}^n \binom{n}{m} P_{\text{a}}(m,n) \text{Pr}(W < w | m),
\end{equation}
where $\text{Pr}(W < w | m)$ is the probability that $W < w$ conditional on the $m$-species endpoint is feasible and non-invasible. Thus,
\begin{equation}\label{eq:cdfW}
  \begin{aligned}
    \text{Pr}(W < w \vert m ) &= \frac{\text{Pr}(W < w, \bm{x}_m > \bm{0}_m, \bm{z}_{n-m} < \bm{0}_{n-m})}{P_{\text{a}}(m,n)} \\
    &= \frac{\text{Pr}(W < w, \bm{z}_{n-m} < \bm{0}_{n-m} \vert \bm{x}_m > \bm{0}_m) P_{\text{f}}(m)}{P_{\text{a}}(m,n)} \\
    &= \frac{\text{Pr}(W < w, \bm{z}_{n-m} < \bm{0}_{n-m}) P_{\text{f}}(m)}{P_{\text{a}}(m,n)},
  \end{aligned}
\end{equation}
the last equality following from the statement in the paragraph above. Now, using the notations introduced in the last section, it holds that
\begin{multline}
\text{Pr}(W < w, \bm{z}_{n-m} < \bm{0}_{n-m}) = \int_{0}^\infty du g(\nu',u) \Theta(u-w^{-1}\bm{1}_m^T\Sigma_{11}^{-1}\bm{1}_m) \\
\times Q^{-}_{n-m}(\bm{1}_{n-m}-\Sigma_{21}\Sigma_{11}^{-1}\bm{1}_m, u^{-1}\bm{1}_m^T\Sigma_{11}^{-1}\bm{1}_m\Sigma_{22.1}).
\end{multline}
Hence, using~\eqref{eq:cdfW} and $P_{\text{a}}(m,n)=P_{\text{f}}(m)P_{\text{ni}}(m,n)$, and taking derivatives with respect to $w$ in~\eqref{eq:cdfW}, the probability density function of the biomass distribution can be expressed as
\begin{multline}\label{eq:biomass_density_attractor}
    g_{\text{a}}(w) = \sum_{m=0}^n \binom{n}{m} P_{\text{f}}(m) \frac{\partial \text{Pr}(W < w, \bm{z}_{n-m} < \bm{0}_{n-m})}{\partial w}\\
    =  \sum_{m=0}^n \binom{n}{m} \frac{\tilde{w}}{w} P_{\text{f}}(m) g(\nu',\tilde{w}) Q^{-}_{n-m}(\bm{1}_{n-m}-\Sigma_{21}\Sigma_{11}^{-1}\bm{1}_m, \tilde{w}^{-1}\bm{1}_m^T\Sigma_{11}^{-1}\bm{1}_m\Sigma_{22.1}),
\end{multline}
where $\tilde{w} := w^{-1}\bm{1}_m^T\Sigma_{11}^{-1}\bm{1}_m$. Figure~\ref{fig:biomass} shows the comparison of \eqref{eq:biomass_density_attractor} with simulations for the constant correlation case in the case in which the interaction matrix is rescaled by the number of traits.

Going back to re-scaling the interaction matrix by $\ell$, total biomass transforms as $w \to \ell w$ (recall that in Section~S3 we considered interaction matrices $A$ as samples of the $\mathcal{W}_n(\Sigma,\ell)$ because scaling $A=GG^T$ by multiplying $GG^T$ by $\ell^{-1}$ does not affect the number of species in the endpoint). By the above calculation,
\begin{multline}
    g_{\text{a}}(w|m) = \frac{1}{P_{\text{ni}}(m,n)} \frac{\partial \text{Pr}(W < w, \bm{z}_{n-m} < \bm{0}_{n-m})}{\partial w}\\
    =  \frac{\tilde{w} g(\nu',\tilde{w})}{w P_{\text{ni}}(m,n)}  Q^{-}_{n-m}(\bm{1}_{n-m}-\Sigma_{21}\Sigma_{11}^{-1}\bm{1}_m, \tilde{w}^{-1}\bm{1}_m^T\Sigma_{11}^{-1}\bm{1}_m\Sigma_{22.1}),
\end{multline}
with $\tilde{w} := w^{-1}\bm{1}_m^T\Sigma_{11}^{-1}\bm{1}_m$. Now, the moments of the distribution of $\ell W$ conditional to $m$ coexisting species, defined as
\begin{equation}
\E[(\ell W)^k|m] = \int_0^{\infty} dw (\ell w)^k g_{\text{a}}(w|m),
\end{equation}
can be calculated by making, in the last integral, the change of variables $w\to \tilde{w}$ defined by $w = \bm{1}_m^T\Sigma_{11}^{-1}\bm{1}_m/\tilde{w}$, giving
\begin{multline}\label{eq:lwk}
\E[(\ell W)^k|m] = \frac{1}{P_{\text{ni}}(m,n)}\int_0^{\infty}  dw g(\nu',w) \\
\times (\ell w^{-1} \bm{1}_m^T\Sigma_{11}^{-1}\bm{1}_m)^k Q^{-}_{n-m}(\bm{1}_{n-m}-\Sigma_{21}\Sigma_{11}^{-1}\bm{1}_m, w^{-1}\bm{1}_m^T\Sigma_{11}^{-1}\bm{1}_m \Sigma_{22.1}),
\end{multline}
where we have written the integration variable as $w$ to ease notation.

Now, particularize to the star phylogeny case, and focus on the average biomass ($k=1$ moment). Observe that the last integral coincides with that of Eq.~\eqref{eq:exact_inv_gen} except for the factor $\ell w^{-1} \bm{1}_m^T\Sigma_{11}^{-1}\bm{1}_m$. Then the saddle point calculation done while computing the expected number of survivors can be reproduced here to approximate the mean of $\ell W|m$ for $\rho \geq 0$, $m = nq $ and $\ell = \gamma n$. Following the same steps leading to Eq.~\eqref{eq:inv_saddle}, the integral we have to evaluate reduces to~\eqref{eq:inv_saddle} up to a multiplication by $\frac{\gamma}{qw^2}\bm{1}_m^T\Sigma_{11}^{-1}\bm{1}_m$. Indeed, observe that the reescaling $w'=\sqrt{w/m}$, given in Eq.~\eqref{eq:inv_rescaling} and used to obtain~\eqref{eq:inv_saddle}, introduces an extra factor $(mw^2)^{-1}$ when substituted into the $w^{-1}$ factor appearing in~\eqref{eq:lwk}, so the combination $\ell w^{-1}$ transforms into $\ell/(mw^2)=\gamma/(qw^2)$. 

Hence the exponent in the integral~\eqref{eq:inv_saddle} does not change so, when the integral is evaluated at the saddle point (at the solution $(y^{\star},w^{\star})$ of \eqref{eq:inv_crit_cond}), the term $\frac{\gamma}{(w^{\star})^2 q}\bm{1}_m^T\Sigma_{11}^{-1}\bm{1}_m$ can be factored out of the integral, yielding
\begin{equation}
\E[\ell W|m] \approx \frac{\gamma\bm{1}_m^T\Sigma_{11}^{-1}\bm{1}_m}{(w^{\star})^2 qP_{\text{ni}}(m,n)}\int_0^{\infty}  dw \left. g(\nu',w)
Q^{-}_{n-m}(\bm{1}_{n-m}-\Sigma_{21}\Sigma_{11}^{-1}\bm{1}_m, w^{-1}\bm{1}_m^T\Sigma_{11}^{-1}\bm{1}_m \Sigma_{22.1})\right\vert_{\text{s.p.}},
\end{equation}
where by s.p. we mean that the integral has to be evaluated at the saddle point. However, the integral trivially reduces to $P_{\text{ni}}(m,n)$ at the saddle point, which implies that
\begin{equation}
\E[\ell W|m] \approx \frac{\gamma}{(w^{\star})^2 q}\bm{1}_m^T\Sigma_{11}^{-1}\bm{1}_m.
\end{equation}
Therefore, neglecting all but the leading order terms in the asymptotic expansion and using that $\bm{1}_m^T\Sigma_{11}^{-1}\bm{1}_m=m/(1-\rho+\rho m)$, we can approximate
\begin{equation}
\E[\ell W|m] \approx \frac{\ell }{(1 -\rho + \rho m)(w^{\star})^2}.
\end{equation}
Assuming that the distribution of survivors is highly peaked at the mode, we can approximate the mean of $W$ by the mean conditional at the mode $q^{\star}$, which we get from Eq.~\eqref{eq:saddle_point_mode_cond}:
\begin{equation}\label{eq:bio_app}
\E[\ell W] \approx \frac{\ell}{(1 - \rho + \rho q^{\star}n)w^{\star}(q^{\star})^2}.
\end{equation}
This is the expression we compared to simulations in Figure~5 of the main text (left panel). Observe that $w^{\star}(q^{\star})$ can be calculated as function of $q^{\star}$ using Eq.~\eqref{eq:sol_sp}.

%%% Local Variables:
%%% mode: latex
%%% TeX-master: "sup_info"
%%% End:

\section{Relative abundances}
For an equilibrium attractor $\bm{x}_m$ with $m$ species, let $\bm{v} := \bm{x}_m/\sum_{i=1}^m x_m^i$ be the relative abundance vector. In particular, $v_m = 1 - \sum_{i = 1}^{m-1} v_i = 1 - \bm{1}^T_{m-1}\widetilde{\bm{v}}$, for $\widetilde{\bm{v}}$ the vector of the first $m-1$ relative abundances. By section~\ref{sec:Wish}, Eq.~\eqref{eq:t}, we know that the vector $\widetilde{\bm{v}}$ follows a multivariate $t$ distribution, so we can write, following the same steps that led to the probability of feasibility~\eqref{eq:pfeas}, the distribution function for $v_m$ conditional on $\bm{x}_m$ being feasible as
\begin{multline}\label{eq:ccdf}
\text{Pr}(v_m < c \vert \bm{x}_m > \bm{0}_m) = 1 - \text{Pr}(v_m > c\vert \bm{x}_m > \bm{0}_m) \\
= 1 - \frac{1}{P_{\text{f}}(m)}\int_{0}^{\infty} du g(\nu,u) \text{Pr}(\bm{y}_u > \bm{0}_{m-1}, \bm{1}_{m-1}^T\bm{y}_u < 1 - c)
\end{multline}
with $\nu=\ell-m+2$. 
%The independence of $\bm{v}$ and invasibility gives us the distribution of $v$ conditional to $\bm{x}_m$ being an attractor of the system with $m$ out of $n$ survivors. Let $\bm{z}_{n-m}$ be defined as in  section~\ref{sec:Wish}.
%Organize $x$ in the first $m$ elements of the equilibrium vector, $A$ in the same way as in the invasibility section and let $z = A_{12}x$.
Then
\begin{multline}
    \text{Pr}(v_m < c \vert m) = \frac{\text{Pr}(v_m < c, \bm{x}_m > \bm{0}_m, \bm{z}_{n-m} < \bm{0}_{n-m})}{P_{\text{a}}(m,n)}  \\
    = \frac{\text{Pr}(\bm{z}_{n-m} < \bm{0}_{n-m} \vert \bm{x}_m > \bm{0}_m, v_m < c) \text{Pr}(v_m < c \vert \bm{x} > \bm{0}_m)}{P_{\text{ni}}(m,n)} 
    = \text{Pr}(v_m < c \vert \bm{x} > \bm{0}_m),
\end{multline}
where we have used the independence of feasibility and invasibility, $P_{\text{a}}(m,n)=P_{\text{f}}(m)P_{\text{ni}}(m,n)$, and the fact that $\text{Pr}(\bm{z}_{n-m} < \bm{0}_{n-m} \vert \bm{x}_m > \bm{0}_m, v_m < c) = \text{Pr}(\bm{z}_{n-m} < \bm{0}_{n-m}) = P_{\text{ni}}(m,n)$. The last expression follows from Eq.~\eqref{eq:indep}, which states that the event $\bm{z}_{n-m} < \bm{0}_{n-m}$ is independent of the event $\bm{x}_m > \bm{0}_m$, from which follows that it is also independent on conditioning on a subset of values of the $m$-th relative abundance, $v_m < c$. Therefore, we can calculate the distribution function $\text{Pr}(v_m < c \vert m)$ of observing the $m$-th relative abundance, $v_m$, conditional on the community having $m$ extant species, using Eq.~\eqref{eq:ccdf}.

In case of a constant correlation $\rho \geq 0$, all species are equivalent so any surviving species $i$ has the same distribution as $x_m$. Applying the same derivation as for the feasibility case, and using the notation of the saddle point calculation with $m = qn$ (see Eq.~\eqref{eq:feas_saddle_init}), we get
\begin{multline}
  \text{Pr}(v_m < c\vert m) = 1 - \frac{i\sqrt{\lambda_q}}{\sqrt{2\pi} P_{\text{f}}(m)} \int_0^{\infty} du g(\nu,u)u^{-1/2} \int_\Gamma d\zeta
    e^{\frac{\lambda_q \zeta^2 }{2u}} \\
    \times \Phi\left(\sqrt{\frac{u}{n\lambda_q}}  + \zeta \sqrt{\frac{\lambda_q}{nu}}\right)^{m-1}
    \Phi\left(\sqrt{\frac{u}{n\lambda_q}}  -c \sqrt{\frac{nu}{\lambda_q}} + \zeta \sqrt{\frac{\lambda_q}{nu}}\right).
\end{multline}
Letting $\tilde{c} = cn$, the integral above can be approximated by the same saddle point calculation we did for feasibility (section~\ref{sec:Wish}) up to a multiplication factor given by
\begin{equation}
 \frac{\Phi\Bigl(\frac{u}{\sqrt{\lambda_q}}(1 - \tilde{c}q) + \frac{\zeta}{u}\sqrt{\lambda_q}\Bigr)}{\Phi\Bigl(\frac{u}{\sqrt{\lambda_q}} + \frac{\zeta}{u}\sqrt{\lambda_q}\Bigr)}.
\end{equation}
Thus, for $(u,\zeta)$ satisfying the system of equations~\eqref{eq:feas_crit_cond} with $\zeta$ real, we get an asymptotic approximation of the integral by neglecting all but the leading terms, which reduces to the following expression for the distribution function:
\begin{equation}
\text{Pr}(v_m < c\vert m) = 1 -   \frac{\Phi\Bigl(\frac{u}{\sqrt{\lambda_q}}(1 - \tilde{c}q^{\star}) + \frac{\zeta}{u}\sqrt{\lambda_q}\Bigr)}{\Phi\Bigl(\frac{u}{\sqrt{\lambda_q}} + \frac{\zeta}{u}\sqrt{\lambda_q}\Bigr)}.
\end{equation}
This distribution was compared to simulations in the main text (Figure~5, right panel). In this expression, the variables $u$, $\zeta$, and $\lambda_q$ are evaluated as functions of the mode $q^{\star}$ via the analytical expressions appearing in the saddle-point calculation subsection.

%%% Local Variables:
%%% mode: latex
%%% TeX-master: "sup_info"
%%% End:

%%% Local Variables:
%%% mode: latex
%%% TeX-master: "sup_info"
%%% End:

\section{Invariant Lotka-Volterra operations}\label{section:lv_ops}
In this section we detail the operations that can be performed in a symmetric stable GLV system without changing the subset of coexisting species.

Let $\bm{r} \in \R^n$ be the vector of growth rates, and $A \in \R^n$ a symmetric and positive definite interaction matrix. Let $\{S\}_m \subset \{1,\ldots, n\}$ be the \emph{unique} subset of $m$ species that form the attractor, with vector of densities $\bm{x}=(x_i)$. Then $\bm{x}$ satisfies:

\vspace*{-3mm}
\begin{equation}\label{eq:end_point_conditions}
  \begin{cases}
    x_i > 0, & i \in \{S\}_m,\\
    x_i(\bm{r}-A\bm{x})_i = 0, & \text{for all } $i$,  \\
    (\bm{r}-A\bm{x})_i  < 0, & i \notin \{S\}_m.
  \end{cases}
\end{equation}
Then we can easily see the effect of the following operations on $A$ and $\bm{r}$ on the attractor $\bm{x}$. Let $\kappa > 0$ and $D$ a positive diagonal matrix. The operations that maintain the identity of the species in the endpoint are:
\begin{enumerate}
\item $A \to \kappa A$: then $\bm{x} \to \kappa^{-1} \bm{x}$.
\item $\bm{r} \to \kappa \bm{r}$: Then $\bm{x} \to \kappa \bm{x}$ .
\item $A \to DAD, \bm{r} \to D \bm{r}$: Then $\bm{x} \to D^{-1}\bm{x}$.
\end{enumerate}
After any of these operations, the set of coexisting species remains \emph{unchanged}.

Additionally, in the case of $\bm{r} = \kappa \bm{1}_n$, for $\kappa > 0$, we can perform an additional operation:

\vspace*{-3mm}
\begin{equation}
  A \to B =  A + \mu \bm{1}_n\bm{1}_n^T.
\end{equation}
Then shifting

\vspace*{-3mm}
\begin{equation}
  \bm{x} \to \bm{y} = \frac{\kappa \bm{x}}{\kappa + \mu \bm{1}_n^T\bm{x}},
\end{equation}
by direct computation of conditions \eqref{eq:end_point_conditions} we see that $\bm{y}$ is a non-invasible equilibrium. If we additionally restrict $\mu > 0$, $\bm{y}$ satisfies the feasibility property and $B$ is positive definite so again the support $\{S\}_m$ of the attractor is unchanged.

Finally, if all the species growth rates $r_i$ are positive, then the GLV model can be re-scaled to make all $r_i=1$ without changing the positive definiteness of matrix $A$. Therefore, all the GLV models with positive growth rates can be studied at once by setting $r_i=1$ for $i=1,\dots,n$.

%%% Local Variables:
%%% mode: latex
%%% TeX-master: "sup_info"
%%% End:

\section{Varying growth rates}

In this section we analyze the effect of growth rates that are not equal for all species. By continuity, we expect our results to hold when $\bm{r} = \bm{1}_n + \bm{\epsilon}_n$ and $\|\bm{\epsilon}_n\| \ll 1$ if $\ell \geq n$. In case $\ell < n$, the matrix $A$ is singular and the solutions of the system can be unbounded. To avoid this unrealistic behavior, we replace the interaction matrix $A$ by $B = A + \mu\bm{1}_n\bm{1}_n^T$, where $\mu$ is a sufficiently large perturbation so that $A_{ij} + \mu > 0$ for every matrix element. In this case $-B=-(A + \mu\bm{1}_n\bm{1}_n^T)$ is negative semidefinite and dissipative~\citep{hofbauer:1998}, so the solutions are always bounded. Still, the solutions can be degenerate in the sense that there is a hyperplane of non-invasible equilibria towards which the system converges. By perturbing the growth rates we can also correct for this. 

Assume now that $\bm{r} = \bm{1}_n + \N(0,\sigma^2)$, where $\sigma \ll 1$ and that $\bm{\hat{x}}$ is a saturated rest point of the system (which exists because $A_{ij}+\mu > 0$). Without lost of generality, we can assume that the first $m$ species survive. Then, we have

\vspace*{-3mm}
\begin{equation}\label{eq:lsys}
\bm{r} - B\bm{\hat{x}} = \begin{pmatrix} \bm{0}_m \\ \bm{z}\end{pmatrix}.
\end{equation}
For $\bm{z} \in \R^{n-m}_-$, if any $z_i = 0$, then for the system considering only the species $\{1,\ldots, m\}\cup\{i\}$ we have that the restriction of $\bm{r}$ to this subset of species (which is a vector of $m+1$ components) must be contained on a plane of dimension $m$: otherwise the linear system above yields the trivial solution $\bm{\hat{x}}=\bm{0}$. Since the distribution of $\bm{r}$ is continuous, the probability of this event is zero almost surely. Hence $z_i < 0$ for any $i$ so that invasibility is \emph{strict}: no species outside the set of survivors can invade. Furthermore, the same argument shows that the rank of $B$ restricted to the survivor subset must be $m$, i.e., the restriction of matrix $B$ to the set of coexisting species is \emph{full rank}. Otherwise, in order to satisfy the linear system, the restriction of vector $\bm{r}$ to the subset of survivors should be contained on a plane of dimension strictly less than $m$, which is a zero-probability event almost surely.

Apply the usual Lyapunov function for the system~\citep{hofbauer:1998},

\vspace*{-3mm}
\begin{equation}
V(\bm{x}) = \sum_{i=1}^n (x_i - \hat{x}_i \log x_i).
\end{equation}
Defined for any $\bm{x} \in \R^n_+$, with a global minimum at $\bm{x}=\bm{\hat{x}}$ and radially unbounded, then we have

\vspace*{-3mm}
\begin{equation}\label{eq:lyap}
\begin{aligned}
\dot{V}(\bm{x}) & = \sum_{i=1}^n \left(1-\frac{\hat{x}_i}{x_i}\right)\dot{x}_i
= \sum_{i=1}^n (x_i-\hat{x}_i)\Bigl(r_i-\sum_{j=1}^n B_{ij} x_j\Bigr)\\
& = -\sum_{ij=1}^n B_{ij}(x_i - \hat{x}_i)(x_j - \hat{x}_j)  
+ \sum_{i=1}^n (x_i - \hat{x}_i) \Bigl(r_i - \sum_{j=1}^n B_{ij} \hat{x}_j\Bigr)\\
& = -\sum_{ij=1}^n B_{ij}(x_i - \hat{x}_i)(x_j - \hat{x}_j)  
+ \sum_{i=m+1}^n x_i z_i.
\end{aligned}
\end{equation}
In the last equality we have used Eq.~\eqref{eq:lsys}, which implies that $r_i - \sum_{j=1}^n B_{ij} \hat{x}_j=0$ for $i=1,\dots,m$, together with the definition $z_i:=r_i - \sum_{j=1}^n B_{ij} \hat{x}_j$ and the equality $\hat{x}_i=0$, both of which hold for $j=m+1,\dots,n$. The first term above is non-positive since the matrix is $-B$ is negative semidefinite, and the second is non-positive because $z_i<0$ and every trajectory satisfies $x_i(t)\ge 0$ because the Lotka-Volterra system leaves invariant the space $\mathbb{R}^n_+$. This proves that $\dot{V}(\bm{x})\le 0$ for an arbitrary trajectory $\bm{x}(t)$.

Moreover, the last sum in~\eqref{eq:lyap} is negative unless $x_i = 0$ for any $i > m$. Given that the restriction of $B$ to the survivors subset is full rank, then $\dot{V} = 0$ only at $\bm{\hat{x}}$, which implies that the equilibrium point $\bm{\hat{x}}$ is globally stable and, in particular, is unique~\citep{hofbauer:1998}. Therefore, in the singular case $\ell < n$, and making the perturbation of the interaction matrix as $A\to B=A+\mu\bm{1}_n\bm{1}_n^T$, the dynamics will unfolds to a unique equilibrium point satisfying Eq.~\eqref{eq:lsys}.

In these cases, while our previous analyses are not exact because of the perturbations introduced in the vector of rates $\bm{r}$ and in interaction coefficients ($A\to B=A+\mu\bm{1}_n\bm{1}_n^T$), we can apply the same machinery that we have developed to provide approximations. This works because we know that the shift of $A \to A + \mu\bm{1}_n\bm{1}_n^T$ does not change properties like feasibility or invasibility (see section~\ref{section:lv_ops}). What changes is that the rank of $A$ goes up by one (see the observation at the end of the section). Forgetting about this, we can use the same machinery as in the non-degenerate case: for feasibility this follows because only full rank subsets are considered, and the restriction of a singular Wishart to a $m\times m$ block ($m \leq \ell$) is a Wishart matrix. Further, the conditional distribution of blocks used for the derivation of the probability of non-invasibility holds in the degenerate case too~\citep{bodnar2008properties}.

Observe that, in the degenerate case, matrix $B$ has rank equal to $\ell+1$, because $B = A + \mu \bm{1}_n\bm{1}_n^T$ and $A=\frac{1}{\ell}GG^T$ has rank $\ell$ since there are $\ell$ trait vectors linearly independent (see also the observation below). Therefore, at most $m=\ell+1$ species can have non-negative densities, according to the linear system~\eqref{eq:lsys}. Thus, the fraction of survivors $q=m/n$ can take, at most, the value $(\ell+1)/n = \gamma +1/n$, which sets $\gamma +1/n$ as an upper bound for the mode $q^*$ of the fraction of survivors. In the singular case it may happen that $q^*$ satisfying Eq.~\eqref{eq:saddle_point_mode_cond} is bigger than $\gamma + 1/n$. Given that we expect the distribution of the number of survivors to be unimodal and increasing with $\gamma$, then our approximation for the mode in those cases is simply $\gamma + 1/n$. Therefore, our analytical upper bound to the expected fraction of survivors $\Omega$ will be

\begin{figure}[!t]
\begin{center}
\includegraphics[width = 0.9\textwidth]{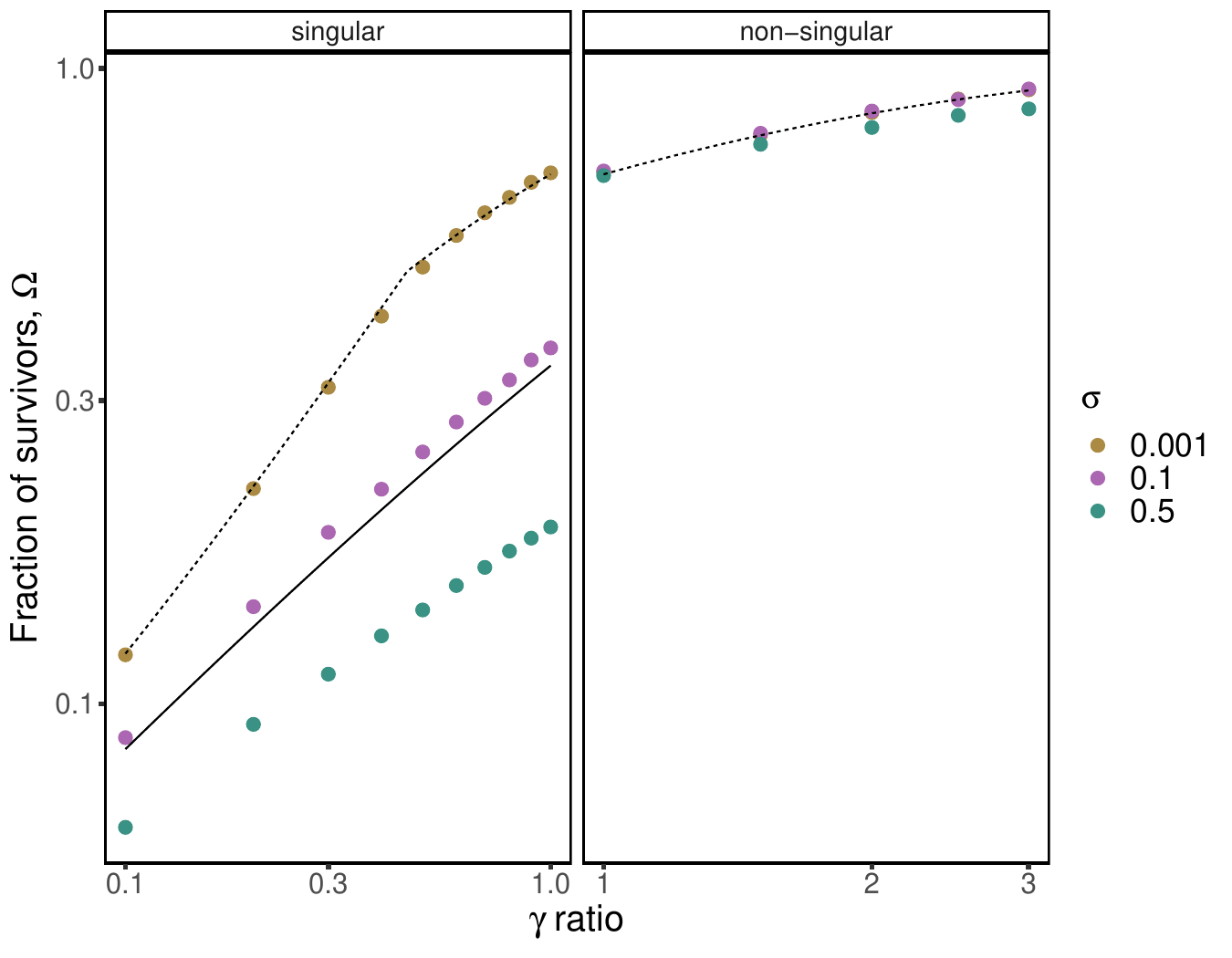}
\end{center}
\caption[Fraction of survivors under distinct levels of growth rate variability]{\label{fig:Sgr}
\textbf{Fraction of survivors under distinct levels of growth rate variability}. Dots mark the average values over simulations with $r \sim \N(1,\sigma^2)$ and $A \sim \W_n(\ell^{-1}I_n,\ell)$. In the singular case, the matrix $A$ was perturbed by $A \to A + (b + 0.01)\bm{1}_n\bm{1}_n^T$ for $b = -\min(A)$. Dotted lines represent our analytical predictions assuming $\sigma =0$. By \Cref{section:lv_ops} the shift in $A$ does not affect $\Omega$ when $\sigma = 0$. The initial decrease of $\Omega$ in the singular case is due to this property not holding when $\sigma \not=0$. The solid line is our analytical prediction for $\sigma = 0$, when $A \sim \W_n(\ell^{-1}\Sigma,\ell)$. $\Sigma$ is a constant correlation matrix with $\rho = \frac{2 \sigma_\ell + 0.01}{1 + 2 \sigma_\ell + 0.01}$ and $\sigma_\ell^2 = \text{Var}(A_{ij})$ for $i\not=j$ which in this case is simply $\sigma_\ell^2= 1/\ell$.}
\end{figure}

\vspace*{-3mm}
\begin{equation}
\Omega =\begin{cases}
\gamma +\frac{1}{n}, & \text{if } \gamma < \gamma_t,\\
q^{\star}(\gamma,0), & \text{if } \gamma \ge \gamma_t,
\end{cases}
\end{equation}
where $q^{\star}(\gamma,0)$ is given implicitly by Eq.~\eqref{eq:saddle_point_mode_cond} for $\rho=0$ (the non-interacting case is the most favorable for coexistence), and $\gamma_t$ is obtained by solving the non-linear equation $\gamma +\frac{1}{n}=q^{\star}(\gamma,0)$ to ensure continuity. These bounds are compared to averages over replicas of the set of coexisting species in Figure~\ref{fig:Sgr}.

{\bf Observation}. The rank of $B = A + \mu \bm{1}_n\bm{1}_n^T$ is equal to the rank of $A$ plus one. Indeed, let $\bm{w} \in \ker B$, then $\bm{w}^TB\bm{w} = \bm{w}^TA\bm{w} + \mu (\bm{1}_n^T\bm{w})^2 = 0$, hence $\bm{w}\in \ker A \cap \bm{1}_n^\perp$, and similarly any $\bm{w} \in \ker A \cap \bm{1}_n^\perp$ is in the kernel of $B$, hence $\ker B = \ker (A \cap 1^\perp)$. Unless $\ker A \subset \bm{1}_n^\perp$, $\dim (\ker B) = \dim (\ker A) - 1$, so the rank increases by one. It remains to show that $\ker A \not\subset \bm{1}_n^\perp$.

Consider then $A =CC^T$ for $C \in \R^{n\times \ell}$, and let $\{\bm{C}_i\}$ be the set of columns of matrix $C$. Then $\ker A$ is simply $U^\perp = \{\bm{C}_i\}^\perp$. As each column $\bm{C}_i$ is sampled independently from a continuous distribution then $W = \{\bm{C}_1,\ldots,\bm{C}_\ell,\bm{1}_n\}$ is a linearly independent set almost surely, then $\dim W^\perp = n - \ell -1$. Since $W^\perp = U^\perp \cap \bm{1}_n^\perp$, and $\dim U^\perp=n-\ell$ then $U^\perp$ cannot be contained in $\bm{1}_n^\perp$.

\subsection*{Deterministic limit}

We can also investigate the effect of varying growth rates in the deterministic limit, without limiting ourselves to the case of small deviations from equal rates. In the deterministic limit, the sample covariance matrix $A$ tends to the tree correlation matrix $\Sigma$, which is no longer random. Assume that the number of traits is large enough to approximate the interaction matrix by $\Sigma$ (see below for a numerical exploration of the rate of convergence as $\gamma\to\infty$). Let us assume also that growth rates $r_i$ are not equal, but drawn independently from a normal distribution $\mathcal{N}(\beta,\sigma^2)$ with mean $\beta$ and variance $\sigma^2$. Then the density function for the vector of rates $\bm{r}$ is

\vspace*{-3mm}
\begin{equation}\label{eq:normal}
f(\bm{r})=(2\pi\sigma^2)^{-\frac{n}{2}}e^{-\frac{1}{2\sigma^2}\Vert \bm{r} - \beta\bm{1}\Vert^2}.
\end{equation}
The case of a deterministic matrix and random (normally distributed) growth rates was considered in a previous paper by some of the authors~\citep{servan2018coexistence}. In that contribution we considered Gaussian rates with a given mean value and variance equal to one. The interaction matrix was deterministic, of the form $A=(\alpha-\mu)I+\mu\bm{1}\bm{1}^T$, with $\bm{1}$ a vector formed by $n$ ones. Therefore, we can extend the results provided in that paper for the case of star-tree phylogenies in the deterministic limit (with $\Sigma = (1-\rho)I + \rho \bm{1}\bm{1}^T$), with normally distributed growth rates with arbitrary variance.

We can directly use the formulae derived by~\cite{servan2018coexistence} because the case of arbitrary variance can be trivially reduced to the case of unitary variance through the change of variables $\bm{r'}=\bm{r}/\sigma$. In that case, the distribution~\eqref{eq:normal} reduces to
\begin{equation}
f(\bm{r'})=(2\pi)^{-\frac{n}{2}}e^{-\frac{1}{2\sigma^2}\Vert \bm{r'} - \beta'\bm{1}\Vert^2}
\end{equation}
with $\beta'=\beta/\sigma$. Therefore, the formulae derived for the probability of coexistence in that reference can be used for the case we are interested here, replacing the average value of growth rates by $\beta/\sigma$ (together with the replacements $\alpha\to 1$ and $\mu\to \rho$ to match the form of the interaction matrix $\Sigma$).

Following~\cite{servan2018coexistence}, the probability of observing an attractor of size $m$ from a pool of size $n$ can be computed as (cf. Eq.~(S26) in the Supplementary Information of that reference)
\begin{equation}\label{eq:Pa}
P_{\text{a}}(m,n)=\frac{m+s}{2\pi}\int_{-\infty}^{\infty}dy\int_{-\infty}^{\infty}dw\,
e^{-\frac{1}{2}(m+2s)y^2+isyw-\frac{1}{2}m w^2}\left[1-\Phi(iw-v)\right]^m
\left[\Phi(y-v)\right]^{n-m},
\end{equation}
where $\Phi(z)=\frac{1}{2}\left[1+\text{erf}\left(\frac{x}{\sqrt{2}}\right)\right]$ is the cdf of the standard normal distribution (which can be extended to the complex plane). In terms of model parameters, constants $s$ and $v$ are defined as
\begin{equation}\label{eq:sv}
s=\frac{1}{\rho}-1,\qquad v = \frac{\beta(1-\rho)}{\sigma(1-\rho+m\rho)}.
\end{equation}
The average fracion of coexisting species, $\Omega$, can be calculated by averaging the distribution
\begin{equation}\label{eq:Patt}
\text{Pr}(\{S\}_m|{n, \rho, \beta, \sigma}) = \binom{n}{m} P_{\text{a}}(m,n),
\end{equation}
i.e., by calculating
\begin{equation}\label{eq:Omega}
\Omega = \frac{1}{n}\sum_{m=0}^n m \text{Pr}(\{S\}_m|{n, \rho, \beta, \sigma}).
\end{equation}
The evaluation of the double integral and the calculation of the average fraction of survivors can be done numerically. We can also check that the limit $\sigma\to 0$, in which growth rates are all equal, leads to a distribution with probability one for $m=n$, and zero otherwise. To verify this, we note that if $\sigma\to 0$, then $v\to\infty$, and consequently $\Phi(y-v)\to 0$, so the only case for which~\eqref{eq:Pa} is nonzero occurs when $m=n$. Moreover, the term $\left[1-\Phi(iw-v)\right]^n\to 1$ as $v\to\infty$. Thus, the probability of the attractor reduces to
\begin{equation}
P_{\text{a}}(n,n)=\frac{n+s}{2\pi}\int_{-\infty}^{\infty}dy\int_{-\infty}^{\infty}dw\,
e^{-\frac{1}{2}(n+2s)y^2+isyw-\frac{1}{2}n w^2}=1,
\end{equation}
which follows by evaluating the Fourier transform in $w$. As mentioned, the limit $v\to\infty$ implies that $P_{\text{a}}(m,n)=0$ for $m<n$. Therefore, in this limit $\Omega=1$ and the model predicts full coexistence in the deterministic limit with equal growth rates (as we already proved in Section S2). However, this is no longer true when variability is considered in growth rates and we allow for non-negative values of $r_i$.

Applying a saddle-point technique to the double integral~\eqref{eq:Pa}, as we did above, one can get an approximation for $\Omega$ in the limit $n\to\infty$, computed as the mode of the distribution~\eqref{eq:Patt} in that limit. This calculation implies that $\Omega$ satisfies approximately the non-linear equation
\begin{equation}\label{eq:nlin}
\frac{\beta(1-\rho)}{\sigma[1 + \rho (n\Omega - 1)]} 
= \frac{e^{-[\Phi^{-1}(1-\Omega)]^2/2}}{\sqrt{2\pi}(\Omega+\frac{1}{n\rho})}
- \Phi^{-1}(1-\Omega),
\end{equation}
for $\Phi^{-1}(q)=\sqrt{2}\text{erf}^{-1}(2q-1)$ the inverse of the distribution function of the standard normal (as defined above). The numerical solution of this non-linear equation yields the parametric dependence of the average fraction of survivors, in the limit of large pool sizes, as function of $\sigma$ and $\rho$.  

\begin{figure}[t!]
\centering
\includegraphics[width = 0.8\textwidth]{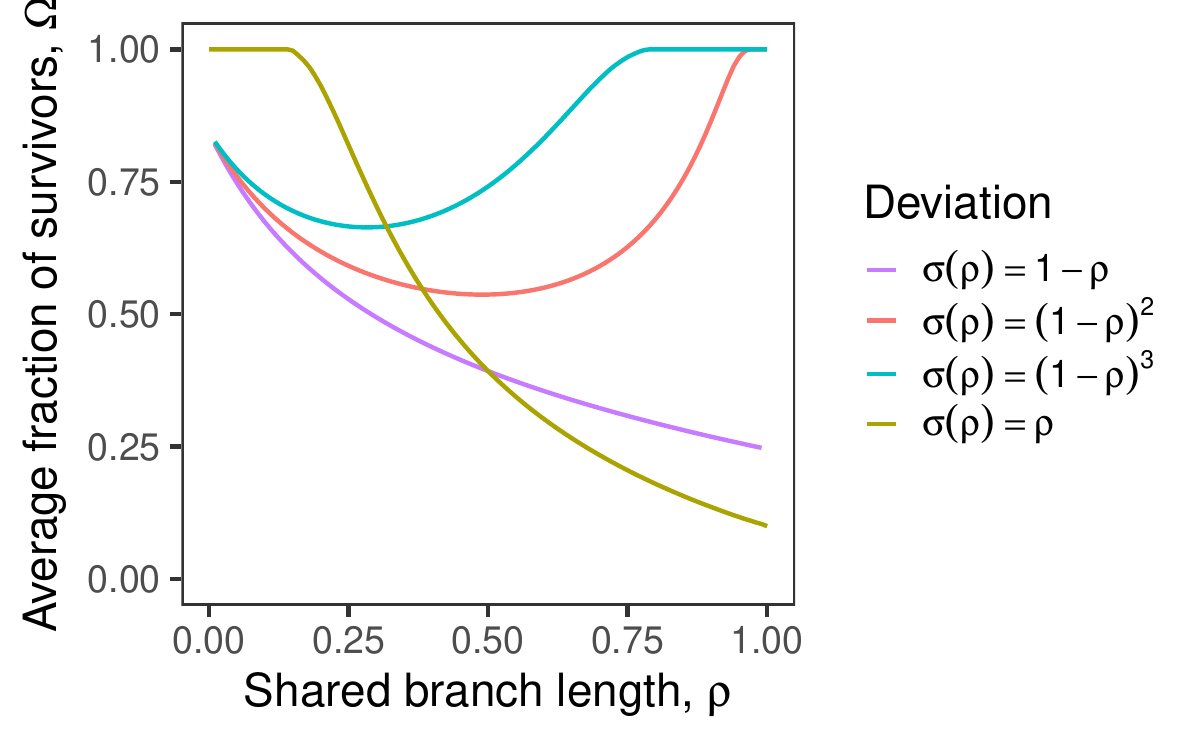}
\caption[Expected fraction of survivors for variable growth rates]{\textbf{Expected fraction of survivors for variable growth rates}. Here the average fraction of survivors ($n=10$), as calculated from Eq.~\eqref{eq:nlin}, is depicted as function of the inter-specific correlation level $\rho$ for a star-tree topology and different assumptions on how the variability of species growth rates, $\sigma$, depends on $\rho$.}
\label{fig:var_sigma}
\end{figure}

Figure~\ref{fig:var_sigma} shows the variation of $\Omega$ with $\rho$ for a mean value of rates $\beta=1$, to match with our previous assumption for $\sigma\to 0$ in the deterministic limit. If we make a correspondence between model parameters and niche/fitness differences (according to modern coexistence theory, see~\cite{chesson2000mechanisms}), we can make predictions, under the assumptions considered in this subsection, about the role of species' niche and fitness differences, and their relation with species phylogenetic proximity. In particular, the tree correlation matrix, written as $\Sigma=(1-\rho)I+\rho\bm{1}\bm{1}^T$, implies that the larger the correlation $\rho$, the more phylogenetically related species are, and the larger the competitive interaction between them because of higher niche overlap. Therefore, increasing species relatedness by augmenting $\rho$ lowers niche differences and increases competition (as in the Darwinian paradigm). Then, we expect higher diversity for small $\rho$ because of larger niche differences.

Once variability has been introduced in growth rates, fitness differences are present as well. One can begin to study an effect of phylogeny on fitness differences by assuming some dependence of the deviation $\sigma$ as function of $\rho$. In Figure~\ref{fig:var_sigma} we made several choices for this dependence: first, we chose a function $\sigma(\rho)=\rho$, so phylogenetic relatedness increases variability and, therefore, augments species fitness differences. In this case we observe that full coexistence is observed for small $\rho$, where niche differences dominate. Above a threshold of species relatedness, the fraction of survivors decreases monotonically. Therefore, this scenario would be compatible with phylogenetic overdispersion in realized communities. 

The second choice we made is $\sigma(\rho)=(1-\rho)^{\omega}$ for $\omega=1,2,3$. In those cases, as species relatedness ($\rho$) increases, variability decreases (and so fitness differences decrease as well). This behavior might be expected if clades of closely related species share similar growth rates. In the particular case $\sigma(\rho)=1-\rho$, we never observe full coexistence. This is because $v$ in Eq.~\eqref{eq:sv} is not divergent for all $0\le \rho \le 1$, and the probability of an attractor saturated with $n$ species satisfies $P_{\text{a}}(n,n)<1$ (this can be seen in Eq.~\eqref{eq:nlin} as well). However, for exponents $\omega>1$ we get full coexistence close to $\rho=1$. Therefore, as relatedness becomes closer to $\rho=1$, coexistence is favored because of equalizing fitness mechanisms (leading to clustering phenomena, as species selected in communities are similar in their growth rates). Interestingly, the behavior of the expected fraction of survivors at the attractor for $\omega>1$ is not monotonic anymore: for small $\rho$, niche differences dominate and decrease diversity, whereas at some point a minimum is reached and diversity starts growing with relatedness $\rho$ to achieve full coexistence once a threshold (close to $\rho=1$) is crossed over.

This subsection illustrates how our framework can help explain potential outcomes of the relation between phylogenetic proximity and species differences, and help elucidate which mechanisms (niche/fitness species differences) may promote diversity in the limit of large number of traits.

\section{Testing model predictions with empirical data}
\label{sec:senna}

\subsection*{Estimation of the number of traits}

Several large experiments have analyzed the interplay bteween biodivesity and ecosystem functioning. In particular, the Biodiversity II experiments~\citep{tilman2001diversity} report species identities and abundances from controlled plots, in which plants chosen from a pool of $n$ species were grown together in a wide variety of different combinations. These experiments have already been used to statistically test for phylogenetic structure in community properties~\citep{lemos2023phylogeny}, and therefore are well suited to test our model predictions. Biodiversity II experiments were carried out annually from 2001 to 2018, and we used them to examine our predictions based on the diversity reported in each realization of the experiment. We used the nine datasets pre-processed in~\citep{lemos2023phylogeny}, in order to exclude low-quality plots containing significant biomass of species not in the experimental pool. Those datasets were filtered to contain only plots for which the complete set of plants were seeded, to comply with our model assumptions (i.e., the entire pool is put together to coexist at arbitrary abundances and the GLV dynamics determines the set of species realized at equilibrium). These nine datasets contain between 13 and 20 species and a substantial number of species assemblages.

For each year the experiment was carried out, the experiment was replicated a number of times. We can consider each replica as a realization of a pool of $n$ species, as in our framework. For each replica, some species go extinct, and the realized diversity after the experiment is reported for each replica. Then we can compute the expected number of survivors as the average realized community size across replicas, $\langle m \rangle$, leading to a target fraction of survivors $\Omega^{\dagger}=\frac{\langle m\rangle}{n}$. We used the molecular phylogenetic tree that can be inferred for each pool of species (which varies depending of the year) using the \emph{V.PhyloMaker} package in R~\citep{jin2019v}. These ``true'' trees are also reported in~\cite{lemos2023phylogeny}.

\begin{figure}[t!]
\centering
\includegraphics[width = 0.9\textwidth]{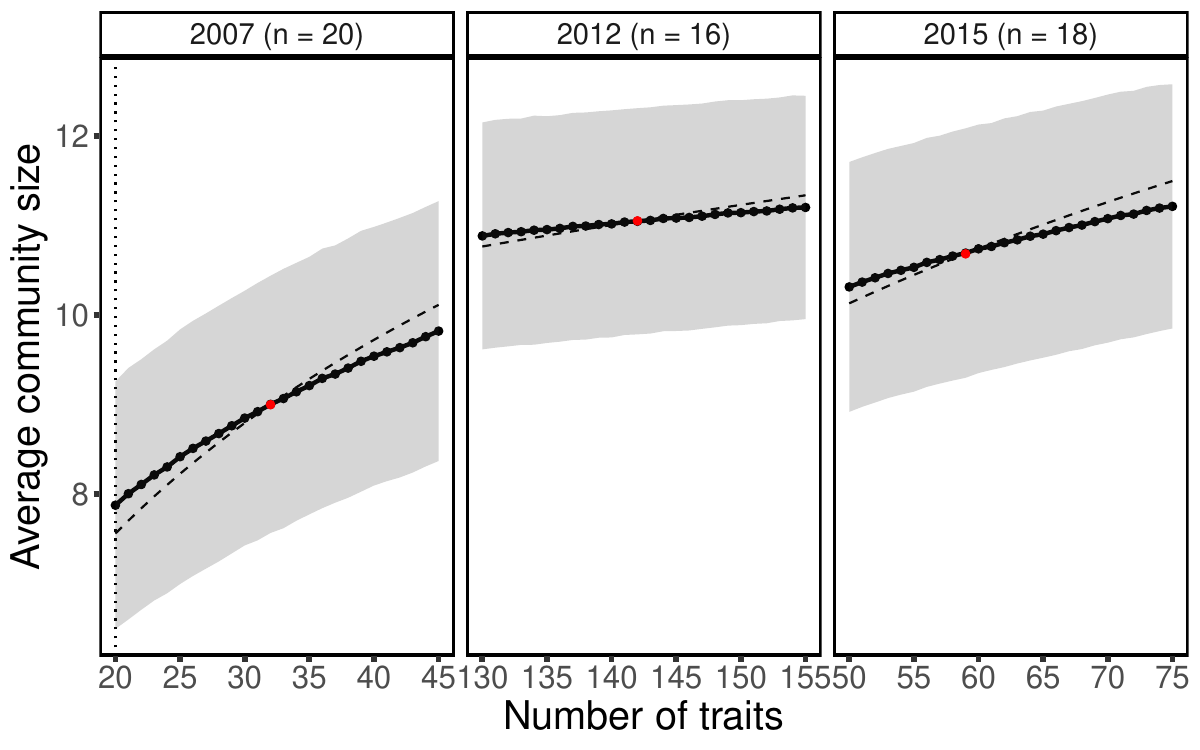}
\caption[Estimation of the number of traits associated to biodiversity data]{\textbf{Estimation of the number of traits associated to biodiversity data}. Model average community size (calculated using empirical trees) as function of the number of traits for three years of the Biodiversity II experiment (black line). The gray ribbon stands for one standard deviation added to simulated average. The red dot marks the empirical average community size, so the number of traits $\ell$ is chosen as the value that best matches model prediction to the target value. The dashed line shows the prediction for a star tree for which parameter $\rho$ is estimated to adjust the empirical average. The vertical, dotted lines marks the line $\ell = n$ ($\gamma=1$). %The empirical diversity in the 2007 experiment is best fit for $\ell = n$ (note the discontinuity in diversity once the threshold $\gamma=1$ is crossed over, which is also observed in Figure~\ref{fig:Sgr}).
}
\label{fig:exp1}
\end{figure}

Using the empirical tree structure, we simulated model communities for a given number of traits $\ell$, and averaged diversity across $10^6$ model realizations, for each of which we recorded the number of survivors at the endpoint of the dynamics. We varied the number of traits to match the observed diversity of each year's set of experiments ---the average diversity was calculated as the average number of species present across experiment's realizations. Then, to estimate an empirical number of traits, we chose $\ell^{\star}$ as the number of traits that best reproduced the target diversity. This procedure is illustrated in Figure~\ref{fig:exp1}. We allowed for singular Wishart matrices ($\gamma<1$) so that the number of traits can be smaller than the size of the pool. The estimate $\ell^{\star}$ never was smaller than the size of the pool and sometimes led to a relative number of traits $\gamma>9$ (see Table~\ref{tab:traits}). A star tree with correlation $\rho$ adjusted to match the empirical diversity at the fitted value $\ell^{\star}$ can reproduce the dependence of the average community size with the number of traits.

\begin{table}[t!]
    \centering
    \begin{tabular}{|l|ccccccccc|}
    \hline
    \textbf{Experiment year} & \textbf{2001} &  \textbf{2006} &  \textbf{2007} &  \textbf{2008} &  \textbf{2011} & \textbf{2012} &  \textbf{2014} &  \textbf{2015} &  \textbf{2017}  \\ \hline
    \textbf{Target diversity, $\Omega^{\dagger}$} & 0.725 & 0.507 & 0.450 & 0.559 & 0.535 & 0.691 & 0.672 & 0.594 & 0.603 \\
    \hline
    \textbf{Pool size, $n$} & 13 & 20 & 20 & 17 & 20 & 16 & 16 & 18 & 19 \\
    \hline
    \textbf{Estimated $\ell^{\star}$} & 128 & 52 & 32 & 58 & 54 & 142 & 120 & 59 & 78 \\ \hline
    \textbf{Estimated $\gamma=\ell^{\star}/n$} & 9.85 & 2.60 & 1.60 & 3.41 & 2.70 & 8.87 & 7.50 & 3.28 & 4.11 \\ \hline
    \end{tabular}
    \caption{Results of the parameter estimation procedure to fit Biodiversity II experimental data. The estimated number of traits $\gamma$ relative to pool size is larger than one.}
    \label{tab:traits}
\end{table}

As Table~\ref{tab:traits} shows, the estimated number of traits is larger than twice the size of the pool $n$ in most of the cases, with some exceptions (years 2001, 2012, and 2014), in which the average fraction of survivors is significantly large in communities and leads to a larger estimate for $\ell^{\star}$. 

\begin{figure}[t!]
\centering
\includegraphics[width = 0.55\textwidth]{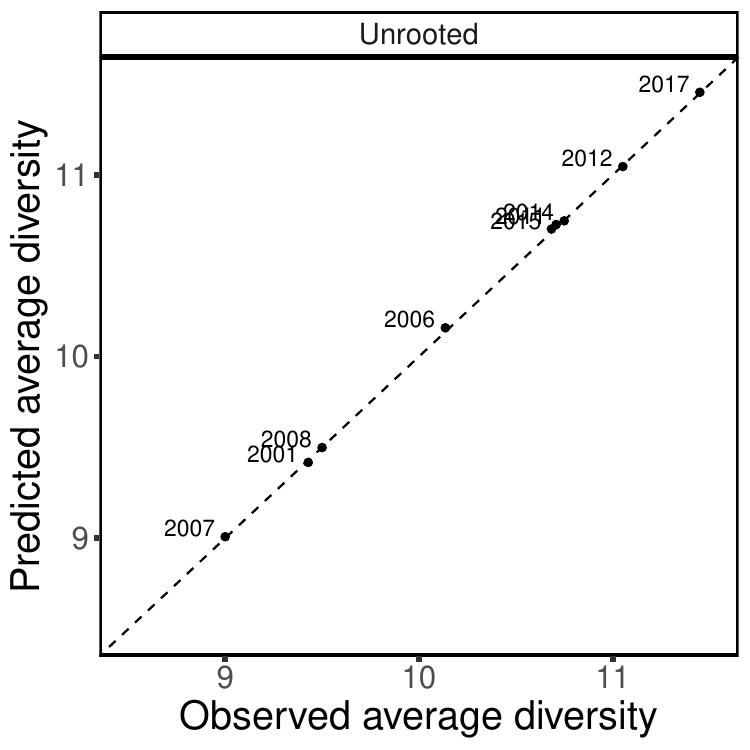}
\caption[Predicted \emph{vs.} observed average diversity across experiments]{\textbf{Predicted \emph{vs.} observed average diversity across experiments}. The number of traits $\ell^{\star}$ is estimated so that model average diversity best matches the empirical average (see also Table~\ref{tab:traits}). Trees used along the fitting procedure are unrooted, as returned by \emph{V.PhyloMaker}. %Right panel: for the 2007 experiment, the tree covariance matrix $\Sigma$ is corrected for the tree to be rooted, and the root length is estimated to improve the fitting of observed average community size.
}
\label{fig:exp2}
\end{figure}

Figure~\ref{fig:exp2} shows predicted \emph{vs.} observed average diversity, according to this fitting procedure. Average diversity in communities is well predicted by this simple estimation procedure. %The exact empirical average diversity for the 2007 experiment cannot be reproduced due to the discontinuity in diversity averages at $\gamma=1$, see Figures~\ref{fig:Sgr} and~\ref{fig:exp1}. The diversity predicted by our model in this case can be improved, though. As empirical trees were obtained as unrooted trees (see Figure~7, main text), and the molecular procedure cannot estimate the root length before the first speciation, we can take it as a free parameter to better estimate observed diversity. For that purpose, we modify the correlation matrix of the tree by adding a constant $z$ to all the matrix elements, as $\Sigma_1 = \Sigma + z \bm{1}\bm{1}^T$, and then normalize $\Sigma_1$ to make it ultrametric again, as $\Sigma_2 = D_1^{-\frac{1}{2}}\Sigma_1 D_1^{-\frac{1}{2}}$, where $D_1=\text{diag}(\Sigma_1)$ and hence $D_1^{-\frac{1}{2}}$ is a diagonal matrix with the inverses of square roots of the diagonal elements of $\Sigma_1$. This modification amounts to adding a branch on length $z$ at the root of the tree. For the 2007 experiment we find a fitted $z^{\star}=16.8$. The right panel of Figure~\ref{fig:exp2} shows that introducing this additional (unknown) root length in the 2007 empirical tree allows to predict almost perfectly observed diversity.

%From now on, since it involves the estimation of one additional parameter, we will use the first fitting procedure based on unrooted trees.

\subsection*{Diversity and biomass}

To test the influence of realistic tree structures on the quantities we have analyzed in this contribution, we fixed phylogenetic trees as those derived from species identities using the \emph{V.PhyloMaker} R package, and simulated the model to compute diversity and biomass. Predicted average community size and total biomass were calculated as function of the number of traits, setting matrix $\Sigma$ as the correlation matrix for the ultrametric empirical phylogenetic tree estimated by molecular methods. Model simulations were carried out by imposing the empirical phylogenetic structure, so that we can compare with the predictions from a star tree approximating the true tree topology. Figure~\ref{fig:exp1} shows this comparison for three years in Biodiversity II experiments, and highlights that a star tree with an average correlation $\rho$ matching the empirical diversity can capture the trend obtained by using the true phylogenetic structure.

\begin{figure}[t!]
\centering
\includegraphics[width = 0.9\textwidth]{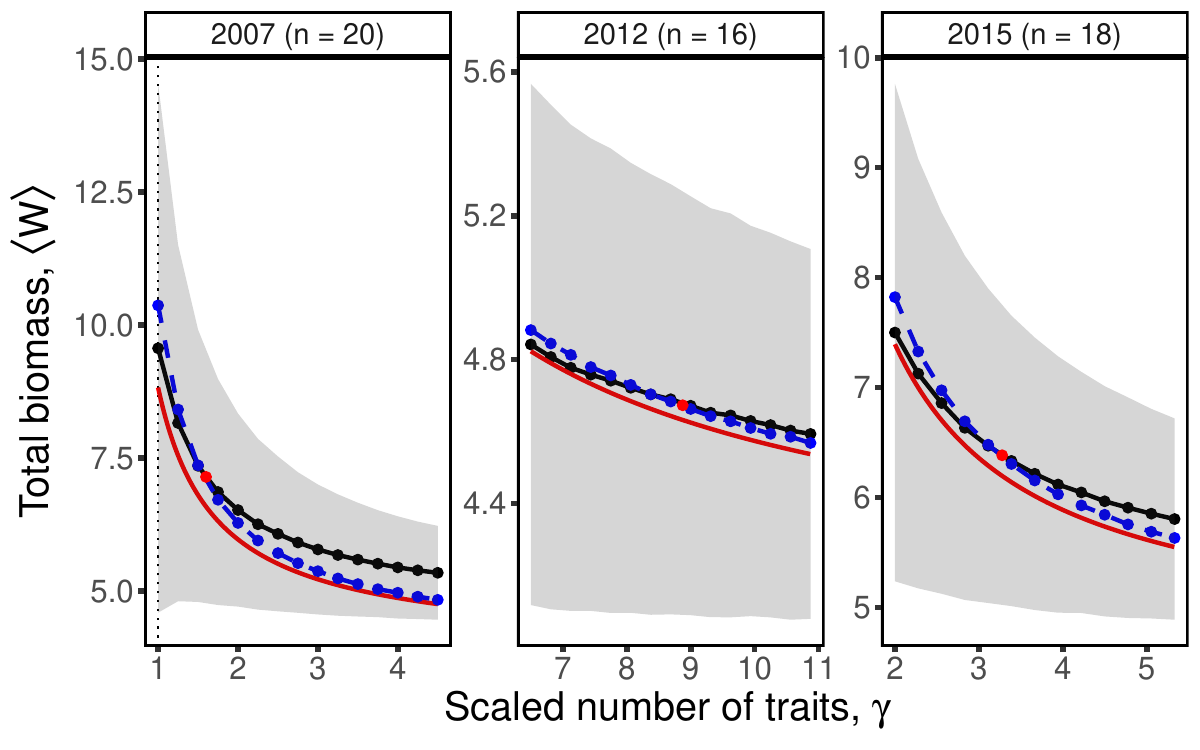}
\caption[Total biomass for fixed, empirical tree structures]{\textbf{Total biomass for fixed, empirical tree structures}. Black lines and dashed areas show the average (plus and minus one standard deviation) across model realizations when the tree structure is fixed to match the molecular tree of each species pool. The blue, dashed line stands for the expectation for a star tree topology with correlation $\rho$ fitted to match model total biomass for the estimated number of traits $\ell^{\star}$ (marked with red dots). The red line shows the analytical approximation given by Eq.~\eqref{eq:bio_app}.}
\label{fig:exp3}
\end{figure}

Total biomass was computed by model simulation based on empirical correlation matrices for the  tree structure derived using molecular methods. The results are summarized in Figure~\ref{fig:exp3}. Again, a star tree topology, with an average correlation $\rho$ fitted to match model expected biomass at the predicted number of traits, captures qualitatively the decreasing pattern of total biomass as the number of traits increase. The analytical approximation given by Eq.~\eqref{eq:bio_app} provides a lower bound to total biomass, for the true phylogenetic structure as well as for the approximated star-tree topology. This approximation, originally constructed for star trees, captures well (up to a translation) the variation of total biomass with $\gamma$ for realistic trees.

\begin{table}[t!]
    \centering
    \begin{tabular}{|c|ccc|}
    \hline
    \textbf{Experiment year} & \textbf{2007} & \textbf{2012} & \textbf{2015} \\ \hline
    \textbf{$\rho$, diversity estimation} & 0.308 & 0.557 & 0.337 \\ 
    \textbf{$\rho$, biomass estimation} & 0.220 & 0.192 & 0.171 \\ \hline
    \end{tabular}
    \caption{Fitted values of star-tree correlation parameter $\rho$, for the estimation of expected diversity and for the estimation of total biomass.}
    \label{tab:rho}
\end{table}

Table~\ref{tab:rho} lists the fitted $\rho$ values, either to capture average diversity or total biomass. Matching the total biomass expected for empirical trees requires correlation values smaller than those required for adjusting expected average diversity.

\subsection*{Deterministic limit}

\begin{figure}[t!]
\centering
\includegraphics[width = \textwidth]{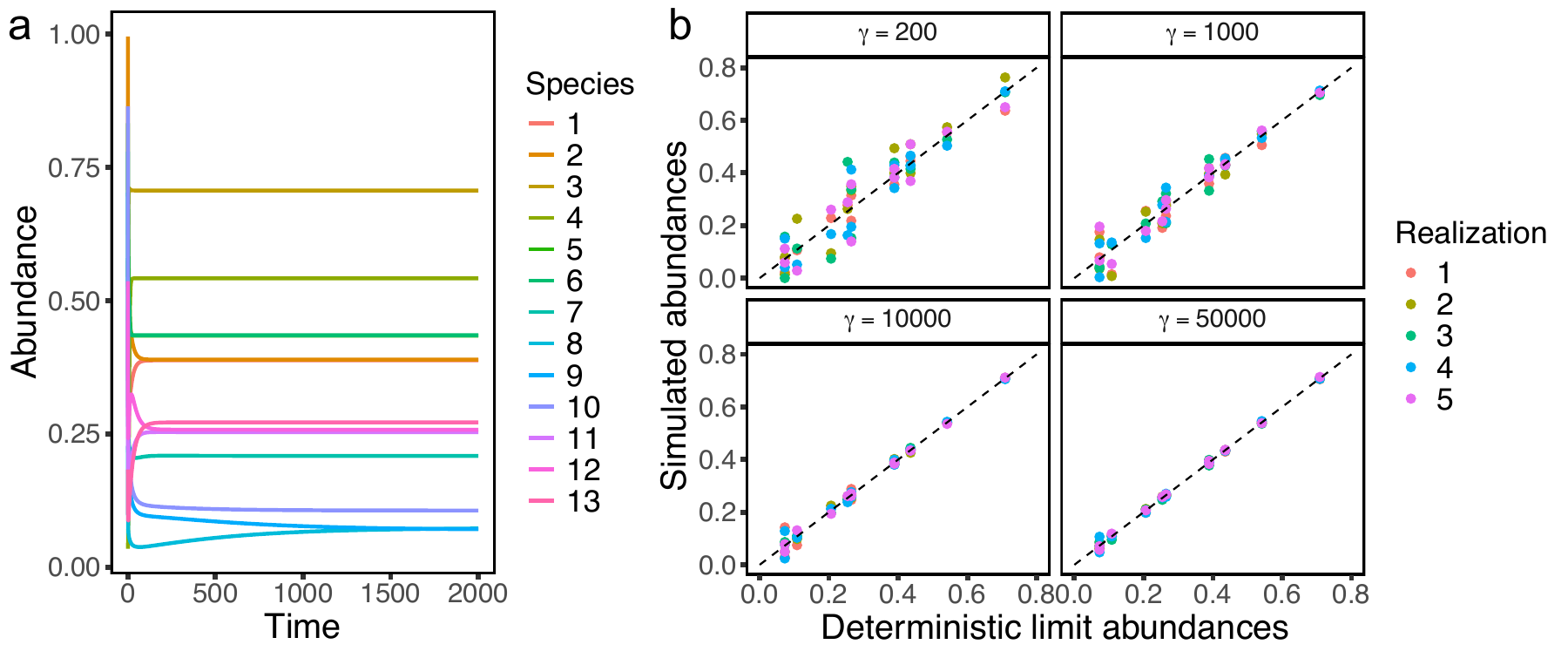}
\caption[Convergence to the deterministic limit abundnaces]{\textbf{Convergence to the deterministic limit abundances}. (a) For the 2001 Biodiversity II experiment, we show a temporal dynamics for the pool of $n=13$ species and $\gamma=5\times 10^4$, showing that full coexistence is ensured in the long term. (b) Deterministic-limit abundances, obtained by solving $\Sigma\bm{x^{\star}}=\bm{1}$ for $\bm{x^{\star}}$, are compared to 5 simulated abundances for increasing values of $\gamma$. As $\gamma\to\infty$, abundances converge to their expectation, although the rate of convergence is slow.}
\label{fig:exp4}
\end{figure}

To illustrate how abundances converge to deterministic-limit abundances as the number of traits increase, we have simulated GLV dynamics using realistic tree structures from Biodiversity II experiments. Figure~\ref{fig:exp4}a shows the temporal variation of abundances for the 2001 experiment, starting from a random initial condition, and for $\gamma=5\times 10^4$ to ensure proximity to the limit $\ell\to\infty$ (in this experiment, the species pool comprises $n=13$ species). As the temporal dynamics shows, full coexistence is ensured at equilibrium.

Predicted abundances in the limit $\gamma\to\infty$ can be obtained simply by solving the linear system $\Sigma\bm{x^{\star}}=\bm{1}$ for vector $\bm{x^{\star}}$, being $\Sigma$ the correlation matrix associated to the 2001 species pool estimated by molecular methods. Figure~\ref{fig:exp4}b shows how the abundances simulated for 5 interaction matrices realizations (for fixed values of $\gamma$) converge to deterministic-limit abundances $\bm{x^{\star}}$ as $\gamma$ increases. The convergence to the vector of abundances takes place for very large values of the scaled number of traits $\gamma$. Do we expect to reach full coexistence for such large values of the number of traits?

\begin{figure}[t!]
\centering
\includegraphics[width = \textwidth]{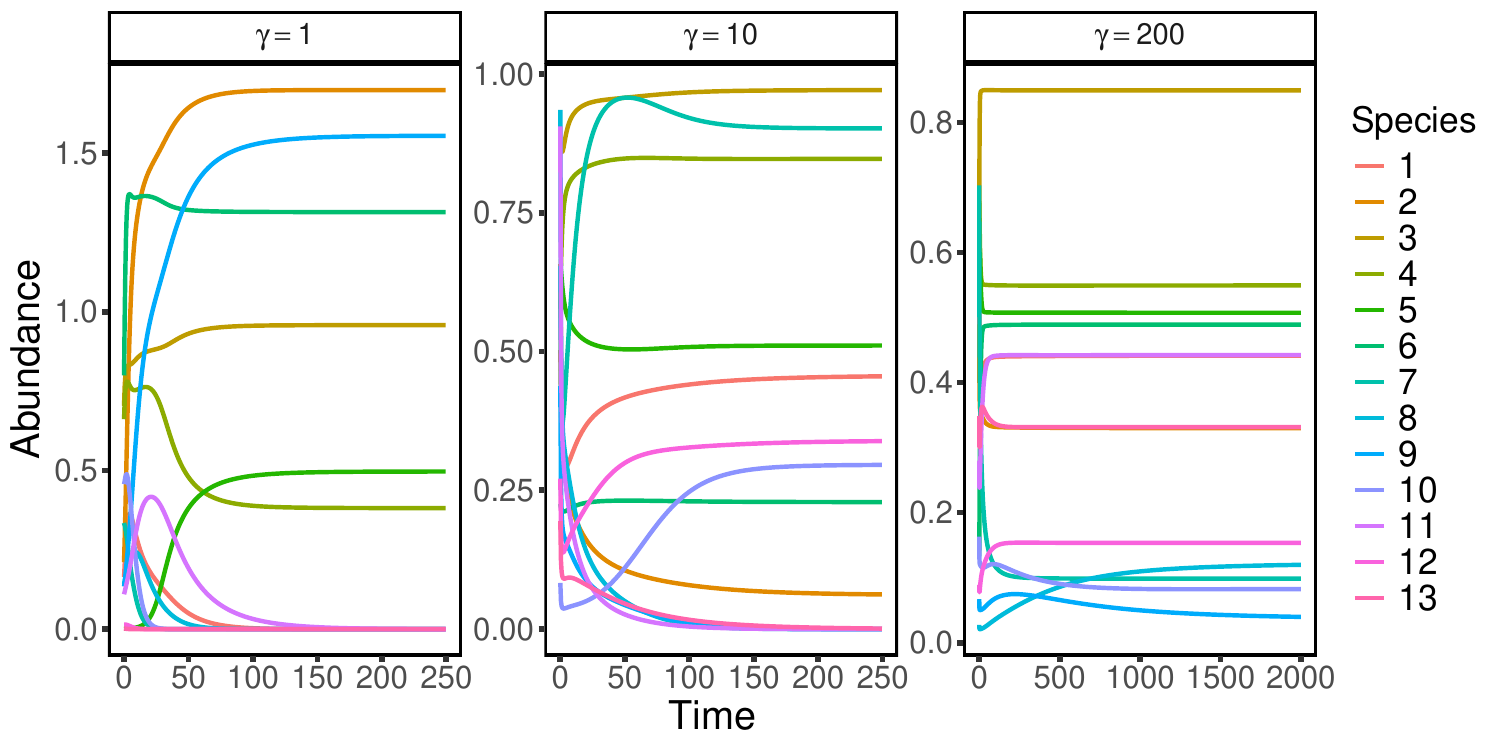}
\caption[Full coexistence in the deterministic limit]{\textbf{Full coexistence in the deterministic limit}. Temporal dynamics the 2001 Biodiversity II phylogeny ($n=13$), as the relative number of traits $\gamma$ increases. Each panel shows the time dynamics for a single realization of the Wishart matrix, given the number of traits $\ell=\gamma n$, setting $\Sigma$ equal to the correlation matrix obtained from the empirical phylogeny. For small number of traits, we observe species extinctions (panels $\gamma=1$ and $\gamma=10$). For larger number of traits ($\gamma=200$), full coexistence is achieved. }
\label{fig:exp5}
\end{figure}

Figure~\ref{fig:exp5} shows three temporal dynamics for increasing values of $\gamma$, using the 2001 experiment phylogeny. We observe how extinctions tend to be less frequent as $\gamma$ increase. Full coexistence is guaranteed well below the values of $\gamma$ at which simulated abundances almost coincide with deterministic-limit abundances (Fig.~\ref{fig:exp4}). As Figure~\ref{fig:exp5} illustrates, for $\gamma=200$ almost any realization of the Wishart interaction matrix $A$ leads to full coexistence. However, abundances at the endpoint of the dynamics do not necessarily match the ones predicted by the deterministic limit.

\subsection*{Survival probability for the Senna tree}

In the main text we have analyzed how species in realistic trees are more ore less frequent to be observed in model communities, and this basically depends on tree structure: species that diverged first in the evolutionary history are more prone to be observed in realized communities. We have shown this phenomenon using the empirical phylogenies used to represent Biodiversity II experiments' species pools (see Figure~7 of the main text). In this subsection, we confirm the pattern using a larger tree.

\begin{figure}
\includegraphics[width = \textwidth]{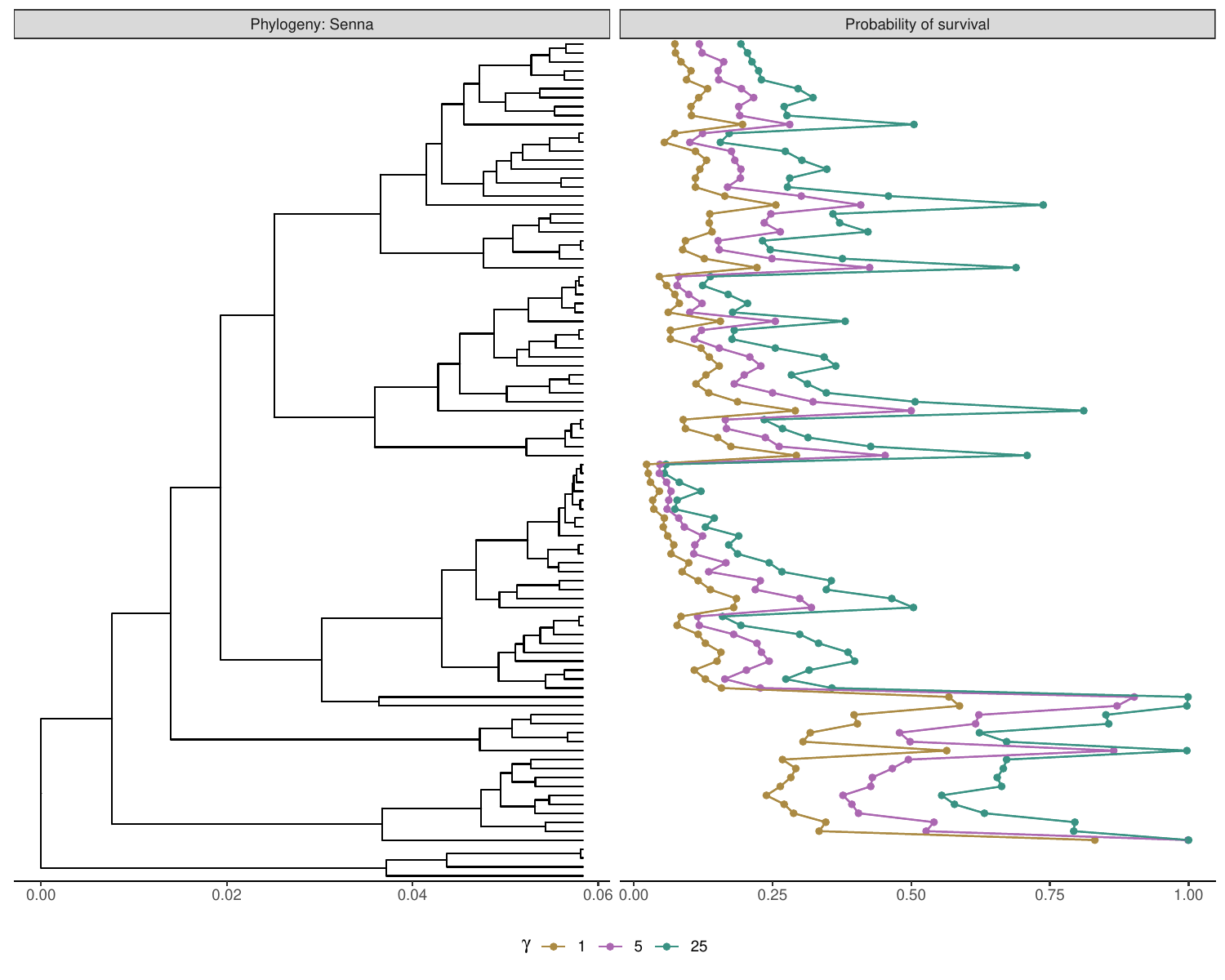}
\caption[Probability of individual species survival for an empirical tree]{\textbf{Probability of individual species survival for an empirical tree}. The probability that a species is observed in the community of coexisting species, $p_s$, out of 5000 simulations, is shown alongside the phylogenetic tree (\emph{Senna} clade) where the outermost group is used to set the root. The values $p_s$ reflect the tree structure and the abundance distribution showed in Figure 3 of the main text: The peaks in $p_s$ correspond to outliers within groups of closely related species, and $p_s$ has a decreasing trend towards the most nested parts of the tree (upward direction). In particular, the model produces phylogenetic overdispersion at multiple cladistic levels in the phylogeny (i.e., for subtrees the overdispersion effect is observed, as well as for the whole tree).}
\label{fig:senna}
\end{figure}

We consider here the Senna phylogenetic tree (Figure~\ref{fig:senna}), being the regional pool formed by $n=90$ species. The covariance matrix of the pool is denoted as $\Sigma_S$. %To construct this covariance matrix we assume equal inter-branching times, so we consider only the effect of tree topology in this section. 
For different values of the number of traits relative to pool's size, $\gamma=\ell/n$, we obtain different interaction matrices $A$ as samples of the Wishart distribution given by Eq.~(3) of the main text, $A\sim \mathcal{W}_n(\ell^{-1}\Sigma_S,\ell)$. 

We can measure the probability of an individual species survives in the saturated equilibrium point, which we name as $p_s$, and estimate it as the frequency of that species appearing with non-zero abundance in every realized community within a sample of replicas of the interaction matrix $A$, see Figure~\ref{fig:senna}. We observe that outliers within groups of closely related species, i.e., those species that diverged first compared with their close relatives in the tree, are the most frequently appearing species in communities among $5000$ replicas of the interaction matrix. Survival probabilities tend to decrease for species that diverged later in the tree, and this pattern is consistent for different values of the number of traits relative to the size of the pool, $\gamma=\ell/n$. This can be interpreted of a signal of phylogenetic overdispersion, because our model implies that closely-related species will compete strongly among each other and, therefore, will be less frequent in realized communities. We have quantified this effect by measuring the (Spearman) correlation $\rho_S$ between $p_s$ and the average phylogenetic distance for each species, defined as the average distance between that species and the remaining ones across the tree. This yields the following results: $\rho_S=0.816$ ($\gamma=1$), $\rho_S=0.817$ ($\gamma=5$), and $\rho_S=0.809$ ($\gamma=25$), all of them statistically significant ($p<10^{-16}$). This means that closely related species are not frequent in realized communities, yielding phylogenetic overdispersion in the set of survivors.

%%% Local Variables:
%%% mode: latex
%%% TeX-master: "sup_info"
%%% End:

\clearpage

\newpage{}
\bibliographystyle{cbe}
\bibliography{references}

\end{document}